\documentclass[a4paper,11pt]{article}
\pdfoutput=1 

\usepackage{jheppub} 

\usepackage[T1]{fontenc} 
\usepackage{amsmath}
\usepackage{subfigure}
\usepackage{verbatim}
\usepackage{setspace}

\usepackage{array,arydshln} 
\notoc
\newcommand{\Kdf}[0]{\cK_{\df,3}}
\newcommand{\K}[0]{\mathcal K}
\newcommand{\cD}[0]{\mathcal D}
\newcommand{\cK}[0]{\mathcal K}
\newcommand{\cM}[0]{\mathcal M}
\newcommand{\cO}[0]{\mathcal O}
\newcommand{\cY}[0]{\mathcal Y}
\newcommand{\df}[0]{\mathrm{df}}
\newcommand{\spect}[0]{\mathrm{spect}}
\newcommand{\fr}[0]{\mathrm{free}}
\newcommand{\iso}[0]{\mathrm{iso}}
\newcommand{\thr}[0]{\mathrm{thr}}
\newcommand{\Kdfiso}[0]{{\cK^{\iso}_{\df,3}}}
\newcommand{\Kiso}[0]{{\cK^{\iso}_{\df,3}}}
\newcommand{\Kisoone}[0]{{\cK^{\iso,1}_{\df,3}}}
\newcommand{\Kisotwo}[0]{{\cK^{\iso,2}_{\df,3}}}
\newcommand{\KA}[0]{{\cK^{(2,A)}_{\df,3}}}
\newcommand{\KB}[0]{{\cK^{(2,B)}_{\df,3}}}
\newcommand{\tr}[0]{\mathrm{tr}}
\newcommand{\wt}[0]{\widetilde}

\newcommand{\HSTH}[0]{Hansen:2016fzj}
\newcommand{\BHSQC}[0]{Briceno:2017tce}
\newcommand{\BHSnum}[0]{Briceno:2018mlh}
\newcommand{\BHSK}[0]{Briceno:2018aml}
\newcommand{\HSQCa}[0]{Hansen:2014eka}
\newcommand{\HSQCb}[0]{Hansen:2015zga}

\newcommand{\Akakia}[0]{Hammer:2017uqm}
\newcommand{\Akakib}[0]{Hammer:2017kms}

\newcommand{\Luscher}[0]{Luscher:1986n2,Luscher:1991n1}
\newcommand{\MD}[0]{Mai:2017bge}
\newcommand{\MDpi}[0]{Mai:2018djl}

\DeclareMathOperator{\Tr}{Tr}
\newcommand{\Z}{\mathbb{Z}}
\newcommand{\iy}{\infty}
\newcommand{\mc}[1]{\mathcal{#1}} 
\newcolumntype{C}{>{$}c<{$}} 

\title{\boldmath Implementing the three-particle quantization condition including higher partial waves}

\author[a]{Tyler D. Blanton,}
\author[b]{Fernando Romero-L\'opez }
\author[a]{ and Stephen R. Sharpe }

\affiliation[a]{Physics Department, University of Washington, Seattle WA 98195-1560, USA}
\affiliation[b]{IFIC, CSIC-Universitat de Val\`encia, 46980 Paterna, Spain}

\emailAdd{blanton1@uw.edu}
\emailAdd{fernando.romero@uv.es}
\emailAdd{srsharpe@uw.edu}

\abstract{We present an implementation of the relativistic three-particle quantization condition
including both $s$- and $d$-wave two-particle channels.
For this, we develop a systematic expansion of the three-particle K matrix, $\Kdf$,
about threshold, 
which is the generalization of the effective range expansion of the two-particle K matrix, $\K_2$.
Relativistic invariance plays an important role in this expansion.
We find that $d$-wave two-particle channels enter first at quadratic order.
We explain how to implement the resulting multichannel quantization condition,
and present several examples of its application.
We derive the leading dependence of the threshold three-particle state on
the two-particle $d$-wave scattering amplitude, and use this to test our implementation.
We show how strong two-particle $d$-wave interactions can lead to significant effects
on the finite-volume three-particle spectrum, including the possibility of a generalized
three-particle
Efimov-like bound state. We also explore the application to the $3\pi^+$ system,
which is accessible to lattice QCD simulations,
where we study the sensitivity of the spectrum to the components of $\Kdf$.
Finally, we investigate the circumstances under which the quantization condition
has unphysical solutions.}

\begin{document} 
\maketitle
\clearpage
\tableofcontents

\flushbottom

\section{Introduction}
\label{sec:intro}

There has been considerable recent progress developing the formalism
necessary to extract the properties of resonances coupling to three-particle 
channels from simulations of lattice QCD, with three different approaches being 
followed~\cite{\HSQCa,\HSQCb,\Akakia,\Akakib,\BHSQC,\MD,\BHSK}.
For a recent review, see Ref.~\cite{HSreview}.
The outputs of this work are quantization conditions,
which relate the finite-volume spectrum with given quantum numbers to
the infinite-volume two- and three-particle interactions.
This development is timely since simulations now have extensive results for the 
finite-volume spectrum above the three-particle threshold;
see, e.g., Refs.~\cite{Dudek:2013yja,Bulava:2016mks,Romero-Lopez:2018rcb} 
and the recent review in Ref.~\cite{Briceno:2017max}.
Turning the formalism into a practical tool remains, however, a significant challenge.
To date, this has been done only for the simplest case, in which all particles are
spinless and identical, the total momentum vanishes,
 the two-particle interaction is purely $s$-wave, and three particles
interact only via a momentum-independent contact 
interaction~\cite{\Akakib,\BHSnum,\MD,Doring:2018xxx,\MDpi}.\footnote{%
There is also an induced three-particle interaction due to the exchange of a virtual
particle between a pair of two-particle interactions. 
This is included in all approaches.}
This is the analog in the three-particle system of the initial implementations of
the two-particle quantization condition of L\"uscher~\cite{\Luscher}, which assumed only
$s$-wave interactions and vanishing total momentum.

In the two-particle case, such an approximation makes sense for levels close 
to the two-particle threshold, since higher partial waves are suppressed by 
powers of the relative momentum. In the meson sector it 
begins to fail for energies around $1\;$GeV. 
Indeed, recent applications of the two-particle
quantization condition use multiple partial 
waves (see, e.g., Refs.~\cite{Andersen:2017una,Woss:2018irj}).
Similar considerations apply for three particles, and we expect
that for many resonances of interest one will need to include higher partial waves.

The aim of this paper is to take the first step in this direction by including the
first higher partial wave that enters in the case of identical, spinless particles, namely the
$d$ wave.\footnote{%
The $p$ wave is absent due to Bose symmetry.}
 In the language of Refs.~\cite{\Akakia,\Akakib,\MD}, we include dimers
(two-particle channels) with both $\ell=0$ and $\ell=2$. At the same time, 
for consistency, we make a corresponding extension of
 the three-particle interaction beyond its local (pure $s$-wave) form.
 We will explain how to implement the formalism in this generalized setting,
 and show examples for which the higher-order terms have a significant impact
 on the finite-volume spectrum.
 
 Three-particle quantization conditions have been developed with three different
 approaches. These use, respectively,  generic relativistic effective field  
 theory analyzed diagrammatically to all orders in perturbation theory 
 (the RFT approach)~\cite{\HSQCa,\BHSQC,\BHSK}, 
 non-relativistic effective field theory (NREFT)~\cite{\Akakia,\Akakib},
 and unitarity constraints on the two- and three-particle S-matrix elements applied to 
 finite-volume amplitudes (the finite-volume unitarity or FVU approach)~\cite{\MD}.
To date, only in the RFT approach has the formalism been worked out explicitly with no
 limitations on the two-particle partial waves, whereas in the other two approaches
 the quantization condition has been written down only for $s$-wave dimers.\footnote{%
 It is expected, however, that there is no barrier to extending to higher waves.
}
Therefore we adopt the RFT approach in this work. 
Specifically, we use the formalism of Ref.~\cite{\HSQCa},
which applies to identical, spinless particles, with a $G$-parity-like $\Z_2$
symmetry that forbids $2\leftrightarrow 3$ transitions.
Another important feature of this approach is that it can be made relativistic~\cite{\BHSQC},
which turns out to simplify the expansion about threshold.
Although we use the RFT approach,
we expect that many of the technical considerations and general
conclusions will apply to all three approaches to the quantization condition.

The formalism of Ref.~\cite{\HSQCa} is restricted to two-particle interactions that
do not lead to poles in $\K_2$, the two-particle K matrix. If there are such poles,
then one should use the generalized, and more complicated,
 formalism derived in Ref.~\cite{\BHSK}.
For simplicity, we consider here only examples in which there are no K-matrix poles.

Since our main goal is to  show how the formalism works
when including higher waves, our numerical examples are mainly
chosen for illustrative purposes and do not represent physical systems.
However,  there is one case in nature for which our simplified setting applies,
namely the $3\pi^+$ system.
Thus, in one of our examples, we
set the two-particle scattering parameters to those measured experimentally for 
two charged pions, and illustrate the dependence of the
resulting three-pion spectrum on the three-particle scattering parameters.
This is similar to the study made in Ref.~\cite{\MDpi} using the
FVU approach, except here we include $d$-wave dimers.

All three-particle quantization conditions involve an intermediate three-particle scattering
quantity that is not physical, but that can be related, in a second step, to the infinite-volume
scattering amplitude by solving integral equations. 
In the RFT formalism this quantity is called $\Kdf$, and the second step is explained
in Ref.~\cite{\HSQCb}.
We do not discuss the implementation
of this second  step in the present work. 
Clearly, it will be important to do so in the future, but the methods required are quite
different from those needed for the quantization condition.

This paper develops the ideas already sketched in Sec.~4 of Ref.~\cite{Blanton:2018guq}. 
It is organized as follows.
In the next section we recall the quantization condition of Ref.~\cite{\HSQCa},
and explain how one can consistently expand $\Kdf$ about the three-particle
threshold, with $d$-wave interactions entering at quadratic order.
In Sec.~\ref{sec:implementation} we describe the implementation of the quantization condition including $d$-wave interactions, focusing on how to make use of the factorization into
 different irreducible representations (irreps) of the cubic group.
Subsequently, in Sec.~\ref{sec:results} we show results illustrating
 the effect of $d$-wave interactions on the three-particle spectrum, 
 including in Sec.~\ref{sec:pipipi} the case of the $3\pi^+$ system with realistic interactions,
which is a target for a potential lattice QCD study.
In addition, in Sec.~\ref{sec:unphysical}, we address the issue of characterizing unphysical solutions to the quantization condition. 
We summarize and close the discussion in Sec.~\ref{sec:conc}. 

We also include seven appendices describing technical details.
 Appendix~\ref{app:defns} is a collection of relevant definitions,
 whereas Appendices~\ref{app:Ft} and \ref{app:projections} provide further details
 concerning the topics of Sec.~\ref{sec:implementation}.
 Appendix~\ref{app:thra2} describes the calculation of the leading contribution
 of $d$-wave scattering to the threshold expansion.
 Finally, the remaining appendices relate to the free solutions discussed in Sec. \ref{sec:free}: Appendix \ref{app:free1} motivates the presence of these solutions in excited states,
 Appendix \ref{app:iso} explains why they are absent in the isotropic approximation of Refs. \cite{\HSQCa, \BHSnum}, and Appendix \ref{app:needquartic} explains in an example
 why removing the free solutions requires higher orders in the
 threshold expansion of $\Kdf$.

\section{Threshold expansion of the three-particle quantization condition}
\label{sec:threshold}



As noted above, we consider a theory of identical, scalar particles, with interactions
constrained only by the imposition of a $\Z_2$ global symmetry that prevents odd-legged
vertices. In such a theory, the spectrum of odd-particle-number states in a cubic
box of length $L$, with periodic boundary conditions, is determined by
solutions to the quantization condition~\cite{\HSQCa}
\begin{equation}
    \det\left[ F_3(E,L)^{-1} + \Kdf(E)  \right] = 0\,.
     \label{eq:QC3}
\end{equation}
This holds up to finite-volume corrections that are exponentially suppressed,
i.e.,  which fall as $\exp(-m L)$ up to powers of $L$, where $m$ is the mass of the particle.
In Eq.~(\ref{eq:QC3}), $F_3$ and $\Kdf$ are matrices with index space $\{\vec k, \ell, m\}$,
where $\vec k\in (2\pi/L) \Z^3$ is the finite-volume momentum assigned to one of the particles
(the ``spectator''), while $\ell$ and $m$ specify the angular momentum of the 
other two (the ``dimer'').\footnote{%
Context determines which meaning of $m$ is intended.}
 This matrix space will be truncated, as explained in
Sec.~\ref{sec:implementation} below, so that the quantization condition (\ref{eq:QC3})
becomes tractable. The matrix $F_3$ is a complicated object given in Eq.~(\ref{eq:F3})
below; all we need to know for now is that it depends on the two-particle K matrix,
$\K_2$. Thus the infinite-volume quantities that enter into the quantization condition
are $\K_2$ and the three-particle quasilocal interaction $\Kdf$.\footnote{%
The subscript ``df'' stands for ``divergence-free'', indicating that a long distance
one-particle exchange contribution that can diverge has been removed. For 
further details, see Ref.~\cite{\HSQCa}. 
}

The quantization condition (\ref{eq:QC3}) is valid only when the CM (center of momentum)
energy lies in the range $m < E^* < 5m$, within which the only odd-particle-number
states that can go on shell involve three particles (rather than one, five, seven, etc.).
Here $E^*=\sqrt{E^2-\vec P^{\;2}}$, with $(E,\vec P)$ the total four-momentum of the
state.
As in the previous numerical studies~\cite{\Akakia,\MD,\BHSnum,Doring:2018xxx},
we further restrict our considerations to the overall rest frame, with  $\vec P=0$,
implying $E^*=E$ henceforth.
We also recall that Eq.~(\ref{eq:QC3}) assumes that there are no poles in $\K_2$ in
the kinematic regime of interest. We discuss the constraints that this places on
the two-particle scattering parameters in Sec.~\ref{sec:implementation}.

The aim of this section is to develop a systematic expansion of $\Kdf$ about the
three-particle threshold at $E=3m$. 
To that end, we make use of the fact that, unlike the matrix $F_3$,
$\Kdf$ is an infinite-volume quantity,
and so is defined for arbitrary choices of the three incoming
and three outgoing on-shell momenta in the scattering process,
and not just for finite-volume momenta.
It is also important that it can be chosen to be relativistically invariant,
if an appropriate choice of the kinematic function $\wt G$ entering $F_3$ is made~\cite{\BHSQC}
[see Eq.~(\ref{eq:defG})].

In the remainder of this section, we first recall the threshold expansion of $\K_2$
and its relation to the partial wave decomposition,
and then describe the generalization of the threshold expansion to $\Kdf$, 
extending an analysis given in Ref.~\cite{\BHSnum}. 
Finally, we show how the terms in this expansion are  decomposed into the matrix form
needed for Eq.~(\ref{eq:QC3}).

\subsection{Warm up: expanding $\K_2$ about threshold}
\label{sec:expandK2}

To illustrate the method that we employ for $\Kdf$, we first consider the simpler,
and well-understood, case of the two-particle K matrix, $\K_2$.
Since $\K_2$ is relativistically invariant, it depends only on
the standard Mandelstam variables $s_2$, $t_2$ and $u_2=4m^2-s_2-t_2$.
It is convenient to use dimensionless variables that vanish at threshold,
\begin{equation}
\wt \Delta_2 = \frac{s_2-4m^2}{4 m^2} = \frac{q_2^{*2}}{m^2}\,,\ \
\wt t_2 = \frac{t_2}{4 m^2} = - \frac{q_2^{*2}}{2 m^2} (1\! -\! c_\theta)\,, \ \
\wt u_2 = \frac{u_2}{4 m^2} = - \frac{q_2^{*2}}{2 m^2}(1\!+\!c_\theta)\,,
\label{eq:stu2}
\end{equation}
where $q_2^*$ is the magnitude of the momentum of each particle in the CM frame,
and $c_\theta$ is the cosine of the scattering angle.
For physical scattering, $\wt \Delta_2$, $-\wt t_2$ and $-\wt u_2$ are all non-negative, 
and satisfy
\begin{equation}
\wt \Delta_2= -\wt t_2 -\wt u_2\,,
\label{eq:constraint2}
\end{equation}
implying that $-\wt t_2$ and $-\wt u_2$ are both bounded by $\wt \Delta_2$.

Since $\K_2$ is known to be analytic near threshold, we can expand it in powers
of $\wt \Delta_2$, $\wt t_2$ and $\wt u_2$. The previous considerations imply
that, for generic kinematics (i.e., $\theta \ne 0$ or $\pi$), all three quantities
are of the same order. Bose symmetry implies that the expression must be
symmetric under $\wt t_2 \leftrightarrow \wt u_2$. Thus, through quadratic order we have
\begin{equation}
\K_2 = \wt c_0 + \wt c_1 \wt \Delta_2 + \wt c_2 \wt \Delta_2^2 
+ \wt c_3 \left(\wt t_2^{\,2} + \wt u_2^{\,2}\right) + \cO(\wt \Delta_2^3)
\,,
\label{eq:K2decomp}
\end{equation}
where the $\wt c_i$ are constants (which are real since $\K_2$ is real),
and we have used the constraint (\ref{eq:constraint2}) to reduce the number of 
independent terms.
We now decompose this result into partial waves, using
\begin{align}
	\mc{K}_2&=\sum_{\ell=0}^\iy (2\ell+1) \mc{K}_{2}^{(\ell)}(\wt \Delta_2)P_\ell(\cos\theta)\,.
	\label{eq:PWE}
\end{align}
All odd partial waves vanish by Bose symmetry, while Eq.~(\ref{eq:K2decomp}) leads to
\begin{align}
\K_{2}^{(0)} &= \wt c_0 + \wt c_1 \wt \Delta_2 + (\wt c_2 + \tfrac23 \wt c_3) \wt \Delta_2^2 
+ \cO(\wt \Delta_2^3)\,,
\label{eq:K20}
\\
\K_{2}^{(2)} &= \tfrac1{15} \wt c_3 \wt \Delta_2^2 +  \cO(\wt \Delta_2^3)\,.
\label{eq:K22}
\end{align}
The first equation  gives the first three terms in the effective range expansion
for $\K_2$, while from the second equation
we recover the well-known result that $\cK_2^{(2)}\propto q_2^{*4}$ near threshold.
By extending this analysis, one can show that $\K_2^{(\ell)}$ only
enters when we include terms of $\cO(\wt \Delta_2^{\ell})$ in 
the threshold expansion~\cite{\BHSnum}.

The threshold expansion has a finite radius of convergence. In particular, we know that
$\K_2$ has a left-hand cut at $\wt \Delta_2= -1$, so that the radius of convergence
cannot be greater than $|\wt \Delta_2|=1$.
In practice, we truncate the expansion at the order shown in Eqs.~(\ref{eq:K20})
and (\ref{eq:K22}) (and set $\K_{2}^{(\ell)}=0$ for $\ell \ge 3$),
use a cutoff function such that $\wt \Delta_2 > -1$,
and restrict $E< 5m$ implying that $\wt \Delta_2 < 3$.
We are thus assuming that the deviations from the truncated threshold expansion
are small over this kinematic range.

\subsection{Invariants for three-particle scattering}
\label{subsec:Mandelstams}
To extend the analysis to the three-particle amplitude $\Kdf$, we begin by listing
the generalized Mandelstam variables,
\begin{equation}
    s \equiv  E^2 \,, \ \
    s_{ij} \equiv (p_i+p_j)^2=s_{ji}, \ \ s_{ij}' \equiv (p_i'+p_j')^2 =s'_{ji}\,, \ \ 
    t_{ij} \equiv (p_i-p_j')^2\,,
    \label{eq:Mandelstams3}
\end{equation}
where $p_i$ ($p'_i$), $i=1-3$, are the incoming (outgoing) momenta.
As in the two-particle case, 
it is convenient to use dimensionless quantities that vanish at threshold,
\begin{equation}
    \Delta \equiv \frac{s - 9m^2}{9m^2}\,, \ \
    \Delta_i \equiv \frac{s_{jk} - 4m^2}{9m^2} \,, \ \
     \quad \Delta_i' \equiv \frac{s_{jk}' - 4m^2}{9m^2}\,, \ \
    \widetilde{t}_{ij} \equiv \frac{t_{ij}}{9m^2}\,,
\end{equation}
where in the definitions of $\Delta_i$ and $\Delta'_i$,
 $(i,j,k)$ form a cylic permutation of $(1,2,3)$.
These sixteen quantities are constrained by the following eight independent relations,
\begin{align}
    \sum_{i=1}^3\Delta_{i} &= \sum_{i=1}^3\Delta_{i}' = \Delta   \label{eq:constraint_1} \\
    \sum_{j=1}^3 \widetilde{t}_{ij} &= \Delta_{i} - \Delta, \quad \sum_{j=1}^3 \widetilde{t}_{ji} = \Delta_{i}' - \Delta. \qquad [i=1,2,3]\,.
     \label{eq:constraint_2}
\end{align}
%
%
Thus only eight are independent: 
the overall CM energy (parametrized here by $\Delta$)
and seven ``angular'' degrees of freedom.\footnote{%
We call these variables angular since they span a compact space.}
This counting is as expected: six on-shell momenta 
with total incoming and outgoing 4-momentum fixed have $3\cdot6-4\cdot2=10$ degrees of freedom, which is reduced to 7 by overall rotation invariance.

For physical scattering,
it is straightforward to show that $\Delta_{i}$, $\Delta_{i}'$, $-\widetilde{t}_{ij}$ 
are all non-negative, and the constraint equations then lead to the inequality
\begin{equation}
    0 \leq \Delta_{i}, \Delta_{i}', -\widetilde{t}_{ij} \leq \Delta\,. 
    \label{eq:thresh_scale}
\end{equation}
Thus all the variables
$\{ \Delta, \Delta_{i}, \Delta_{i}', \widetilde{t}_{ij} \}$ 
can be treated as being of the same order in an expansion about threshold.

\subsection{Expanding $\mc{K}_{\text{df},3}$ about threshold}
\label{subsec:K3_expansion}
By construction, $\Kdf$ is a smooth function for some region around threshold.\footnote{%
More precisely, what is shown in Ref.~\cite{\HSQCa} is that $\Kdf$ has no kinematic singularities
at threshold, a result that is checked by the explicit perturbative calculations
of Refs.~\cite{Hansen:2015zta,Sharpe:2017jej}.
There can be dynamical singularities due to a three-particle resonance, but, generically,
this will lie away from threshold.}
Thus it can be expanded in a Taylor series in the variables
$\{ \Delta, \Delta_{i}, \Delta_{i}', \widetilde{t}_{ij} \}$, which are all treated as being
of $\cO(\Delta)$.
Since $\Kdf$ is real, the coefficients in this expansion must also be real.
The expansion must also 
respect the symmetries of $\mc{K}_{\text{df},3}$, which is invariant 
under~\cite{\BHSQC}:\footnote{%
The first two symmetries hold because we are considering identical bosons. 
They would not hold in the more general case of nonidentical particles, 
allowing additional terms to be present in $\Kdf$.
}
\begin{itemize}
    \item Interchange of any two incoming particles: $p_i\leftrightarrow p_j \Rightarrow \Delta_{i} \leftrightarrow \Delta_{j}$ and $\widetilde{t}_{ik}\leftrightarrow \widetilde{t}_{jk}$
    \item Interchange of any two outgoing particles: $p_i'\leftrightarrow p_j' \Rightarrow \Delta_{i}' \leftrightarrow \Delta_{j}'$ and $\widetilde{t}_{ki}\leftrightarrow \widetilde{t}_{kj}$
    \item Time reversal: $p_i\leftrightarrow p_i' \ \ (\forall i) \ \Rightarrow\  \Delta_{i} \leftrightarrow \Delta_{i}'$ and $\widetilde{t}_{ij}\leftrightarrow \widetilde{t}_{ji} \ \ (\forall ij)$
\end{itemize}
It is then a tedious but straightforward exercise to write down the allowed terms
at each order in $\Delta$, and simplify them using
the constraints \eqref{eq:constraint_1}--\eqref{eq:constraint_2}.
Through quadratic order we find
\begin{align}
    m^2 \Kdf &= \K^{\iso}
    + \KA\Delta^{(2)}_A + \KB \Delta^{(2)}_B + \mc{O}(\Delta^3)\,, 
    \label{eq:termsKdf}
    \\
    \K^{\iso} &= \Kiso + \Kisoone\Delta + \Kisotwo\Delta^2
    \label{eq:Kisoterm}
    \\
    \Delta^{(2)}_A &= \sum_{i=1}^3 (\Delta_{i}^2 + \Delta_{i}'^{\,2}) - \Delta^2, \qquad		\label{eq:K3Aterm}
    \\
    \Delta^{(2)}_B &= \sum_{i,j=1}^3 \widetilde{t}_{ij}^{\;2} - \Delta^2 \,,
    \label{eq:K3Bterm}
\end{align}
where $\Kiso$, $\Kisoone$, $\Kisotwo$, $\KA$ and $\KB$ are real, dimensionless constants.
We thus see that there is a single term both at leading (zeroth) order and at first order,
while there are three independent terms at quadratic order.
The particular linear combinations of the quadratic terms that appear in
Eqs.~(\ref{eq:K3Aterm}) and (\ref{eq:K3Bterm})
(and in particular the subtraction of $\Delta^2$ in  $\Delta^{(2)}_A$ and $\Delta^{(2)}_B$)
are chosen based on our numerical experiments described below in order to ensure that
their contributions to the finite-volume spectrum are distinct.

As noted in Ref.~\cite{\BHSnum},
the leading order contribution to $\Kdf$ in Eq.~(\ref{eq:termsKdf}) is 
independent of momenta $p_i$ and $p'_j$.
This shows that the isotropic approximation to $\Kdf$,
defined as independence of the seven angular variables,
arises naturally in the same way as the $s$-wave approximation to $\K_2$.
What we add here is the result that $\Kdf$ remains isotropic at $\cO(\Delta)$,
having only an overall linear dependence on $s$.
Furthermore, at quadratic order, we find only two terms that depend on angular
variables ($\Delta^{(2)}_A$ and $\Delta^{(2)}_B$), compared to
the seven angular variables that are needed to fully characterize three-particle scattering.
Thus, if it is a good approximation to truncate the threshold expansion at $\cO(\Delta^2)$,
the number of parameters needed to describe $\Kdf$ is smaller than one might
naively have expected.

For most of our numerical investigations, we have restricted ourselves to quadratic
order in the expansion of $\Kdf$. It is interesting, however, to  push the
classification to higher order for at least three reasons. 
First, in order to know how rapidly the number of parameters grows;
second, to see which dimer partial waves enter; and, third, to investigate the
issue of solutions to the quantization condition with energies given by those
of three noninteracting particles (see Sec.~\ref{sec:free}).
Thus we have classified all terms of cubic order.
We find eight independent  terms: three that are just $\Delta$ times each of the
terms of  quadratic order, plus five new angular terms,
\begin{gather}
\begin{alignat}{3}
	\Delta^{(3)}_{A} &= \sum_i \left( \Delta_{i}^3 + \Delta'^{\,3}_i \right), \qquad	
	&&\Delta^{(3)}_{B} = \sum_{i,j} \widetilde{t}_{ij}^{\,\,3}	
	\\
	\Delta^{(3)}_{C} &= \sum_{i,j} \Delta_{i} \widetilde{t}_{ij} \Delta_{j}', \qquad 
	&&\Delta^{(3)}_{D} = \sum_{i,j} \widetilde{t}_{ij}^{\,\,2} \left( \Delta_{i} + \Delta_{j}' \right)
\end{alignat}
\\
	\Delta^{(3)}_{E} = 
	\sum_{\sigma\in S_3} \widetilde{t}_{1\sigma(1)} \widetilde{t}_{2\sigma(2)}
	 \widetilde{t}_{3\sigma(3)},
\end{gather}
where $\sigma\in S_3$ is a permutation of the indices $(1,2,3)$.
Thus the number of terms is growing fairly rapidly with order.\footnote{%
We do not think that there is any significance to the fact that the number of terms
depending on angular variables through cubic order, i.e. $2+5=7$,
equals the number of independent angles in three-particle scattering. The dependence
on these angles can be arbitrarily complicated, so there is not a one-to-one correspondence
between variables and functions.}
%

\subsection{Decomposing $\mc{K}_{\text{df},3}$}
\label{subsec:K3_decomp}

In order to use $\Kdf$ in the quantization condition, we need to decompose it into
the variables used in its matrix form. This is the analog of the partial wave decomposition
of $\K_2$, described in Sec.~\ref{sec:expandK2} above. 

The steps in this decomposition were presented in Ref.~\cite{\HSQCa} and we recall
them here. The total four-momentum $P^\mu$ is fixed, in our case to $(E,\vec 0)$.
One each of the initial and final particles is designated as the spectator, with
three-momenta denoted $\vec k$ and $\vec p$, respectively. Since $\Kdf$ is symmetric
separately under initial and final particle interchange, it does not matter which particles
are chosen as the spectators, and we take $\vec k=\vec p_3$ and $\vec p= \vec{ p}_3\,\!'$.
The remaining two particles form the (initial and final) dimers. The total momenta
of both dimers are fixed, e.g. to $P-p_3$ in the initial state. For each dimer,
we can boost to its CM frame,
and the only remaining degree of freedom is the direction of one of the particles 
in the dimer in this frame.
We take this particle to be $p_1$ in the initial state, and denote its direction
in the dimer CM frame by $\hat a^*$. Similarly, the direction of $p'_1$
in the final-state-dimer CM frame is called $\hat a'^*$.
Using these variables we can write\footnote{%
As above, the $2\cdot(3+2)=10$ momentum components are reduced to seven 
independent angular variables by rotation invariance.}
\begin{equation}
\Kdf = \Kdf(\vec p,\hat a'^*;\vec k,\hat a^*)\,.
\label{eq:Kdfnewvariables}
\end{equation}

The next step is to set each spectator momentum to one of the allowed finite-volume values,
e.g. $\vec k= \vec n (2\pi/L)$, with $\vec n$ a vector of integers.
The final step is then to
decompose the dependence on $\hat a^*$ and $\hat a'^*$ into spherical harmonics
\begin{equation}
\Kdf(\vec p,\hat a'^*;\vec k,\hat a^*)
=
4\pi Y^*_{\ell' m'}(\hat a'^*) \K_{\df,3;p\ell' m';k\ell m} Y_{\ell m}(\hat a^*)\,,
\label{eq:Kdfharmonics}
\end{equation}
where there is an implicit sum over all angular-momentum indices.
This defines the entries in the matrix form of $\Kdf$.\footnote{%
Note that we follow Ref.~\cite{\HSQCa} and drop the vector symbol on the momenta
in the matrix indices, in order not to overly clutter the notation.}
In practice, we use the real version of spherical harmonics, so the complex
conjugation in Eq.~(\ref{eq:Kdfharmonics}) has no impact.

The simplest example of this decomposition is for the isotropic terms in $\Kdf$,
namely $\K^\iso$ in Eq.~(\ref{eq:Kisoterm}).
Recalling that $E$, and thus $\Delta$, is fixed, $\K^\iso$ is simply a constant.
This implies that the matrix form of $\K^\iso$ vanishes unless $\ell'=\ell=0$, and
is independent of $\vec p, \vec k$:
\begin{equation}
\K^{\iso}_{\df,3;p\ell' m';k\ell m} = \K^{\iso} \delta_{\ell' 0} \delta_{m' 0} \delta_{\ell 0} \delta_{m 0}\,.
\label{eq:Kisodecomp}
\end{equation}
The approximation $\Kdf=\K^{\iso}$ is studied in Ref.~\cite{\BHSnum}.

We next work out the decomposition of $\Delta_A^{(2)}$, Eq.~(\ref{eq:K3Aterm}),
which is conveniently written as
\begin{equation}
\Delta_A^{(2)} = \left[\Delta_3^2 + \Delta'^{\,2}_3 - \Delta^2\right]
+ \left[\Delta_1^2+\Delta_2^2\right] + \left[\Delta'^{\,2}_1+\Delta'^{\,2}_2\right]\,.
\label{eq:Delta2A}
\end{equation}
The first term depends on $\vec k^{\,2}$ and $\vec p^{\,\,2}$, but not on $\hat a^*$ or $\hat a'^*$.
This can be seen from
\begin{equation}
9 m^2 \Delta _3 = (P- p_3)^2 - 4 m^2
= E^2 - 2 E \omega_k - 3 m^2\,,
\end{equation}
with $\omega_k=\sqrt{\vec k^2+m^2}$, and the corresponding result for $\Delta'_3$.
Thus the first term in Eq.~(\ref{eq:Delta2A}) leads to a purely $s$-wave 
($\ell'=\ell=0$) contribution to $\Kdf$, although now with nontrivial dependence on
$\vec k$ and $\vec p$, so this differs from an isotropic contribution.

The second term in Eq.~(\ref{eq:Delta2A}) can be rewritten using
\begin{equation}
\frac{9 m^2}2 \left[\Delta_1^2+\Delta_2^2\right] 
=
( p_+\cdot p_3 - 2 m^2)^2 + (p_-\cdot p_3)^2
=
(E \omega_k - 3 m^2)^2 + \frac{4E^2}{E_{2,k}^{*2}} (\vec a^{\, *}\cdot \vec k)^2
\,,
\label{eq:Delta1plus2}
\end{equation}
where $p_\pm = p_1\pm p_2$, and $E_{2,k}^{*2}= (P-p_3)^2$. 
To obtain the second form one must
explicitly boost to the dimer CM frame, in which $\vec p_-$ equals $2\vec a^{\,*}$,
with $a^{*2}=9 m^2 \Delta_3/4$.
The first term on the right-hand side of Eq.~(\ref{eq:Delta1plus2}) is independent of
$\hat a^*$, and thus again contributes only an $s$-wave component.
The second term, however, depends quadratically on $\hat a^*$, and thus,
through the addition theorem for spherical harmonics,\footnote{%
Again, in practice, we use real spherical harmonics, so the complex conjugation
is not needed.}
\begin{equation}
(\hat a \cdot \hat k)^2 = \frac13 + \frac{8 \pi}{15} \sum_m Y^*_{2m}(\hat a) Y_{2 m}(\hat k) 
\,,
\end{equation}
leads to both $s$- and $d$-wave contributions. In other words, both
$\K_{\df,3; p 0 0; k 0 0}$ and $\K_{\df,3;p 0 0; k 2 m}$ are nonvanishing.
These contributions are straightforward to work out from the
above equations, and we do not display them explicitly.

The final term in Eq.~(\ref{eq:Delta2A}) differs from the second term only by changing
unprimed quantities to their primed correspondents. Thus one finds contributions
both to $\K_{\df,3; p 0 0; k 0 0}$ and $\K_{\df,3;p 2 m'; k 0 0}$. Overall, we conclude that the angular dependence in $\Delta_A^{(2)}$ leads
to both $s$- and $d$-wave dimer interactions, although there are no terms with
both $\ell=2$ and $\ell'=2$. The latter result arises
 from the fact that there are no terms in $\Delta_A^{(2)}$ that
depend on both incoming and outgoing momenta.

Finally, we consider $\Delta_B^{(2)}$, given in Eq.~(\ref{eq:K3Bterm}).
This is more complicated to decompose because $\wt t_{ij}$ contains both incoming
and outgoing momenta, but this same property leads to contributions with
$\ell=\ell'=2$. We provide only a sketch of the decomposition, as the details
are tedious, lengthy, and straightforward to automate.
Expanding $\Delta_B^{(2)}$, one finds terms that are similar to those dealt with
in $\Delta_A^{(2)}$, which lead to additional contributions to
$\K_{\df,3; p 0 0; k 0 0}$, $\K_{\df,3;p 0 0,k 2 m}$, and $\K_{\df,3;p 2 m'; k 0 0}$,
and a term proportional to
\begin{equation}
(p_- \cdot p'_-)^2 = a_i^* a_j^* S_{ij,rs} a'^*_r a'^*_s\,,
\end{equation}
where $p'_\pm=p'_1\pm p'_2$,
$i$, $j$, $r$, and $s$ are now spatial vector indices, and $S$ is a tensor
that depends on $\vec k$ and $\vec p$ and is symmetric 
separately under $i\leftrightarrow j$ and $r\leftrightarrow s$.
By decomposing $S$ into the spherical tensor basis one finds contributions
to the $\ell=\ell'=2$ part of $\Kdf$, 
$\K_{\df,3;p 2 m'; k 2 m}$,
as well as to the other three components. 

In summary, because the terms of $\cO(\Delta^2)$ in $\Kdf$ are
at most quadratic in $\vec a\,^*$ and/or $\vec a\,'^{\,\!*}$, they give rise to
dimer interactions that are either $s$- or $d$-wave. This is the analog
of the result derived in Sec.~\ref{sec:expandK2} that, at the same order,
only $\K_2^{(0)}$ and $\K_2^{(2)}$ are present.

The generalization to higher order is straightforward. 
Terms of $\cO(\Delta^3)$, can, in principle
be cubic in $\vec a\,^*$ and/or $\vec a\,'^{\,\!*}$, but Bose symmetry forbids odd powers.
Thus $\cO(\Delta^3)$ terms lead only 
to $s$- and $d$-wave contributions to $\Kdf$, as we have checked explicitly.
In order to obtain contributions with $\ell=4$ or $\ell'=4$ one must work at
$\cO(\Delta^4)$ in the threshold expansion. The pattern continues 
similarly at higher order.

\section{Implementing the quantization condition}
\label{sec:implementation}


In this section we describe how we numerically implement the quantization condition, 
Eq.~\eqref{eq:QC3}, when working to quadratic order in the threshold expansion.
The expression for $F_3$ is\footnote{%
This is the form given in Appendix C of Ref.~~\cite{\HSQCa}, with $\wt F=F/(2\omega)$
and $\wt G=G/(2\omega)$. The matrix $H$ should not be confused with the cutoff function
$H(\vec k)$, which is always shown with an argument.}
\begin{gather}
    F_3 = \frac{1}{L^3}\left[ \frac{\widetilde{F}}{3} - \widetilde{F}H^{-1}\widetilde{F}  \right],
	\label{eq:F3}
    \\
    H = \frac{1}{2\omega\mathcal{K}_2} + \widetilde{F} + \widetilde{G},\label{eq:H}
\end{gather}
where all quantities are matrices with indices $\{k,\ell,m\}$.
$\K_2$ is a diagonal matrix
\begin{equation}
	\left[\frac{1}{2\omega\mc{K}_2}\right]_{p\ell'm';k\ell m} = \delta_{pk}\delta_{\ell'\ell}
	\delta_{m'm} \frac{1}{2\omega_k\mc{K}_{2;k}^{(\ell)}}\,,
	\label{eq:K2mat}
\end{equation}
where the only nonzero elements are the $s$- and $d$-wave terms
\begin{align}
	\frac{1}{\mc{K}_{2;k}^{(0)}} &= \frac{1}{16\pi E_{2,k}^*}
	\left\{ -\frac{1}{a_0} + r_0\frac{q_{2,k}^{*2}}{2} + P_0(r_0)^3 q_{2,k}^{*4} + |q_{2,k}^*| [1-H(\vec{k})] \right\}\,,
	\label{eq:K2swave}
	\\
	\frac{1}{\mc{K}_{2;k}^{(2)}} &= \frac{1}{16\pi E_{2,k}^*} \frac{1}{q_{2,k}^{*4}}
	\left\{ -\frac{1}{a_2^5} + |q_{2,k}^{*5}| [1-H(\vec{k})] \right\}\,. 
	\label{eq:K2dwave}
\end{align}
Here $E_{2,k}^{*2}=(P-k)^2$ is the invariant mass of the dimer,
while $q_k^*=\sqrt{E_{2,k}^{*2}/4-m^2}$ is the momentum of each particle composing the
dimer in its CM frame.\footnote{%
These quantities were denoted $s_2$ and $q_2^*$, respectively, in Sec.~\ref{sec:expandK2},
but here we need to make explicit that they depend on $\vec k$.
The notation here is the same as in Ref.~\cite{\HSQCa}.}
The expression (\ref{eq:K2swave}) is the standard form for the effective range expansion
through quadratic order, with $a_0$ the $s$-wave scattering length, $r_0$ the effective
range, and $P_0$ the shape parameter. Expanding the overall factor
of $E_{2,k}^*$ about threshold, and for now ignoring the $1-H(\vec k)$ term,
one recovers the form given in Eq.~(\ref{eq:K20}).
Similarly, aside from the $1-H$ term,
the expression for $\K_{2;k}^{(2)}$, Eq.~(\ref{eq:K2dwave}),
is equivalent to the earlier result, Eq.~(\ref{eq:K22}).
Here the leading order term is parametrized in terms
of the $d$-wave scattering length $a_2$.\footnote{%
This expansion is often written with a different definition of $a_2$,
in which $a_2^5$ is replaced by $a_2$. We prefer the present form since then
$a_2$ has dimensions of length.}

The $1-H$ terms in the expressions (\ref{eq:K2swave})  and (\ref{eq:K2dwave}) 
arise from the need to introduce a smooth cutoff function $H(\vec k)$ that vanishes for
$E_{2,k}^{*2} \le 0$. We refer the reader to Refs.~\cite{\HSQCa,\HSTH} for further
explanation of both the need for this cutoff and the manner in which it enters these expressions.
It is sufficient to note here that the $1-H$ term turns on smoothly only well below the dimer
threshold at $E_{2,k}^*=2m$. The explicit form of $H(\vec k)$ that we use is given in
Appendix~\ref{app:defns}.

As noted above, the quantization condition holds only if
there are no poles in $\K_2$ in the kinematic regime under study. 
The kinematic range of $q_{2,k}^*$
is  given by $-m^2 < q_{2,k}^{*2} <  3 m^2$ (corresponding to $0 < E_{2,k}^{*2} < 16 m^2$).
The parameters in Eqs.~(\ref{eq:K2swave}) and (\ref{eq:K2dwave}) are thus
constrained so that neither right-hand side vanishes for this range of $q_{2,k}^{*2}$.
In our numerical investigations, we always use values of the scattering
parameters that satisfy these constraints.
For $a_2$ the constraint is that $m a_2 < 1$, with arbitrarily negative values allowed.

The other two quantities appearing in $F_3$ are the finite-volume kinematic functions
$\wt F$ and $\wt G$. The former is essentially a two-particle quantity,
and thus is diagonal in spectator momenta, though not in the angular-momentum 
indices:\footnote{%
We are abusing notation here,
but the two versions of $\wt F$ will always be distinguishable by the presence or
absence of the argument $\vec k$.}
\begin{equation}
	\widetilde{F}_{p\ell'm';k\ell m} \equiv 
	\delta_{pk} H(\vec k) \widetilde{F}_{\ell'm';\ell m}(\vec{k})\,.
	\label{eq:Ft}
\end{equation}
$\wt G$ is a kinematic function that arises from one-particle exchange between dimers,
and is thus a quantity that involves all three particles.
In particular, it is not diagonal in any of the indices.
We give the explicit forms of $\wt F$ and $\wt G$ in Appendix~\ref{app:defns},
and provide some details of their numerical evaluation of $\wt F$ in Appendix~\ref{app:Ft}.

An important property is that  $\wt G_{p\ell' m';k\ell m}$ is proportional to
$H(\vec p) H(\vec k)$, and is thus truncated 
to the finite number of values of spectator momenta for which $H(\vec k) \ne 0$.
We call this number $N_{\spect}(E,L)$.
The same truncation applies to $\wt F$, due to the factor of $H(\vec k)$ in Eq.~(\ref{eq:Ft}).
Both matrices are, however, infinite-dimensional in angular-momentum space.
This is to be contrasted to $\K_2$ and $\Kdf$, which are (by approximation) truncated
in angular momenta but not in spectator-momentum space. In angular momentum
space the dimension is $1+5=6$ when keeping both $s$ and $d$ waves.

Nevertheless, it turns out that these two truncations are sufficient to reduce
the quantization condition, Eq.~(\ref{eq:QC3}), to a determinant of a $6N_\spect$-dimensional
matrix.
To show this, we first write the quantization condition as
\begin{equation}
\det\left[F_3^{-1}\right] \det\left[ 1 + F_3 \Kdf\right] = 0\,.
\label{eq:QC3a}
\end{equation}
It appears from this rewriting that there will be solutions to the quantization condition
when $\det[F_3] \to \infty$, i.e., when $F_3$ has a diverging eigenvalue.
However, in that case, the second determinant will, for a general $\Kdf$, also diverge,
leading to a finite product.
Thus we expect that the only solutions of the quantization condition (\ref{eq:QC3})
for general $\Kdf$ will be those that also satisfy 
\begin{equation}
\det[1+F_3 \Kdf]=0\,.
\label{eq:QC3b}
\end{equation} 
This also makes sense intuitively,
since we expect all finite-volume energies to depend upon the three-particle interaction.
The advantage of the form (\ref{eq:QC3b}) is that it has been shown in Ref.~\cite{\HSQCa}
that it effectively truncates all  matrices that appear 
(i.e., $\wt F$, $\wt G$, $\K_2$ and $\Kdf$)
to $N_\spect$ entries in 
spectator-momentum space and to $s$ and $d$ waves in angular-momentum
space. By ``effectively'' we mean that elements of the matrices that lie outside the
truncated space do not contribute to the determinant.

In the following, we also consider at times the further truncation to only $s$-wave
dimer interactions. This is effected by setting to zero all entries in the matrices having
$\ell=2$, so that their dimension becomes $N_\spect$.

We have now explained how all the matrices contained in the quantization condition 
Eq.~(\ref{eq:QC3}) are constructed, for given values of $E$ and $L$.
We combine these matrices to form $F_3^{-1}+\Kdf$, 
and calculate its eigenvalues. For a given choice of $L$, 
the finite-volume spectrum is then given by those values of $E$
for which an eigenvalue vanishes.

The practical calculation of this spectrum is facilitated by decomposing into
irreducible representations (irreps) of the symmetry group of finite-volume
scattering. For a cubic box with $\vec P=0$, this is the cubic group, $O_h$.
For the case of pure $s$-wave dimers, this decomposition has been worked out
for the NREFT and  FVU quantization conditions in Ref.~\cite{Doring:2018xxx}.
It has also been used implicitly in the numerical study of the isotropic approximation to
the RFT quantization condition in Ref.~\cite{\BHSnum}, since the isotropic
approximation automatically involves a projection onto the trivial ($A_1^+$) 
irrep.\footnote{%
For a more detailed discussion of the isotropic approximation,
see Appendix~\ref{app:iso}.}
The new result that we now present is the generalization of the decomposition
to the case in which one has both $s$- and $d$-wave dimers.

\subsection{Projecting onto cubic group irreps}
\label{sec:proj}

We begin by recalling some useful properties of the cubic group, $O_h$.
It has dimension $[O_h]=48$, and ten irreps.
Its character table can be found, e.g. in Ref.~\cite{atkins1970tables}.
The labels for, and dimensions of, the irreps can be seen in Table~\ref{tab:dI} below.
Each finite-volume momentum, $\vec k = (2\pi/L)\vec n_k$,
lies in a ``shell'' (also known as an orbit) 
composed of all momenta related to $\vec k$ by cubic group transformations.
We refer to this shell as $o_k$. There are seven types of shell, differing by
the symmetry properties of the individual elements. We label these by the form of $\vec n_k$:
$(000)$, $(00a)$, $(aa0)$, $(aaa)$, $(ab0)$, $(aab)$ and $(abc)$, 
where $a$, $b$ and $c$ are all different, nonzero components. 
They have dimensions $N_o=1$, $6$, $12$, $8$, $24$, $24$ and $48$, respectively.
For example, $\vec n_k=\hat x $ lies in the $(001)$ shell of type $(00a)$,
and $\vec n_k=\hat x+ 2\hat z$ lies in the $(120)$ shell of type $(ab0)$.
Each element in a shell is invariant under rotations in a subgroup of $O_h$,
called its little group, $L_k$. The little groups for all elements in a shell
are isomorphic, with dimension $[L_k]=[O_h]/N_o$.

The four matrices that enter the quantization condition Eq.~(\ref{eq:QC3}),
namely $2\omega\K_2$, $\Kdf$, $\wt F$ and $\wt G$, 
are all invariant under a set of orthogonal transformations $U(R)$, where $R\in O_h$. 
Specifically, if $M$ is one of these matrices, then
\begin{align}
M &= U(R)MU(R)^T\,, \qquad U(R) U(R)^T = \mathbf{1}\,,
\label{eq:invariance}
\\
	U(R) &= S(R) \otimes \mathcal{D}(R)^T,
    \\
    U(R)_{p\ell'm';k\ell m} &= \delta_{o_p o_k} S^{(o_p)}_{pk}(R) 
    \delta_{\ell'\ell} \mathcal{D}^{(\ell)}_{mm'}(R). \label{eq:U}
\end{align}
Here the Wigner D-matrix is defined in Eq.~(\ref{eq:Wigner_D_1}),
while $S(R)$ permutes the spectator momenta within shells: 
\begin{gather}
    S(R)_{pk} = \delta_{o_p o_k} S^{(o_p)}_{pk}(R) = \delta_{p_R k} 
    \equiv \begin{cases}
    1, & R\vec{p} = \vec{k} \\
    0, & \text{otherwise}\,.
    \end{cases} \label{eq:S}
\end{gather}
For $2\omega\K_2$ and $\Kdf$ this result follows because they are
invariant under rotations, while for $\wt F$ and $\wt G$ it follows from the
fact that they are form-invariant under cubic-group rotations if the quantization
axis that defines the spherical harmonics is rotated along with the spectator momenta.

The matrices $\{U(R)^T\}_{R\in O_h}$ furnish a representation of $O_h$:
\begin{equation}
U(R_2R_1)^T = U(R_2)^TU(R_1)^T\,, \quad \forall R_1,R_2\in O_h\,,
\ \ {\rm and}\ \ U(\textbf{1}_3)^T = \textbf{1}_{k\ell m}\,.
\label{eq:UTrep}
\end{equation}
One may decompose this reducible representation into irreps $I$ of the cubic group using
projection matrices (see, e.g., Ref.~\cite{Georgi:1982jb})
\begin{gather}
    P_I = \frac{d_I}{[O_h]} \sum_{R\in O_h} \chi_I(R) U(R)^T\,, 
    \label{eq:P_I}
\end{gather}
where $d_I$ is the dimension of $I$ and $\chi_I(R)$ its character.\footnote{%
Normally one would write $\chi_I(R)^*$ in Eq.~\eqref{eq:P_I}, 
but since $O_h$ only involves real orthogonal transformations, 
all characters are real and the conjugation is trivial.}
%
%
%
%
An important simplifying property of $U(R)$, which carries over to $P_I$,
is that it is block-diagonal.
For the spectator-momentum indices, this follows because
\begin{gather}
    U(R)^T_{pk} = S(R)_{kp}\otimes \mathcal{D}(R) = \delta_{k_{\!R}\, p}\mathcal{D}(R)
     = \begin{cases}
     \mathcal{D}(R), & R\vec{k} = \vec{p} \\
     0, & \text{otherwise}\,,
     \end{cases}
     \label{eq:URT}
\end{gather}
which implies that each $U(R)$ is block diagonal in shells, $o$.
We label the resulting ``shell blocks'' of $P_I$ as $P_{I,o}$.
These shell blocks inherit from $\mc{D}(R)$ the property of
being block diagonal in $\ell$, and we label the corresponding sub-blocks
as $P_{I,o(\ell)}$, with $\ell=0$ or $2$. The result is that
we can write $P_I$ in the form
\begin{equation}
P_I = \text{diag}(P_{I,o_1},~P_{I,o_2},\ldots)\,,\qquad 
P_{I,o} = \text{diag}(P_{I,o(0)},~P_{I,o(2)})\,.
\label{eq:Pblocks}
\end{equation}
This simplified structure allows for more efficient computation of the
$P_I$ matrices, as explained in Appendix~\ref{app:P^I_comp}.

Using these projectors, we can decompose the quantization condition into separate
conditions for each irrep. From Eq.~(\ref{eq:invariance}) we know that $[P_I,M]=0$,
for each of the four matrices $M$,
from which it follows that
\begin{equation}
[P_I,F_3^{-1}+\Kdf]=0\qquad (\forall I)\,.
\end{equation}
Using $\sum_I P_I = \mathbf{1}$, and the orthogonality of the projectors onto different
irreps, one can then show that the determinant factorizes into that for each irrep
\begin{equation}
\det[F_3^{-1}+\Kdf] = \prod_I \det_{\text{sub},I}[P_I(F_3^{-1}+\Kdf) P_I]\,,
\label{eq:factordet}
\end{equation}
where the subscript indicates that the determinant is taken only over the subspace
onto which $P_I$ projects. 
Thus the quantization condition for irrep $I$ becomes
\begin{equation}
 \det_{\text{sub},I}[P_I(F_3^{-1}+\Kdf) P_I] = 0\,,
 \label{eq:QCirrep}
\end{equation}
If desired, one can also apply the projectors to all the matrices contained in $F_3$,
Eq.~(\ref{eq:F3}), so that the entire evaluation of the quantization condition involves
matrices of reduced dimensionality.

The number of eigenvalues in a given irrep is given by
the dimension of the projected subspaces, $d(P_I)$.
This is obtained by summing the dimensions of the sub-blocks,
\begin{equation}
d(P_I) = \sum_{o} \sum_{\ell=0,2} d(P_{I,o(\ell)})\,,
\label{eq:dimPI}
\end{equation}
where the sum over $o$ runs over all shells that are ``active'', i.e., that lie below the cutoff.
We explain how the $d(P_{I,o(\ell)})$ are calculated in Appendix~\ref{app:subspace},
and list the results in Table~\ref{tab:dI}.
From this we  learn, for example, that the $\vec k=\vec 0$ shell contains one
$A_1^+$ irrep for $\ell=0$, and one each of the $E^+$ and $T_2^+$ irreps for $\ell=2$.
Note that shells can contain multiple versions of a given irrep, e.g., the
$(00a)$ shell-type with $\ell=2$ contains two versions each of the $E^+$, $T_2^+$,
$T_1^-$ and $T_2^-$ irreps.

\begin{table}[tb]
\centering
\caption{Dimension of irrep projection sub-blocks for each shell-type
and angular momentum, $(d(P_{I,o(0))}, d(P_{I,o(2)}) )$.
Each row corresponds to an irrep of the cubic group $O_h$, 
whose dimension is also listed for completeness.
\label{tab:dI}}
\vspace{0.1in}
\begin{tabular}{C|C|CCCCCCC}
&  &\multicolumn{7}{c}{shell types}\\
\text{irrep} & \text{dim}   & (000)    &  (00a)   & (aa0) & (aaa) & (ab0) & (aab) & (abc) \\
 \hline
 A_1^+ & 1& (1,0) & (1,1) & (1,2)  & (1,1) & (1,3)  & (1,3)   & (1,5)   \\
 A_2^+ & 1& (0,0) & (0,1) & (0,1)  & (0,0) & (1,3)  & (0,2)   & (1,5)   \\
 E^+   & 2& (0,2) & (2,4) & (2,6)  & (0,4) & (4,12) & (2,10)  & (4,20)  \\
 T_1^+& 3 & (0,0) & (0,3) & (0,9)  & (0,6) & (3,21) & (3,21)  & (9,45)  \\
 T_2^+ &3 & (0,3) & (0,6) & (3,12) & (3,9) & (3,21) & (6,24)  & (9,45)  \\
 A_1^- & 1 & (0,0) & (0,0) & (0,1)  & (0,0) & (0,2)  & (0,2) & (1,5)     \\
 A_2^- & 1 & (0,0) & (0,1) & (0,1)  & (1,1) & (0,2)  & (1,3) & (1,5)     \\
 E^-   & 2 & (0,0) & (0,2) & (0,4)  & (0,4) & (0,8)  & (2,10) & (4,20)   \\
 T_1^-  & 3 & (0,0) & (3,6) & (3,12) & (3,9) & (6,24) & (6,24) & (9,45)   \\
 T_2^- & 3 & (0,0) & (0,6) & (3,12) & (0,6) & (6,24) & (3,21) & (9,45)     
\end{tabular}
\end{table}

\begin{figure}[tb!]
\centering 
\includegraphics[width=.7\textwidth]{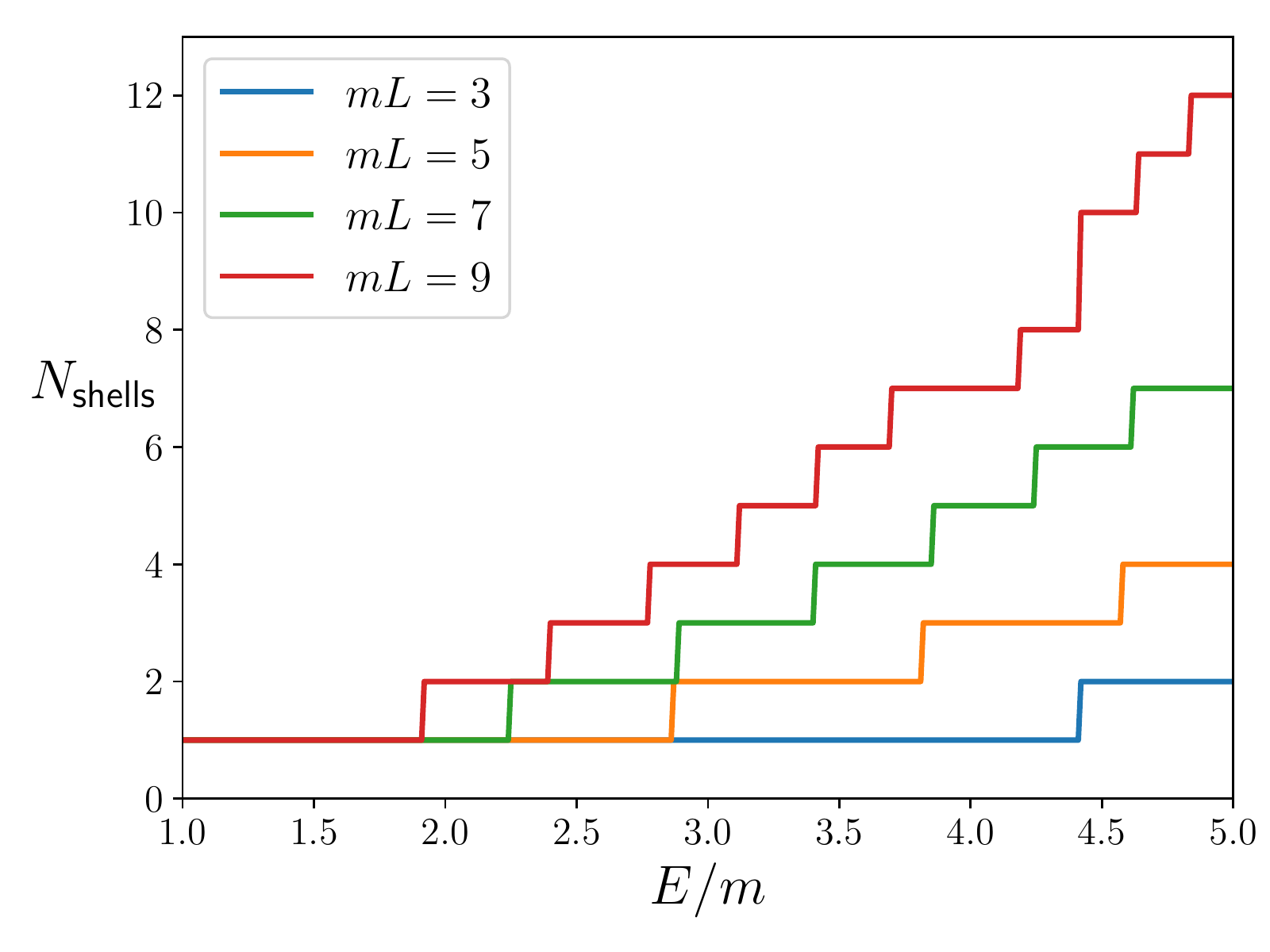}
\caption{Number of active momentum shells for fixed $mL$ as a function of $E$.}
\label{fig:nshell} 
\end{figure}

At this stage it is useful to give an example of how shells become active
as $E$ and $L$ are increased. With our cutoff, described in Appendix~\ref{app:defns},
the maximum value of $|\vec n_k|$, $n_{k,\rm max}$,
is determined by the vanishing of $E_{2,k}^{*2}$:
\begin{equation}
E_{2,k}^{*2}=0 \ \ \Rightarrow\ \  n_{k,\rm max} = \frac{L}{2\pi} 
\left(\frac{E^2-m^2}{2E}\right)\,.
\end{equation}
This can be easily converted into the number of active shells,
an example being shown in Fig.~\ref{fig:nshell}.
The first fifteen shells are $(000)$, $(001)$, $(110)$, $(111)$,
$(002)$, $(120)$, $(112)$, $(220)$, $(221)$, $(003)$, $(130)$, $(113)$,
$(222)$, $(230)$ and $(123)$, at which point examples of all seven types have appeared.

Although each $P_I$ is block diagonal in $o$ and $\ell$, $F_3^{-1}+\Kdf$ is generally not.
Thus even though each eigenvector of $F_3^{-1}+\Kdf$ lies in a single irrep, 
it will generally be a nontrivial linear combination of vectors 
lying in the subspaces projected onto by $P_{I,o(\ell)}$.
However, we can still use Table~\ref{tab:dI} to determine 
how many eigenvalues will be present in a given irrep for a given choice of $E$ and $L$.
For example, suppose we have both $s$- and $d$-wave interactions turned on and we are in the $E,L$ regime where only the first two momentum shells, $(000)$ and $(001)$, are active,
so that $N_\spect=1+6=7$.
Then the table tells us that $F_3^{-1}+\Kdf$ has 3 eigenvalues in $A_1^+$ since
\begin{multline}
 d(P_{A_1^+}) =
 d(P_{A_1^+,000(0)})
 +d(P_{A_1^+,000(2)})
 +d(P_{A_1^+,001(0)})
 +d(P_{A_1^+,001(2)}) \\ 
 = 1+0+1+1 = 3 \,. 
 \label{eq:A1+count}
 \end{multline}
Looking at the other irreps, we see that in this regime there is 1 eigenvalue in 
$A_2^+$, 8 in $E^+$, 3 in $T_1^+$, 9 in $T_2^+$, 0 in $A_1^-$, 1 in $A_2^-$, 2 in $E^-$, 9 in $T_1^-$, and 6 in $T_2^-$ giving the correct total of  $6 N_\spect=42$ 
eigenvalues. We stress that eigenvalues lying in a given irrep always come in
degenerate multiplets corresponding to the dimension of the irrep. 
Thus, for example, the eight
eigenvalues in the $E^+$ irrep in the two-shell regime
consist of four two-fold-degenerate pairs.

A point that may lead to confusion when we present 
results in the following section is that the number of eigenvalues of $F_3^{-1}+\Kdf$
bears no direct relation to the number of solutions to the quantization condition.
For there to be a solution, an eigenvalue must vanish, and this occurs only for a subset
of the eigenvalues in the energy range of interest. This point can be seen explicitly
if the interactions $\K_2$ and $\Kdf$ are weak, for then we expect the number of
states to be the same as for noninteracting particles. We quote in Table~\ref{tab:freeirreps}
the irreps that appear in the first few three-particle levels for noninteracting particles.
These states have energies
\begin{equation}
E^{\rm free}(\vec n_1,\vec n_2)=\sum_{i=1}^3 \sqrt{m^2 + (2\pi/L)^2 \vec n_i^2}\,,
\qquad
\vec n_3= -\vec n_1-\vec n_2\,,
\label{eq:Efree}
\end{equation}
where $\vec n_i$ are integer vectors.
As an example of the difference between the dimensions of $F_3^{-1}+\Kdf$ and
the number of solutions, we consider $mL=5$ and the $A_1^+$ irrep, 
and focus on the energy range $E/m=3-5$.
From Fig.~\ref{fig:nshell} we see that the number of active momentum shells 
begins at $2$ for $E=3m$, increases to $3$ at some point, and then reaches $4$ below
$E=5m$. From Table~\ref{tab:dI} we deduce that the corresponding number of
eigenvectors in the $A_1^+$ irrep are $3$, $6$ and $8$.
By contrast, the free levels in this irrep occur at $E=3m$, $E=4.21m$, $E=5.08m$, \dots.
For weak interactions, we expect solutions to the quantization
condition only near these three values, and thus we find that, in all cases, the
number of eigenvalues of $F_3^{-1}+\Kdf$ significantly exceeds the number of solutions
at, or below, the given energy.

\begin{table}[tb]
\centering
\caption{Irreps appearing in the lowest energy levels of three identical noninteracting particles.
The first column gives the level number (for values of $mL\sim 5$), starting at zero.
The states are labeled by the squares of the three vectors $\vec n_i$ that determine
the momenta of the particles---see Eq.~(\ref{eq:Efree})---and these are given in the
second column. The third column gives the degeneracy, and the final column the irreps that
appear.
\label{tab:freeirreps}}
\vspace{0.1in}
\begin{tabular}{C|C|C|C}
\text{level} & (\vec n_1^2,\vec n_2^2,\vec n_3^3) & \text{degen.} & \text{irreps}\\
 \hline
0 &( 0,0,0) & 1& A_1^+ \\
1 & (1,1,0) & 3 & A_1^+ + E^+ \\
2 & (2,2,0) & 6 &  A_1^+ + E^+ + T_2^+ \\
3 & (2,1,1) & 12 & A_1^+ + E^+ + T_2^+ + T_1^- + T_2^-\\
4 & (3,3,0) & 4 & A_1^+ + T_2^+  \\
\end{tabular}
\end{table}

We close this section by noting that the components of $\Kdf$, 
given in Eq.~(\ref{eq:termsKdf}),
can themselves be decomposed into different irreps.
While it is clear that $\Kiso$, Eq.~(\ref{eq:Kisoterm}), lies purely in the $A_1^+$ irrep,
we also find that the same is true for the $\KA$ term.
The $\KB$ term, however, has components that lie in the $A_1^+$, $E^+$, $T_2^+$ and $T_1^-$
irreps. For components lying in the remaining irreps one must go to cubic or higher order
in the threshold expansion.

\section{Results}
\label{sec:results}

The goal of this section is to illustrate the impact of including $d$-wave interactions
in the quantization condition. In particular, we aim to determine
which energy levels and which irreps are particularly sensitive to such interactions.
We begin, however,  with a case where the impact of $d$-wave interactions is small,
namely the ground state energy with a weak two-particle interaction.
This allows us to test of our implementation of the quantization condition in a
regime where we can make an analytic prediction.
We then consider the impact of a strong $d$-wave interaction, $m |a_2| \sim 1$,
comparing its effect on the ground and excited states, and for different irreps.
Next we study the sensitivity of the finite-volume spectrum of the physical $3 \pi^+$ state,
with $\K_2$ taken from experiment, to the various terms in $\Kdf$.
And, finally, we discuss the different types 
of unphysical solutions to the quantization condition that appear.

\subsection{Threshold expansion including $a_2$ \label{sec:thresholda2}}

In this section we consider the energy of the lightest two- and 
three-particle states in the case of weak two-particle interactions, and with the
three-particle interaction $\Kdf$ set to zero. The energy of these states (called
$E_2^{(0)}$ and $E_3^{(0)}$, respectively) lie close
to their noninteracting values, and we define the differences as
\begin{equation}
\Delta E_n = E_n^{(0)} - n\times m\,.
\label{eq:DEn}
\end{equation}
These can be expanded in powers of $1/L$ (up to logarithms), the results being
called the threshold expansions.
These expansions have been worked out in a relativistic
theory to $\cO(L^{-6})$ in 
Refs.~\cite{Luscher:1986n2,Hansen:2015zta,Hansen:2016fzj}:\footnote{%
The terms up to $\cO(L^{-5})$ agree with those obtained previously using nonrelativistic
QM~\cite{Beane:2007qr,Tan:2007bg}.}
\begin{equation}
\Delta E_2=\frac{4\pi a_0}{mL^3}\,\biggl\{1+c_1\biggl(\frac{a_0}{\pi L}\biggr)
+c_2\biggl(\frac{a_0}{\pi L}\biggr)^2
+c_3\biggl(\frac{a_0}{\pi L}\biggr)^3
+\frac{2\pi r_0(a_0)^2}{L^3}-\frac{\pi a_0}{m^2L^3}\biggr\}
+\cO(L^{-7})\, ,  
\label{eq:thresholdE2}
\end{equation}
\begin{align}
  \begin{split}
    \Delta E_3 =&  \frac{12 \pi a_0}{m L^3} \biggl\{ 1 + d_1 \biggl(\frac{a_0}{\pi L}\biggr)
    + d_2 \left(\frac{a_0}{\pi L} \right)^2 +  \frac{64 \pi^2 (a_0)^2 \mathcal{C}_3}{m L^3}
    + \frac{3\pi a_0}{ m^2L^3}  + \frac{6\pi r_0 (a_0)^2}{L^3} \\
    &+ \left( \frac{a_0}{\pi L} \right)^3 \left(d_3 + c_L \log \frac{m L}{2\pi} \right)  \biggr\}- \frac{\mathcal{M}_{3,\text{thr}}}{48 m^3 L^6}  + \cO(L^{-7})\,. 
    \label{eq:thresholdE3}
\end{split}
\end{align} 
Here $c_L$, $\mathcal{C}_3$, and the $c_i$ and $d_i$,
are numerical constant available in the aforementioned references,
and $\cM_{3,\thr}$ is a subtracted three-particle threshold scattering amplitude,
whose definition will be discussed in Appendix~\ref{app:thra2}.

What we observe from these results is that they depend, through $\cO(L^{-5})$, only
on the $s$-wave scattering length, $a_0$, with the effective range $r_0$ first entering
at $\cO(L^{-6})$. There is no explicit dependence on the $d$-wave scattering amplitude
at this order. We can understand this pattern qualitatively as follows.\footnote{%
See also Appendix C in Ref. \cite{Luu:2011ep}.}
The typical relative momentum, $q$, satisfies $\Delta E\sim q^2/m$, and thus, since 
$\Delta E\sim a_0/L^3$, we learn that $q^2\sim a_0/L^3$. 
Using the effective range expansion, Eq.~(\ref{eq:K2swave}),
we then expect that the relative contribution from the $r_0$ term
will be $r_0 a_0 q^2\sim r_0 a_0^2/L^3$, and this is indeed what is seen
in Eqs.~(\ref{eq:thresholdE2}) and (\ref{eq:thresholdE3}).
By the same argument, we expect the $q^4$ terms proportional
to both $P_0$ and $a_2^5$ to appear first at
relative order $\cO(L^{-6})$, and thus contribute to $\Delta E_n$ at $\cO(L^{-9})$.
If this were the case, it would be very challenging to see the dependence of
the threshold energies on $a_2$.

However, it turns out that there is an additional contribution of $\cO(L^{-6})$
to $\Delta E_3$ that depends on $a_2$, and indeed on all higher partial waves,
hidden in $\mathcal{M}_{3,\text{thr}}$. 
In Appendix~\ref{app:thra2} we calculate the leading dependence 
on $a_2$ in a perturbative expansion in the scattering amplitudes, finding
\begin{equation}
m^2 \cM_{3,\thr} \supset {d_\thr} (m a_0)^2 (m a_2)^5 
\left[1 + \cO(a_0) + \cO(a_2^5) \right]\,,\quad
d_\thr=-14110\,.
\label{eq:M3thra2}
\end{equation}
The appearance of $a_2^5$, rather than $a_2$, follows from our parametrization of the
$d$-wave K matrix, Eq.~(\ref{eq:K2dwave}).
In order to isolate the $a_2$ dependence of $\Delta E_3$, we consider the difference
\begin{equation}
\delta E^d(L,a_0,a_2) = \Delta E_3(L,a_0, a_2) -\Delta E_3(L,a_0, a_2=0)\,.
\label{eq:deltaEd}
\end{equation}
Substituting Eq.~(\ref{eq:M3thra2}) into the expression for $\Delta E_3$,
Eq.~(\ref{eq:thresholdE3}), we obtain the theoretical prediction
\begin{equation}
\frac{\delta E^d}{m} = -\frac{d_\thr}{48} \frac{\left(ma_0\right)^2 \left(m a_2\right)^5}{(mL)^6}
 \left[1 + \cO(a_0) + \cO(a_2^5) \right]
+ \cO\left(L^{-7}\right)
\,.
\label{eq:analyticala2}
\end{equation}

\begin{figure}[tb!]
\centering 
\includegraphics[width=.7\textwidth]{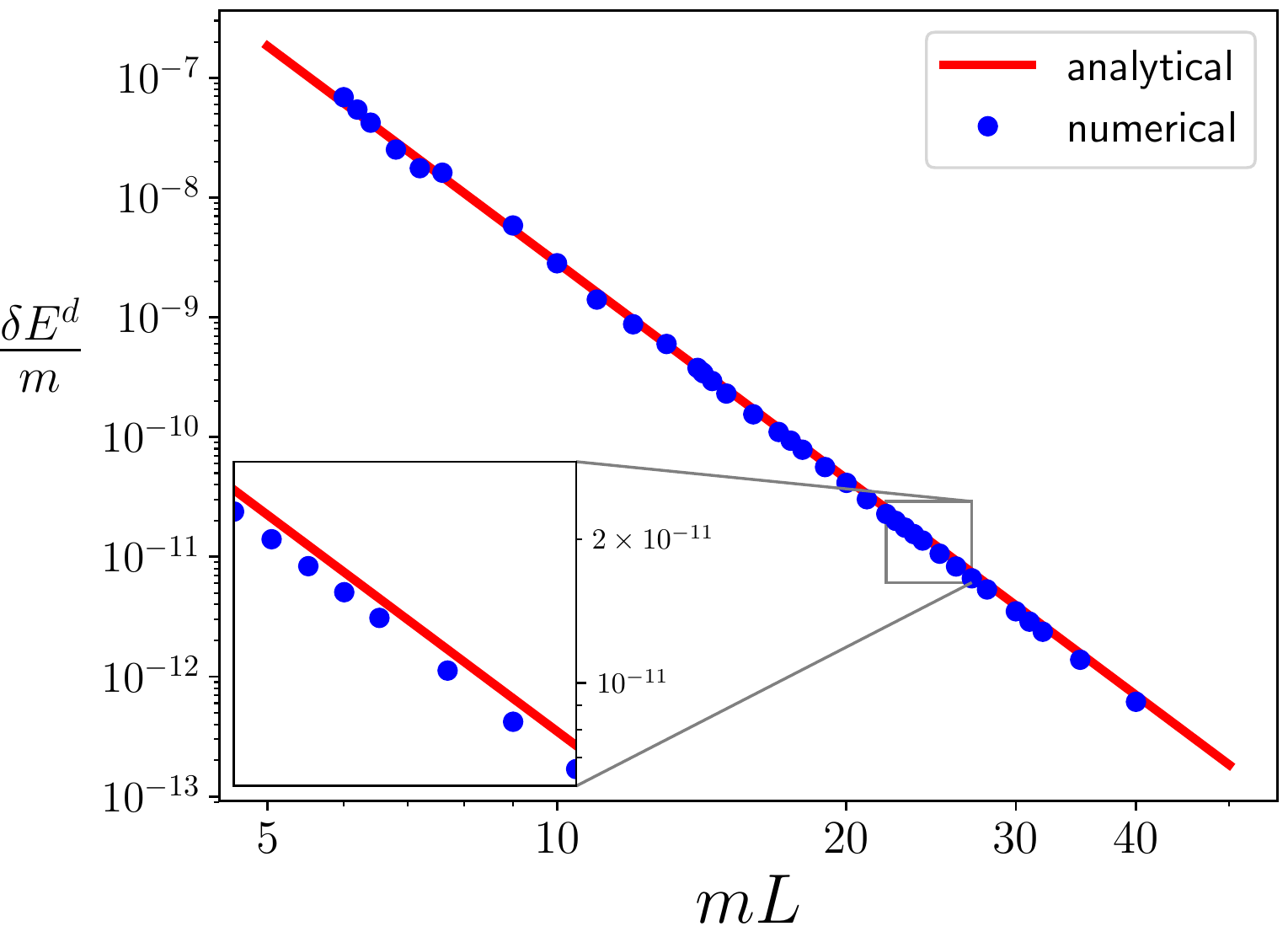}
\caption{\label{fig:thresholda2} Comparison of the analytical prediction 
(which is absolutely normalized) with the results from a numerical solution of the quantization condition. The parameters are $ma_0 = 0.1$, $ma_2=0.25$, and $r_0=P_0=\Kdf=0$.
The lack of linearity for smaller values of $mL$ is related to the opening up of new 
momentum shells.}
\end{figure}

We have checked that the results from numerically solving the quantization condition 
are consistent  with Eq.~(\ref{eq:analyticala2}). 
In particular, we have verified that the leading dependence  on $a_0$, $a_2$ and $1/L$ is
as predicted. An example of the comparison, showing the $L$ dependence,
is given in Fig.~\ref{fig:thresholda2}.
Agreement at the 10\% level holds over many orders of magnitude. 
Based on our tests, we find that the major source of this small discrepancy arises from
terms of higher order in $a_0$.

This comparison provides a strong cross-check of our numerical implementation.
However, for weakly interacting system, such as mesons in QCD, 
one cannot achieve, using lattice calculations,
results for the spectrum with the precision shown in
the figure, nor can one work at such large values of $mL$.
We  now turn to situations in which
$a_2$ has a numerically more significant effect.

 \subsection{Effects of $a_2$ on the three-particle spectrum \label{sec:effecta2} }   
 
We begin by studying the strongly interacting regime, where $m |a_2| \sim 1$. 
 This regime, although hardly conceivable in particle physics, 
 represents an interesting academic problem that is relevant in the physics
 of cold atoms~\cite{PhysRevA.95.032707,PhysRevA.86.062511}.

\begin{figure}[tb!]
\centering 
\subfigure[ \label{fig:specta2A} $A_1^+$ irrep]{\includegraphics[width=.49\textwidth]{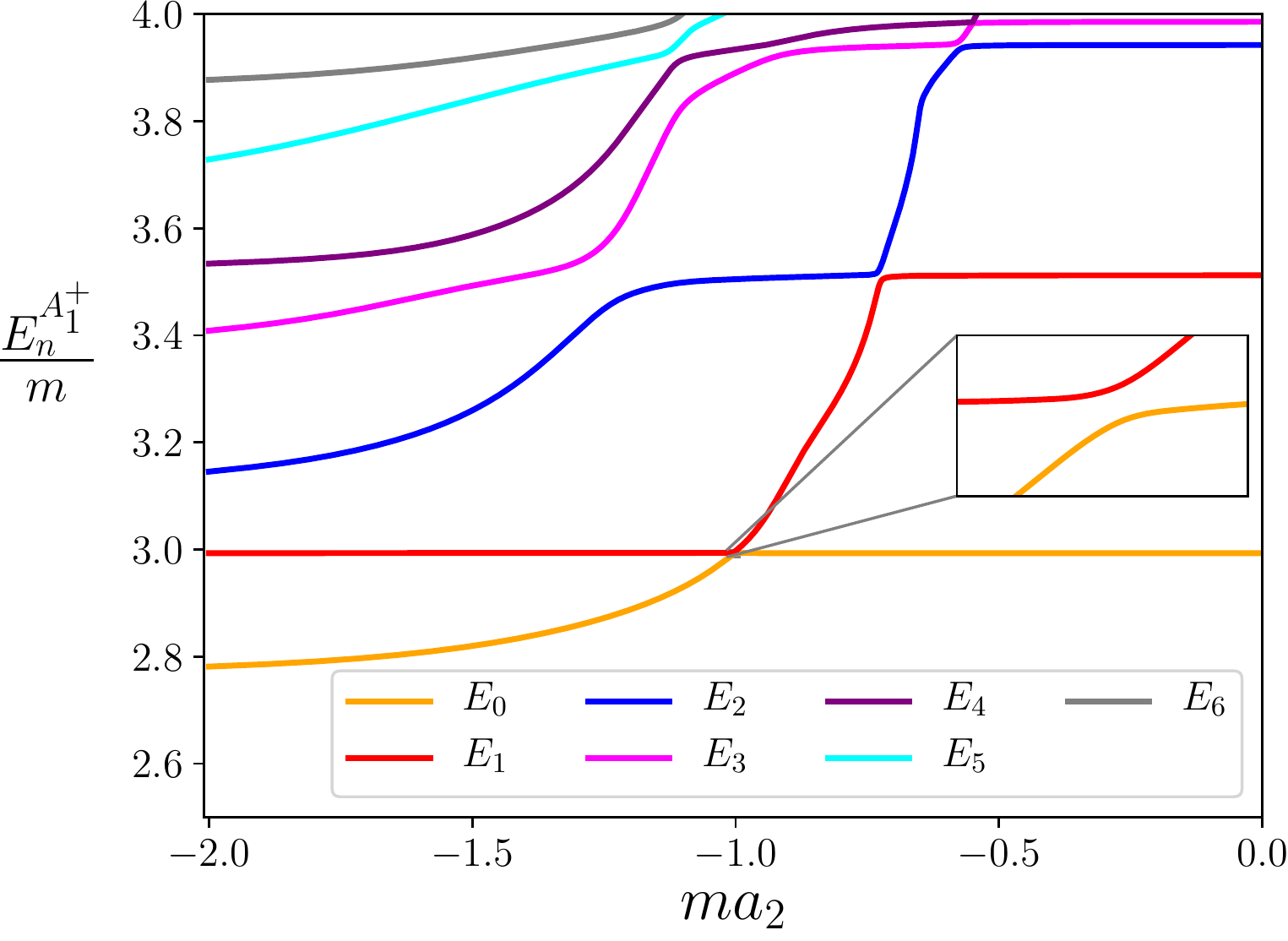}}
\hfill
\subfigure[ \label{fig:specta2B} $E^+$ irrep]{\includegraphics[width=.49\textwidth]{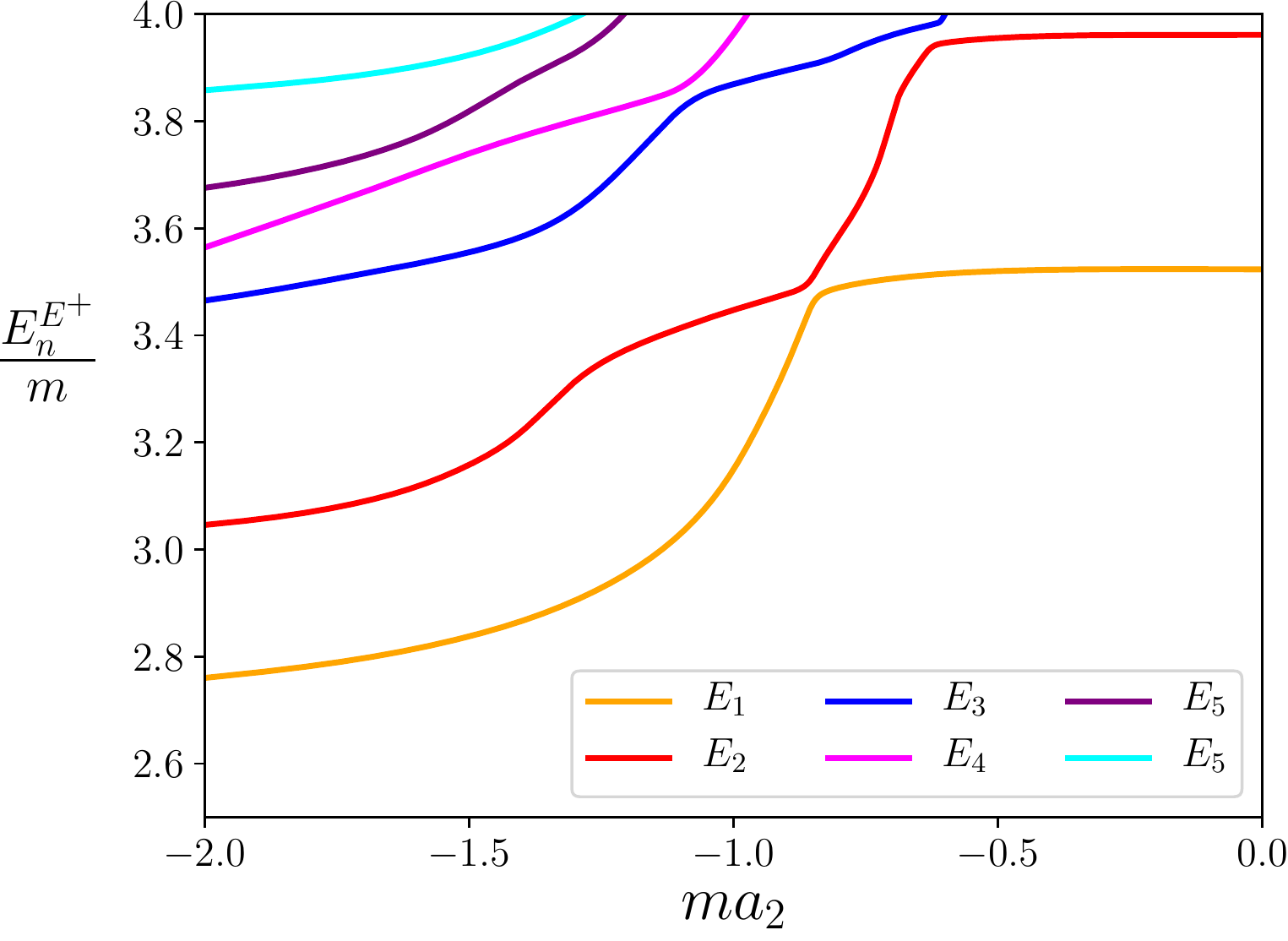}}
\caption{\label{fig:specta2} Energy levels as a function of $ma_2$ in the region $E<4m$ with $mL=8.1$ and $m a_0=-0.1$, $r_0=P_0=\Kdf =0$ in the $A_1^+$ irrep (left) and the $E^+$ irrep (right). }
\end{figure}

In Fig. \ref{fig:specta2}, we show the three particle spectrum for $E<4 m$ in two irreps, 
$A_1^+$ and $E^+$, as a function of negative $m a_2$.
Here we have fixed the volume to $mL=8.1$, and chosen a weakly attractive
$s$-wave interaction, $m a_0=-0.1$, with other scattering parameters set to zero.
We choose negative values for $m a_2$ in order to avoid the possibility of a pole
in $\K_2^{(2)}$, Eq.~(\ref{eq:K2dwave}), for which our formalism breaks down.
Note that negative $a_2$ corresponds, at least for small magnitudes,
to an attractive interaction, as seen from the result for $\delta E^d$, Eq.~(\ref{eq:analyticala2}).
Since we use a small value of $m |a_0|$, the energy levels at the right-hand edges of both
plots (where $a_2=0$) lie close to the energies of three noninteracting particles
(which are $E/m=3$, $3.53$, $3.97$, $4.02, \dots$ for $mL=8.1$).
As $m |a_2|$ increases, the energies are almost flat, until at a value 
$m |a_2| \sim 1$, the levels shift rapidly downwards. 
This shift begins at smaller values of $m|a_2|$ for excited states. 
As the magnitude of $a_2$ increases, the excited states approach lower-lying states until an avoided level crossing occurs.   We also observe that 
 states in the $E^+$ irrep are more sensitive to $d$-wave interactions, 
 which seems to be a general feature, as will be seen in the following section.

\begin{figure}[tb!]
\centering 
\includegraphics[width=.49\textwidth]{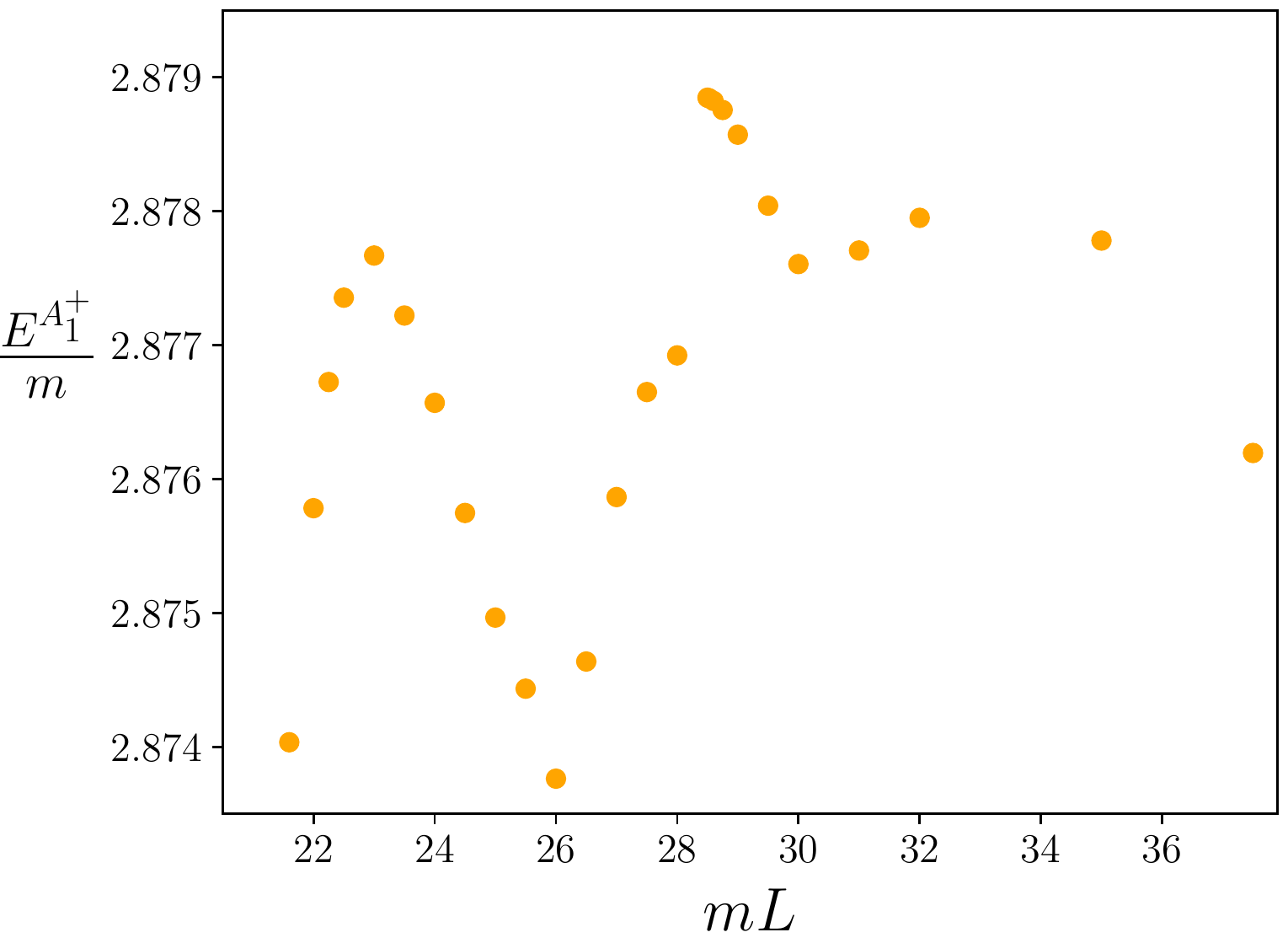}
\caption{\label{fig:boundL} Energy of the subthreshold state in the $A_1^+$ irrep as a function of $mL$. The parameters are $m a_0 = -0.1$, $m a_2 =-1.3$ and $r_0=P_0=\Kdf=0$. 
Note the highly compressed vertical scale.}
\end{figure}

Another interesting observation from Fig.~\ref{fig:specta2}
is the presence of a deep subthreshold state for $m|a_2| > 1$.
This resembles the Efimov effect, which describes a three-particle bound state arising 
from an attractive two-particle interaction $m|a_0| \gg 1$~\cite{EFIMOV1970563}. 
The Efimov bound state has been reproduced numerically with only $s$-wave interactions present, both in the NREFT approach~\cite{Hammer:2017kms,Doring:2018xxx} and in the isotropic approximation of the RFT formalism~\cite{Briceno:2018mlh}. 
Moreover, there is some theoretical work regarding the existence of this generalized Efimov 
scenario in the presence of $d$-wave interactions~\cite{PhysRevA.86.062511},
although there is no result concerning the asymptotic volume dependence, 
unlike in the $s$-wave case~\cite{Meissner:2014dea}. 
We have been able to solve the quantization condition numerically
up to $mL = 37.5$ and the bound state energy barely changes, 
which strongly suggests that it is indeed an infinite volume bound state. 
Results for $m a_2=-1.3$ are shown in Fig.~\ref{fig:boundL}.
The volume dependence of the energy is dominated by effects of the UV cut-off,
which manifest themselves as small oscillations when a new shells become active.
These are similar to oscillations observed in several quantities in Ref.~\cite{\BHSnum}.

We close by commenting on the impact of using a relativistic formalism, as opposed to
a NR approach, on the results of this section. We expect that the qualitative
features of the results will be unchanged, but that the quantitative energy levels will be
changed once they differ significantly from $3 m$. Thus,
for example, we expect that the energy of the subthreshold state will be only slightly changed,
since it lies at the border of the NR regime.

\subsection{Application: spectrum of $3\pi^+$ on the lattice \label{sec:pipipi}}

The simplest application in QCD for the three-particle quantization condition is the 
$3\pi^+$ system, not only from the theoretical point of 
view---no resonant subchannels---but also from the technical side---no 
quark-disconnected diagrams and a good signal/noise ratio. 
Here we use our formalism to predict the $3\pi^+$ spectrum,
using values for the two-body scattering parameters determined from experiment,
and a range of choices for the parameters in $\Kdf$.\footnote{%
We ignore QED effects, which are numerically small,
and, in any case, cannot be incorporated into the present formalism.}
Our focus will be on how to differentiate effects arising 
from the different components of $\Kdf$, listed in Eq.~(\ref{eq:termsKdf}).

An important point in the following is that
that there is no natural size for the parameters in $\Kdf$:
the magnitudes of
the dimensionless coefficients $\Kiso$, $\Kisoone$, $\Kisotwo$, $\KA$, and $\KB$
are not constrained.
Strictly speaking, we know this only for $\Kiso$, because, in the nonrelativistic limit,
it is related to the three-particle contact interaction in NREFT (a relation given
explicitly in Ref.~\cite{HSreview}), and it is well known that the latter interaction varies
in a log-periodic manner  from $-\infty$ to $\infty$ as the cutoff varies~\cite{Bedaque:1998kg}.
But we see no reason why this should not also apply to the other coefficients.
In particular, we note that the physical three-particle scattering amplitude, $\cM_3$,
does not diverge when $\Kdf$ does~\cite{\HSQCb,\BHSnum}.

We take the parameters describing isospin-2 $\pi \pi$ scattering from Ref.~\cite{Yndurain:2002ud}:
\begin{equation}
m_\pi a_0 = 0.0422, \ \ \ m_\pi r_0 = 56.21, \ \ \ P_0 = -3.08 \cdot 10^{-4}, \ \ \ m_\pi a_2 = -0.1867\,. \label{eq:I2parameters}
\end{equation}
In a  lattice simulation, these parameters would be extracted from the two-pion
spectrum, using the two-particle quantization condition.
Indeed, there is considerable recent work on the $2\pi^+$ system using lattice
methods, in some cases incorporating $d$-wave 
interactions~\cite{Beane:2011sc,Dudek:2012gj,Fu:2013ffa,Kurth:2013tua,Helmes:2015gla,Bulava:2016mks}.
We emphasize that one must determine these parameters with high precision
in order to disentangle the two- and three-body effects in the three-particle spectrum.


For the relatively weak two-particle interactions of Eq.~(\ref{eq:I2parameters}),
the energy levels lie close to the noninteracting energies of Eq.~(\ref{eq:Efree}).
For the regime of box sizes available in current lattice simulations,
$4 \lesssim m_\pi L \lesssim 6 $,
there are at most three such levels  below the five-particle threshold,  $E=5 m_\pi$
(above which the quantization condition breaks down). 
For these levels,  the solutions lie in three irreps:
$\Gamma = A_1^+, E^+, T_2^+$ (see Table~\ref{tab:freeirreps}).
We denote the difference between the actual energy and its noninteracting value as
\begin{equation}
\Delta E_n^\Gamma = E_n^\Gamma - E_n^{\text{free}} 
\end{equation}
where $n=0,1,\dots$ labels the levels following the numbering scheme of Table~\ref{tab:freeirreps}. 
It is known that, asymptotically,~\cite{Akakiprivate}
\begin{equation}
\Delta E_n^\Gamma \propto \frac{a_0}{m L^3} + \cO(L^{-4})\,.
\end{equation}
We stress, however, that the asymptotic result is
not numerically accurate for the range of $mL$ that we consider.

\begin{figure}[tb!]
\centering 
\subfigure{\includegraphics[width=.49\textwidth]{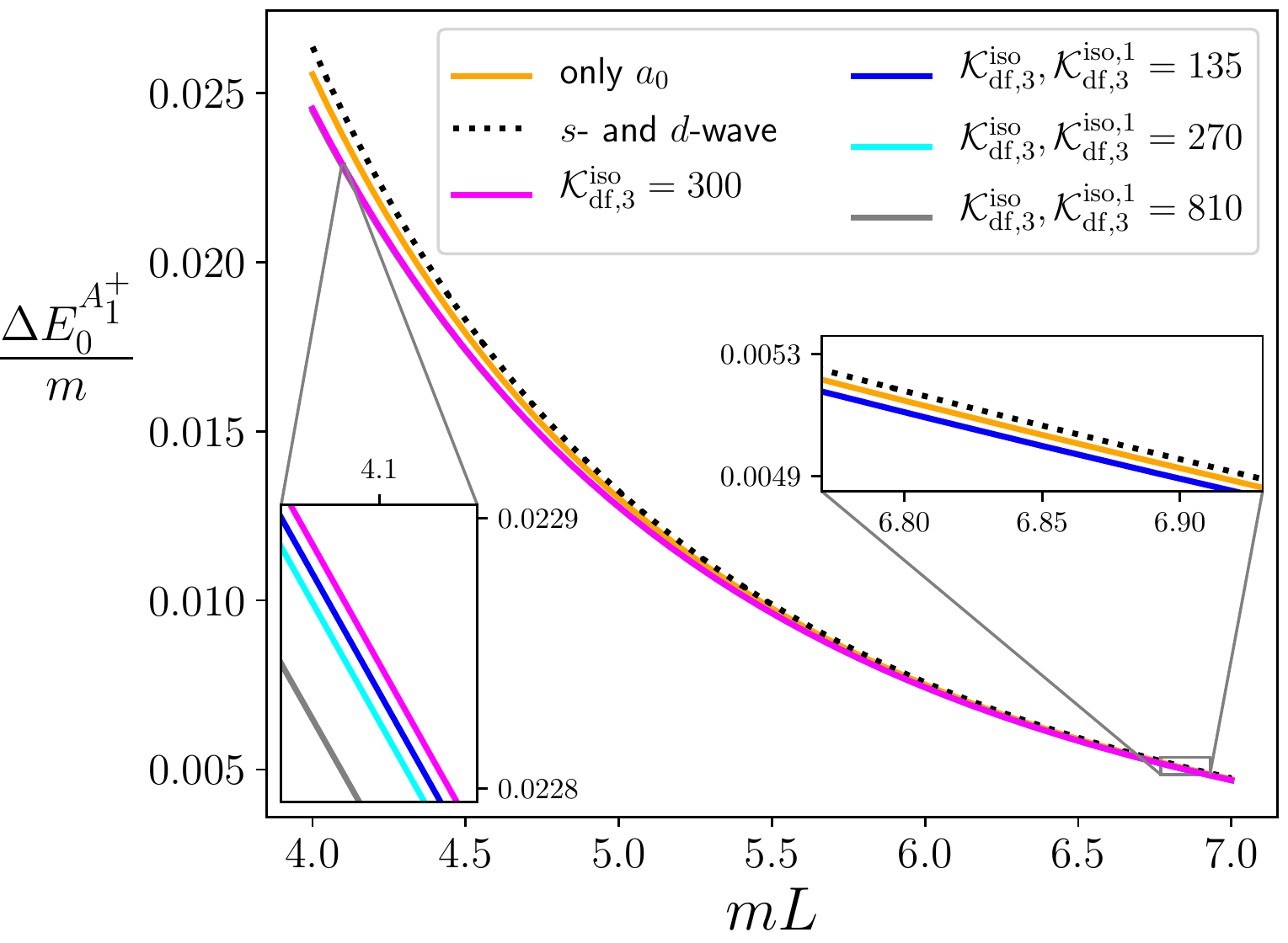}} 
\hfill
\subfigure{\includegraphics[width=.49\textwidth]{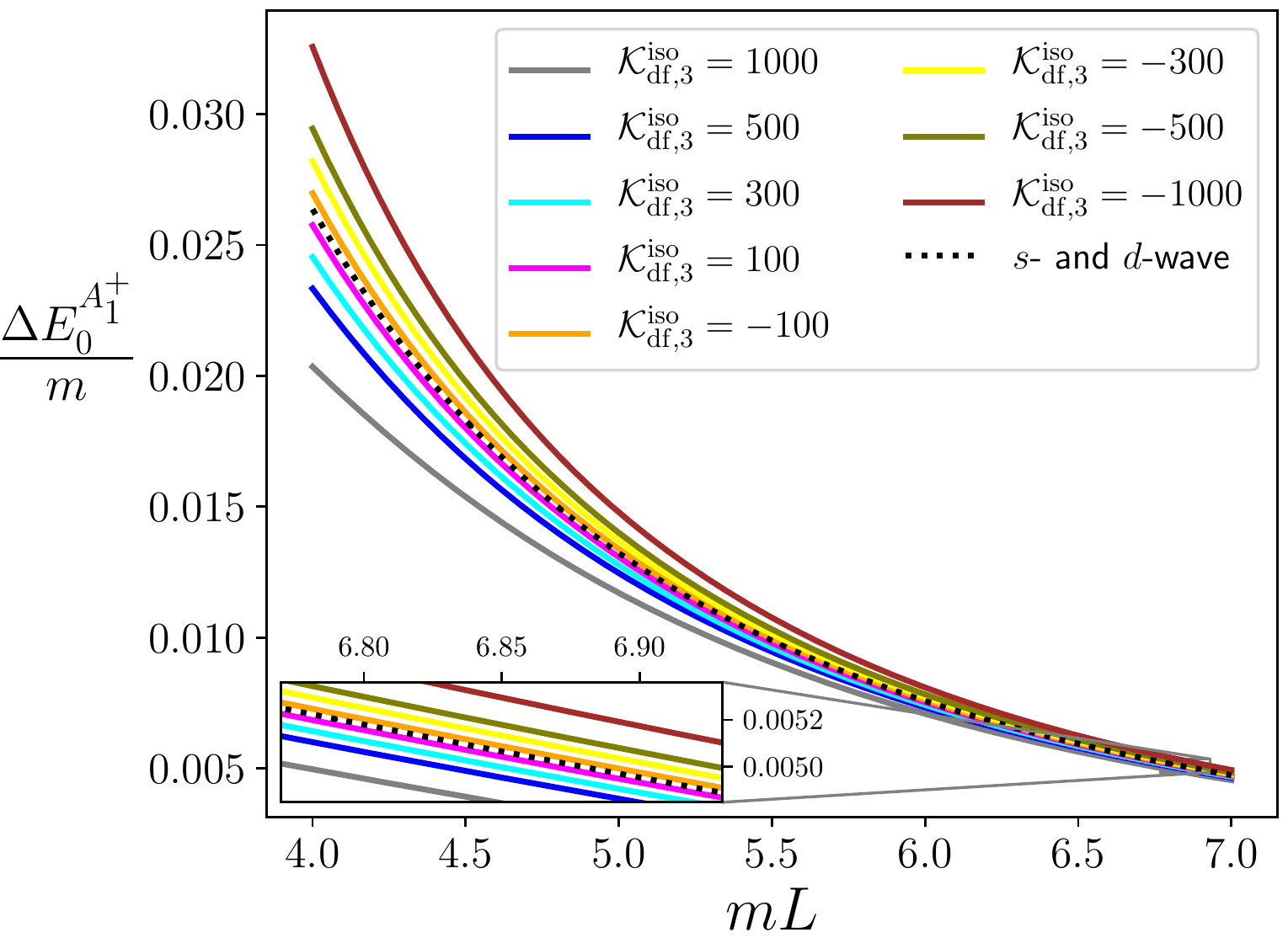}} 
\caption{\label{fig:ground} Energy shift for the 
ground state in the $A_1^+$ irrep, for which $E_0^{\rm free}=3m$.
The two-particle scattering parameters are those in 
Eq.~\ref{eq:I2parameters}, aside from the orange curve in the left panel,
where only $a_0$ is nonzero. The three particle scattering parameters are
as indicated in the legend, and explained further in the text. We use the convention that
a parameter value not given explicitly is set to the value given earlier. 
For example, the blue line in the left panel
has the parameters set to $\Kiso=300$ and $\Kisoone=135$,
while $\Kisotwo=\KA=\KB=0$.
}
\end{figure}

Let us start from the ground state, which lies in the $A_1^+$ irrep.
Here our expectations are guided by the threshold expansion,
Eq.~(\ref{eq:thresholdE3}). In addition to explicit dependence on $a_0$ and $r_0$,
and the implicit dependence on $a_2$ worked out in Sec.~\ref{sec:thresholda2},
the energy depends on $\Kdf$ through the $\cM_{3,\thr}/L^6$ term.
Following the arguments given in Sec.~\ref{sec:thresholda2}, we expect that
only $\Kiso$ will enter at this order, with dependence on $\Kisoone$ suppressed by 
$1/L^3$ and that on $\Kisotwo$, $\KA$ and $\KB$ by $1/L^6$.
This is borne out by our numerical results, shown in Fig.~\ref{fig:ground}.
The left panel compares results with several choices of parameters:
(i) those of Eq.~(\ref{eq:I2parameters}) plus $\Kdf=0$ (labeled
``$s$- and $d$-wave''---black, dotted line);
(ii) the same as (i) but with $\Kiso=300$ and all other parameters
in $\Kdf$ vanishing (magenta);
(iii) the same as (ii) but with $\Kisoone$ also turned on, taking the
three values $135$ (blue), $270$ (cyan) and $810$ (grey);
and (iv) the isotropic approximation, i.e., with only $s$-wave interactions,
and $a_0$ the only nonzero scattering parameter (orange).
We see that adding $d$-wave two-particle interactions has a similar
impact to adding $\Kiso=300$, but that adding $\Kisoone$ with a similar
magnitude has almost no impact.

The right panel shows the dependence on $\Kiso$, with other parameters
fixed at the values in Eq.~(\ref{eq:I2parameters}). The range we consider is
$\Kiso=[-1000,+1000]$. In order to have sensitivity to $\Kiso$ in this range,
a determination of $\Delta E_0/m$ with an error  of $\approx 0.01$ is needed.
Such an error can be achieved with present methods.
Thus, as noted in Ref.~\cite{\BHSnum}, if one has a sufficiently accurate knowledge
of the two-particle scattering parameters, one can use the ground state energy
to determine the leading three-particle parameter $\Kiso$. Indeed, this approach
has been carried out successfully in
Refs.~\cite{Detmold:2008fn,Romero-Lopez:2018rcb}.

\begin{figure}[tb!]
\centering 
\includegraphics[width=.49\textwidth]{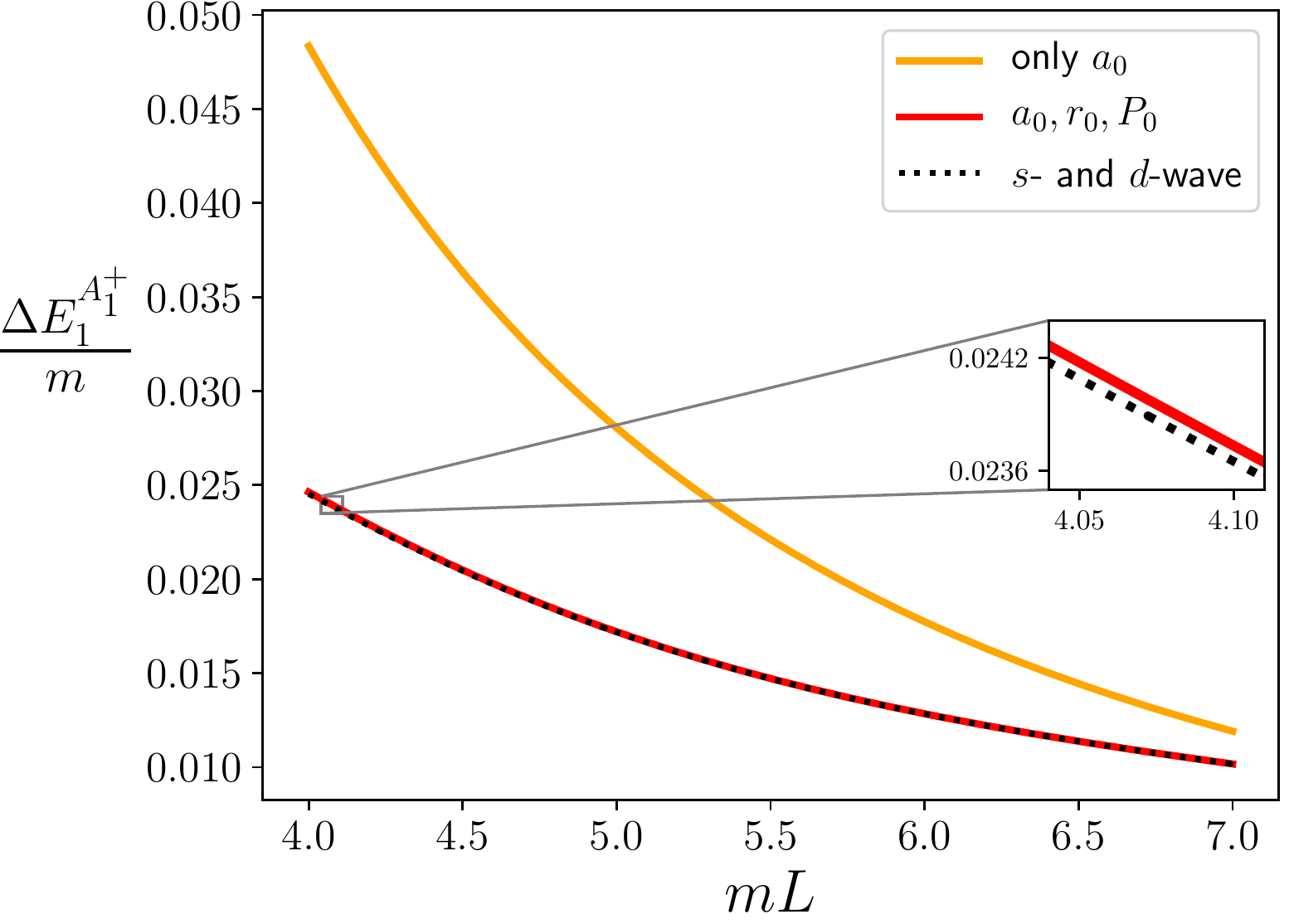}
\hfill
\includegraphics[width=.49\textwidth]{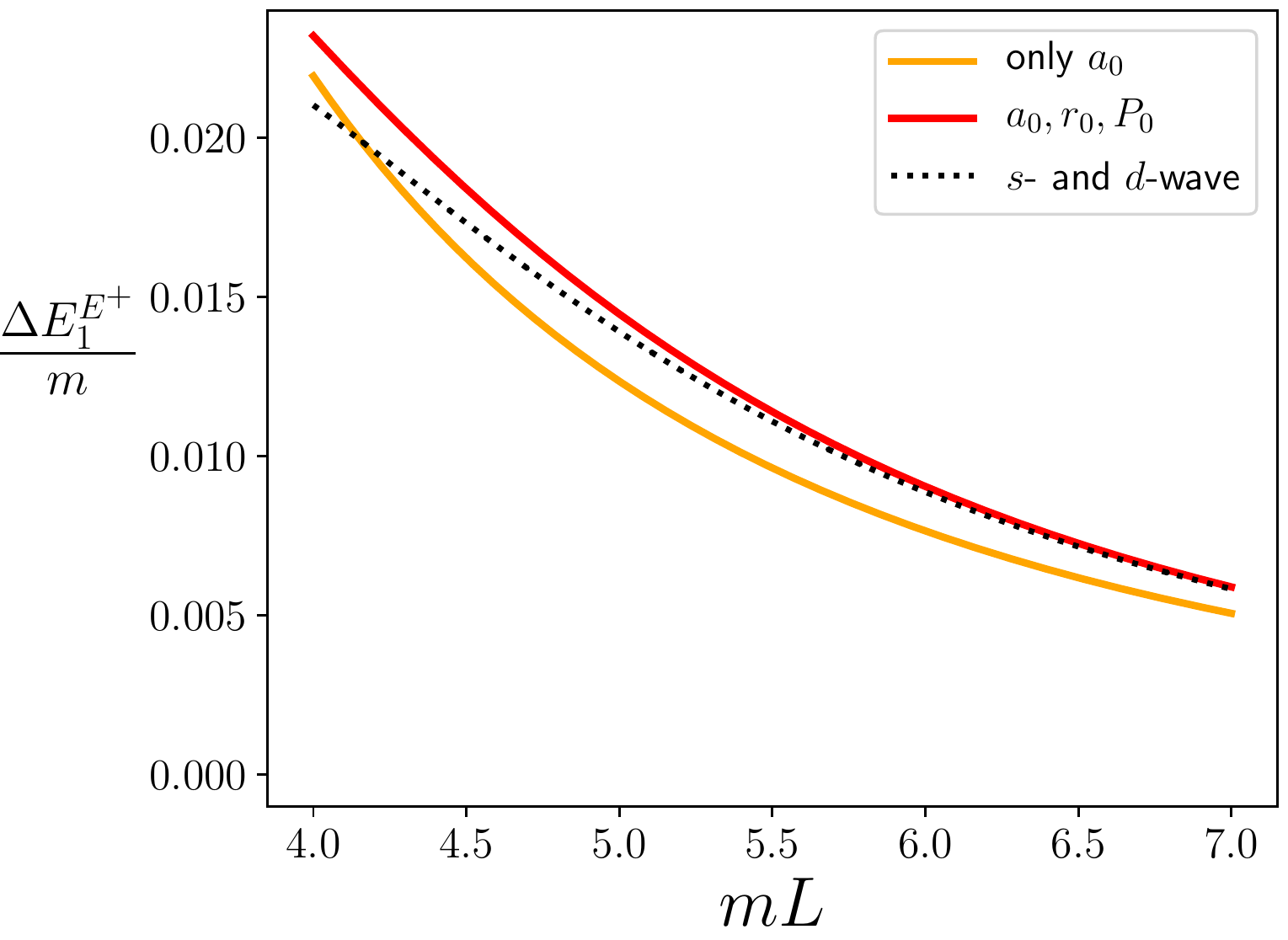}
\caption{\label{fig:1exc} Energy shift of the first excited state in the $A_1^+$ irrep (left) and $E^+$ irrep (right). In the range of $mL$ shown, $E_1^{\rm free}/m= 4.7-3.9$.
The quantization condition is solved with
only two-particle scattering parameters being nonzero, while $\Kdf=0$.
When a parameter is nonzero, its value is given by Eq.~(\ref{eq:I2parameters}).
The solid orange and red curves include only $s$-wave dimers, the former having only
$a_0$ turned on (``only $a_0$''), with the latter having all three $s$-wave parameters 
in $\K_2$ nonzero (``$a_0, r_0, P_0$''). 
The dotted black line shows the impact of adding $d$-wave dimers, with $a_2$ nonzero
(``$s$- and $d$-wave'').
}
\end{figure}

In Fig.~\ref{fig:1exc}, we investigate the sensitivity of the energy of the first excited state 
to the various two-particle scattering parameters, comparing the two irreps that are present.
The magnitude of the energy shifts are comparable to those for the ground state, but
the dependence on the scattering parameters differs markedly.
This can be understood because the relative momenta between the particles is
nonvanishing for the excited state.
Denoting generically the relative momenta by $q$, 
this satisfies $q/m \approx 2\pi/(mL) \sim \cO(1)$.
Because of this we expect that the
higher-order terms in the effective range expansion, i.e. $r_0$ and $P_0$, should play
a much more significant role. This is borne out by the results in the figure,
particularly for the $A_1^+$ irrep. We observe that the effect of these additional terms
is opposite in the two irreps, which is consistent with the prediction of the threshold
expansion generalized to excited states~\cite{Akakiprivate}.
We also see that adding $d$-wave dimers has almost no impact on the $A_1^+$ irrep
(indeed, the effect is smaller than for the ground state)
while the impact is comparable to that of $r_0$ and $P_0$ for the $E^+$ irrep.
Qualitatively, this is as expected, since the averaging over orientations in the $A_1^+$
irrep suppresses the overlap with $d$-wave dimers.

\begin{figure}[tb!]
\centering 
\subfigure{\includegraphics[width=.49\textwidth]{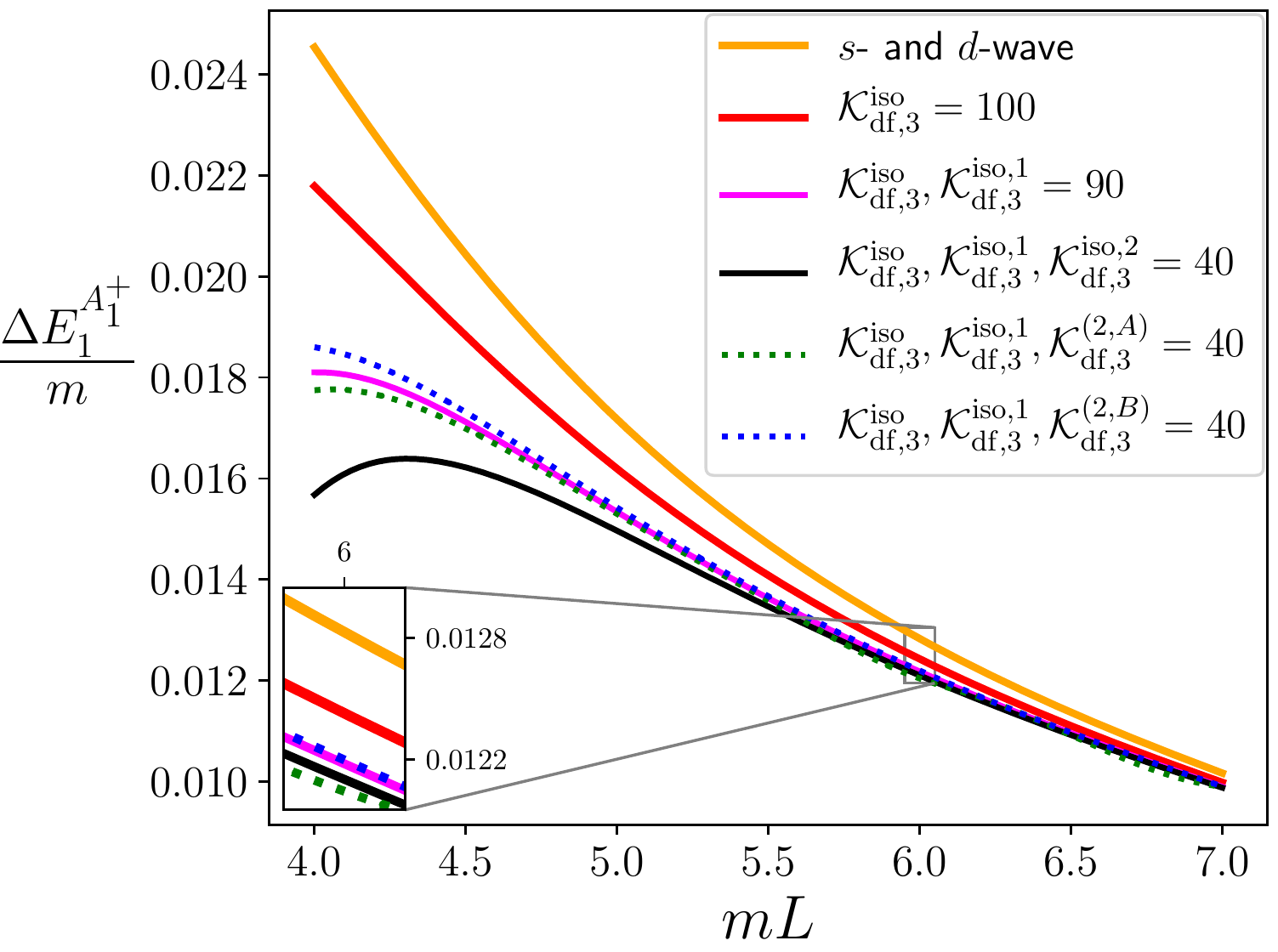}} 
\hfill
\subfigure{\includegraphics[width=.49\textwidth]{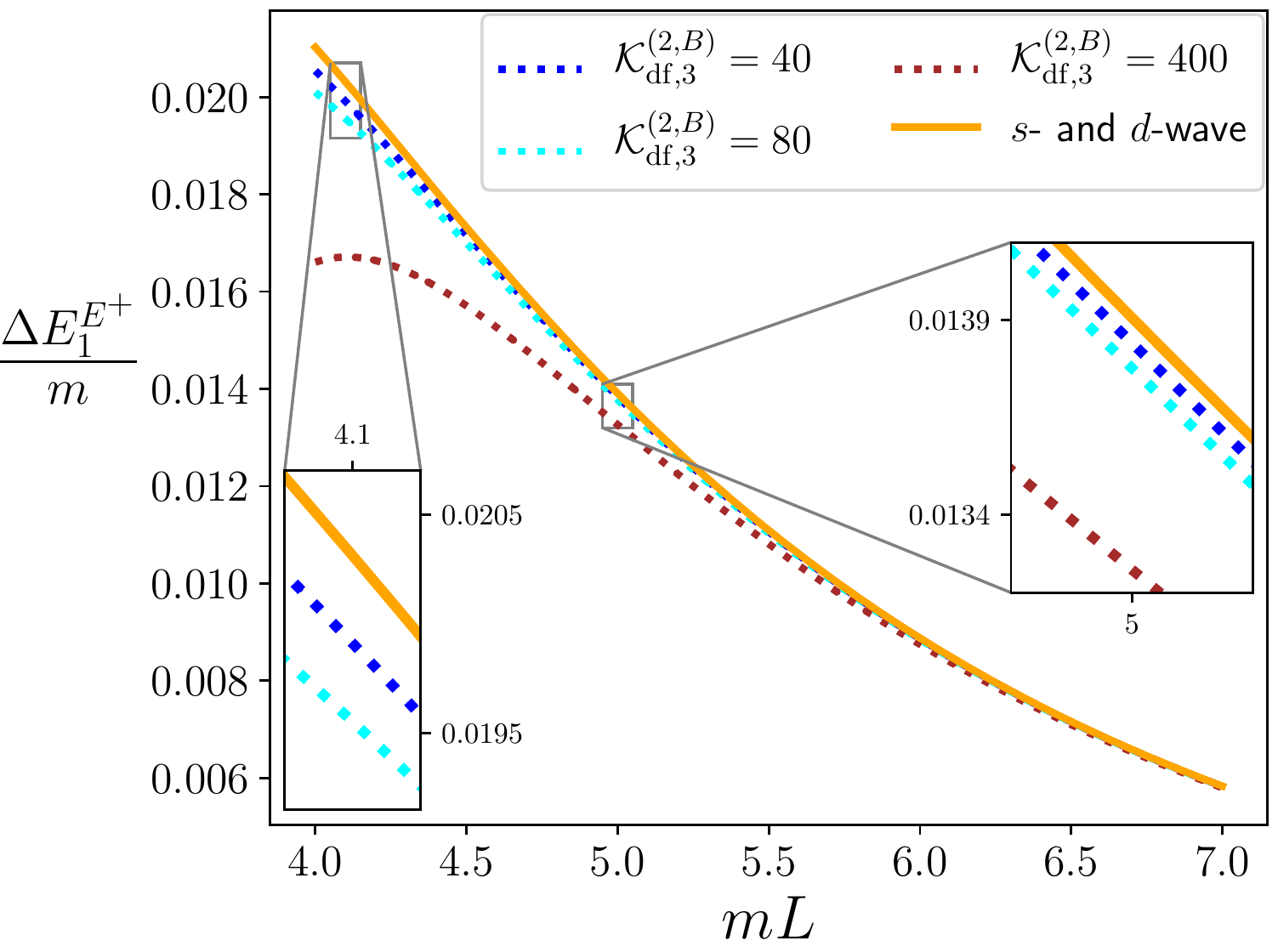}} 
\caption{\label{fig:1excK} Energy shift of the first excited state in the $A_1^+$ irrep (left) and $E^+$ irrep (right) with various choices of the parameters in $\Kdf$. 
The two-particle scattering parameters are given by Eq.~(\ref{eq:I2parameters}) for all curves.
The choices of $\Kdf$ parameters is explained by the legend, with the convention that
a parameter value not given explicitly is set to the value given earlier.
For example, the black line has the parameters set to $\Kiso=100$, $\Kisoone=90$,
and $\Kisotwo=40$, while $\KA=\KB=0$.
}
\end{figure}

In Fig.~\ref{fig:1excK} we illustrate the dependence of the same two excited states
on the five parameters in $\Kdf$, Eq.~(\ref{eq:termsKdf}).
Because $q/m\sim\cO(1)$ we expect that, unlike for the ground state,
the energy should be sensitive to all five parameters, and not just to $\Kiso$.
This is borne out for the $A_1^+$ irrep, where there is strong sensitivity to
all three isotropic parameters, and a somewhat weaker dependence on 
$\KA$ and $\KB$.
As noted above, only $\KB$ affects the $E^+$ irrep, and Fig.~\ref{fig:1excK} 
illustrates this dependence.

\begin{figure}[tb!]
\centering 
\subfigure{\includegraphics[width=.49\textwidth]{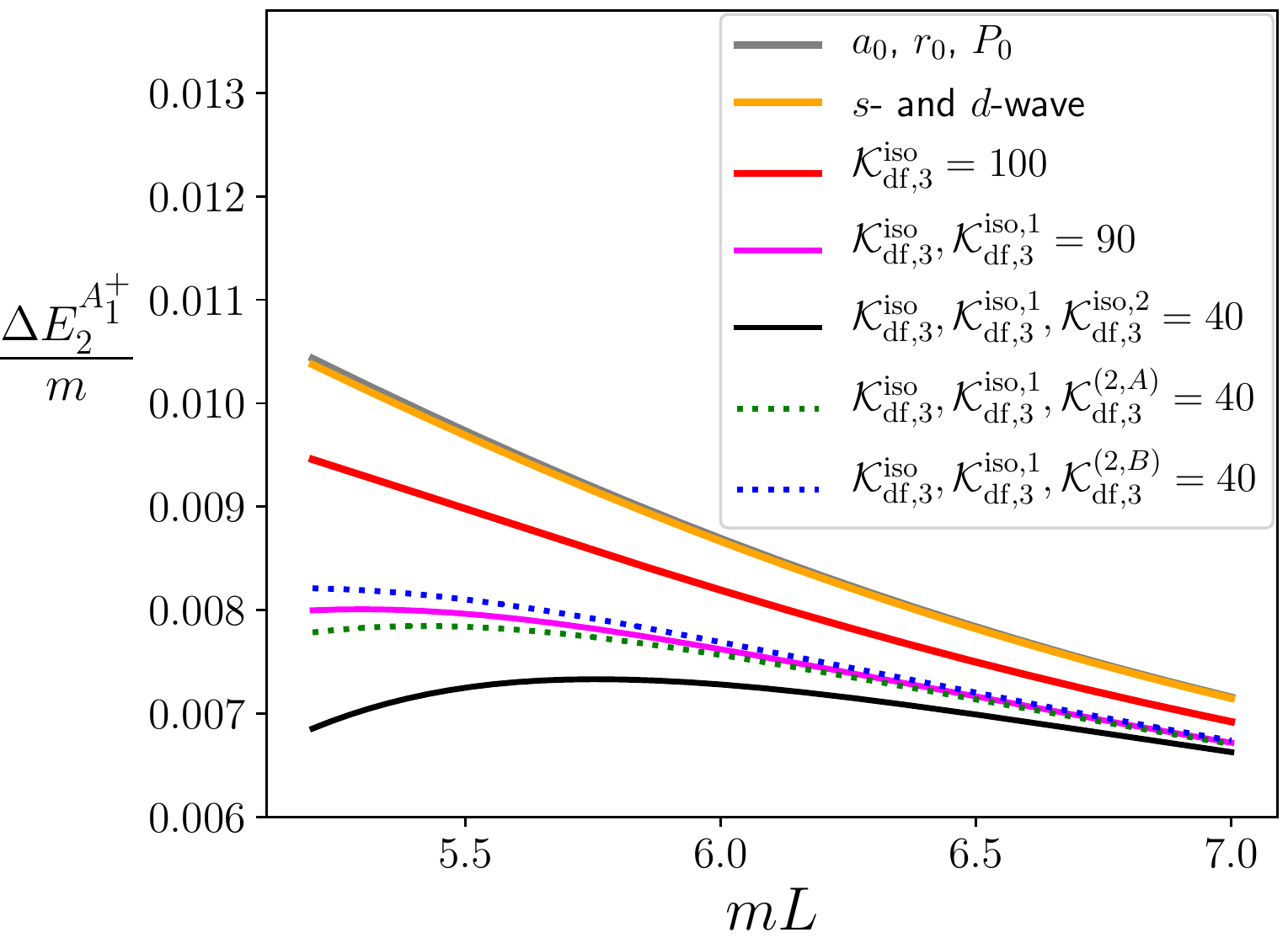}} 
\hfill
\subfigure{\includegraphics[width=.49\textwidth]{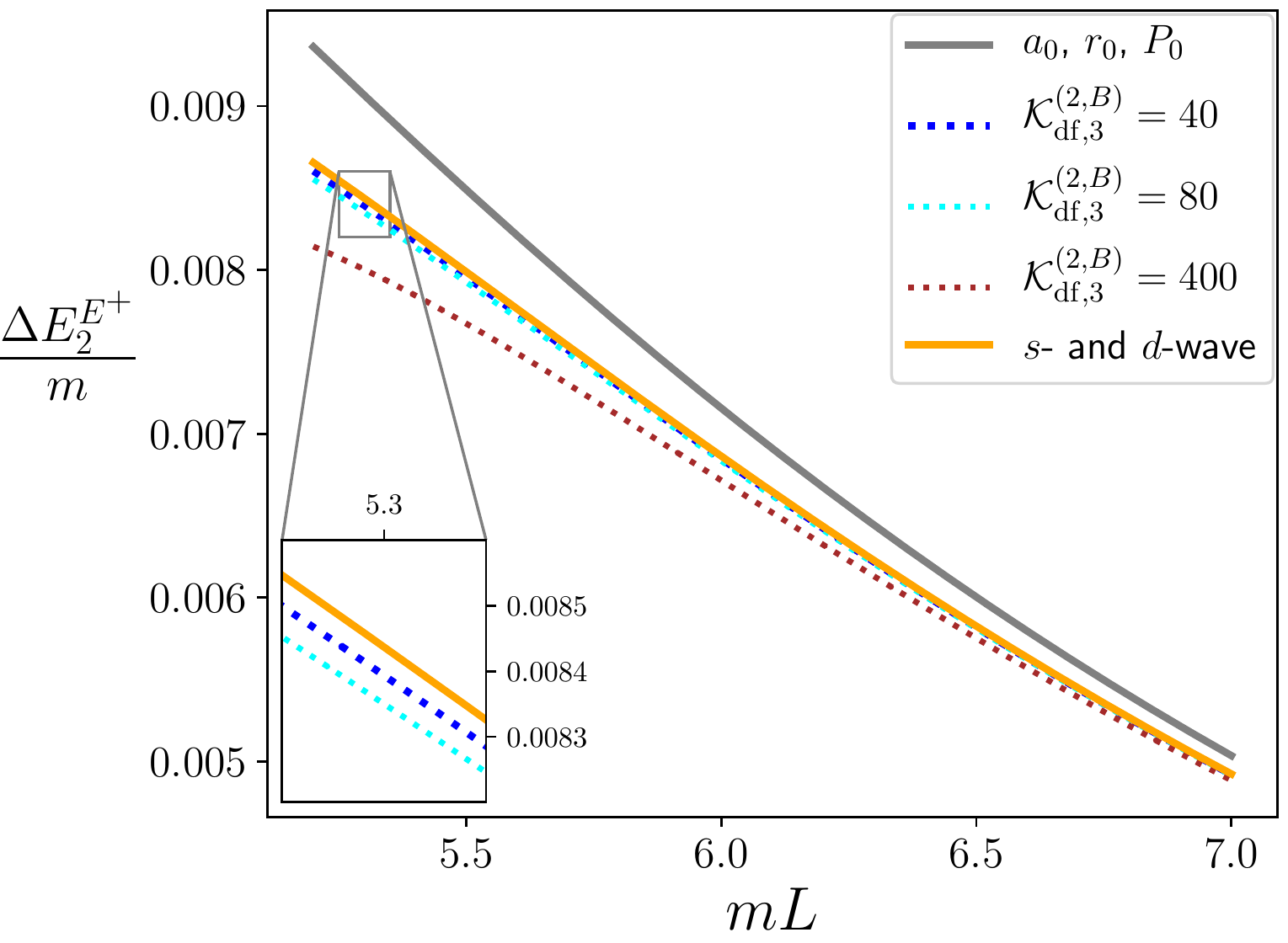}} 
\hfill
\subfigure{\includegraphics[width=.49\textwidth]{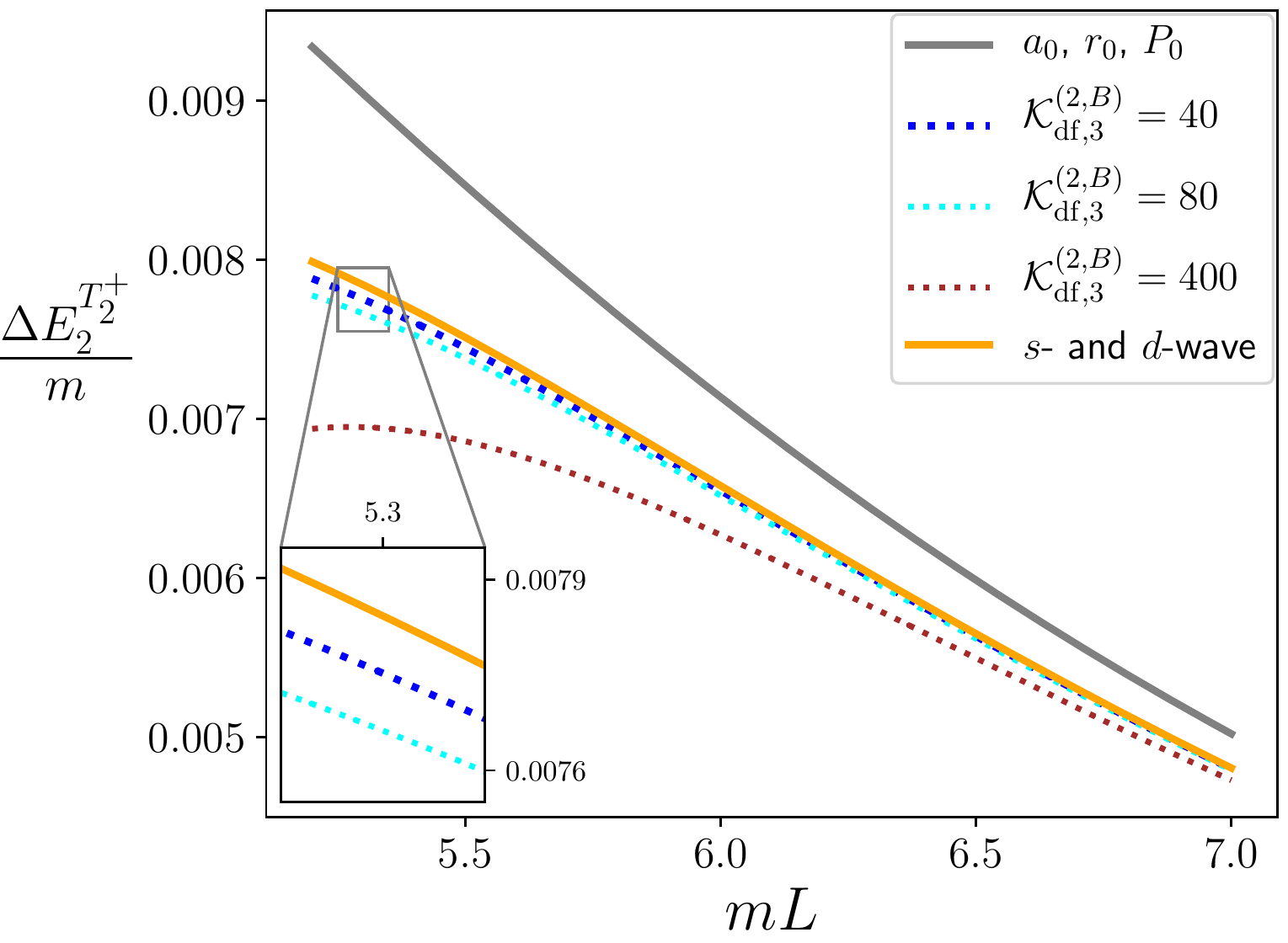}} 
\hfill
\caption{\label{fig:2excK} Energy shift of the second excited states in the $A_1^+$ irrep (top left), the $E^+$ irrep (top right) and $T_2^+$ irrep (bottom). The meaning of the legend
is as in previous figures.}
\end{figure}

The energy shift for the second excited states are shown in Fig.~\ref{fig:2excK}.
We show results only for those volumes for which the states
lie below the five-particle threshold, which requires $mL\gtrsim 5.2$.
The $A_1^+$ energy-shift depends on all parameters in $\Kdf$, 
while the $E^+$ and $T^+_2$ irreps depend only on $\KB$. 
The results show a similar dependence on parameters as for
the first excited states.
We also find that the $E^+$ and $T_2^+$ irreps show the greatest sensitivity to $a_2$
of all the states considered.

To sum up, a possible program for determining the coefficients in $\Kdf$ 
up to quadratic order in the threshold expansion is as follows:
\begin{enumerate}
\item Determine $a_0$, $r_0$, $P_0$, and $a_2$ from the two-body sector using 
standard two-particle methods.
\item Extract $\Kdfiso$ from the threshold state.
\item Use states in the $E^+$ and $T_2^+$ irreps to calculate $\KB$.
\item Use the excited states in the $A_1^+$ irrep to obtain the rest of the parameters.
The most difficult parameter to determine would be $\KA$, because
its contribution to the energy is smaller. 
\end{enumerate}
Further information could be obtained using moving frames,
as has been done very successfully in the two-particle case.
The formalism of Ref.~\cite{\HSQCa} is still valid, but the detailed implementation 
along the lines of this paper has yet to be worked out.

We close by commenting on the importance of using a relativistic formalism for the results
that we have presented in this section. We note that the excited states whose energies
we consider lie in the relativistic regime. For example, at $mL=5.5$, the relativistic
noninteracting energy of the second excited state is $E_2^{\rm free}=4.80 m$,
to be compared to the nonrelativistic energy $3 m+ 2 m(2\pi/(mL))^2=5.61 m$.
Nevertheless, it may be that the energy splittings $\Delta E_n^\Gamma$ are  much less
sensitive to relativistic effects, and it would be interesting to implement the NREFT
approach including $d$ waves in order to study this. We do expect, however, that
the parametrization of the three-particle interaction will require additional terms
once the constraints of relativistic invariance are removed.



\subsection{Unphysical solutions}
\label{sec:unphysical}

In this section we describe solutions to the quantization condition that are,
for various reasons, unphysical. 
These fall roughly into two classes (although there is some overlap): 
solutions that occur at the energies of three noninteracting particles 
(which we refer to as ``free solutions'', occurring at ``free energies''), 
and solutions that correspond to poles in the finite-volume correlator that
have the wrong sign of the residue.
The latter were first observed in Ref.~\cite{\BHSnum} within the isotropic approximation.
In the following, we begin with a general discussion of
the properties of physical solutions,
and then discuss the two classes of unphysical solutions in turn.

\subsubsection{General properties of physical solutions}
\label{sec:physical}

We recall here the properties that physical solutions to the quantization condition,
Eq.~(\ref{eq:QC3}), must obey. This extends the analysis presented in Ref.~\cite{\BHSnum}
for the isotropic approximation.

The key quantity is the two-point correlation function in Euclidean time,
\begin{equation}
\wt C_L(\tau) = \langle 0| \cO(\tau) \cO^\dagger(0)|0 \rangle\,,
\label{eq:CLtaudef}
\end{equation}
where the operator $\cO^\dagger$ has the correct quantum numbers to
create three particles (and here also has $\vec P=0$). 
We stress that its hermitian conjugate is used to destroy the states.
Inserting a complete set of finite-volume states with appropriate quantum numbers, 
we find the standard result
\begin{equation}
\widetilde C_L(\tau) = \sum_j \frac{c_j}{2 E_j} \exp(-E_j |\tau|)\,,
\end{equation}
where $E_j>0$ are the energies relative to the vacuum, and the $c_j$ are real and positive.
Fourier transforming to Euclidean energy and Wick rotating yields
\begin{equation}
C_L(E) = \sum_j  c_j \frac{i}{E^2- E_j^2} = \sum_j \frac{i c_j}{(E+E_j)(E-E_j)}\,,
\label{eq:CLE}
\end{equation}
where $E$ is the Minkowski energy that appears in the quantization condition.
Thus $C_L(E)$ is composed of single poles whose residues, for $E>0$,
are given by $i$ times real, positive coefficients.

Next we recall from the analysis of Ref.~\cite{\HSQCa} that the correlator can also
be written as
\begin{equation}
C_L(E) = A^\dagger \frac{i}{F_3^{-1} + \Kdf} A = \sum_j
|A^\dagger \cdot v_j(E)|^2 \frac{i}{\lambda_j(E)} \,,
\label{eq:CLEa}
\end{equation}
where $A$ is a column vector, and to obtain the second form we have
decomposed $F_3^{-1}+\Kdf$ in terms of its eigenvalues $\lambda_j(E)$
and eigenvectors $v_j(E)$.\footnote{%
For the sake of brevity, we do not show explicitly that the quantities also depend on $L$.}
Since $F_3^{-1}+\Kdf$ is real and symmetric, the eigenvalues are real.

It follows from comparing Eqs.~(\ref{eq:CLE}) and (\ref{eq:CLEa}) that
\begin{itemize}
\item[(a)]
$\lambda_j(E)$ cannot have double zeros. 
This is because, in the vicinity of a double zero at $E_j$,
$C_L(E)$ would have a double pole, $C_L(E) \propto 1/(E-E_j)^2$.
The same prohibition applies to higher-order zeros.
\item[(b)]
Eigenvalues of $F_3^{-1}+\Kdf$ that pass through zero (and thus lead to
solutions to the quantization condition) must do so from below as $E$ increases.
To understand this, note that, if
 $\lambda_j(E)$ has a single zero at $E=E_j$, then
\begin{equation}
C_L(E) = |A^\dagger \cdot v_j(E_j)|^2 \frac{i}{ \lambda'_j(E_j) (E-E_j)} + \textrm{non-pole}\,.
\label{eq:CLEb}
\end{equation}
Comparing to Eq.~(\ref{eq:CLE}) we learn that
\begin{equation}
\lambda'_j(E_j)\equiv \frac{d\lambda_j(E)}{dE}\bigg|_{E=E_j} > 0\,.
\label{eq:derlambda}
\end{equation}
This is the generalization of a condition found in Ref.~\cite{\BHSnum} for the
isotropic approximation (where there is only a single relevant eigenvalue).
\end{itemize}
Any solutions to the quantization condition that do not satisfy both of these
conditions we refer to as unphysical. 

We are aware of only three possible sources for unphysical solutions.
First, they can arise from the truncation
of the quantization condition to a finite-number of partial waves.
Second, they could be the result of an unphysical parametrization of
$\K_2$ and $\Kdf$; for example, the truncation of the threshold expansion for $\Kdf$
could be unphysical.
And, finally, the exponentially-suppressed terms that we have dropped could be large
in some regions of parameter space, particularly for small $mL$.
We now present examples of unphysical solutions that we have found in our
numerical investigation.


\subsubsection{Solutions with the wrong residue}
\label{sec:wrongresidue}



In this section we give examples of unphysical solutions to the quantization condition that
do not satisfy Eq.~(\ref{eq:derlambda}), i.e. which lead to single poles whose residues
have the wrong sign. 
These were observed in the isotropic approximation in Ref.~\cite{\BHSnum},
where it was found that they occurred only when $|\Kiso|$ was very large.
Here we investigate how this result generalizes in the presence of $d$-wave dimers.

\begin{figure}[tb!]
\centering 
\subfigure[ \label{fig:unphysA} $mL = 8.1 $]{\includegraphics[width=.49\textwidth]{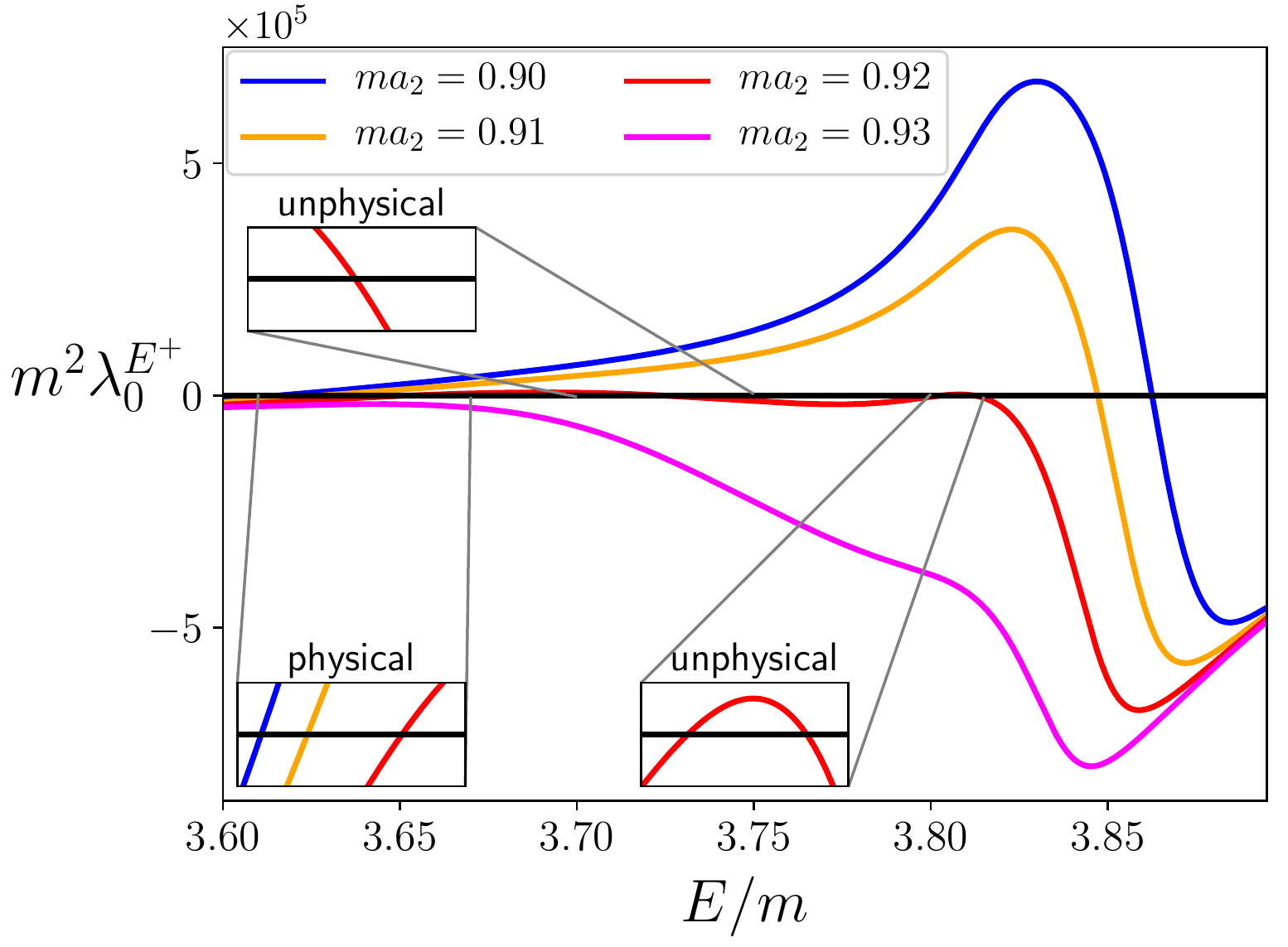}}
\hfill
\subfigure[ \label{fig:unphysB}  $mL = 10$]{\includegraphics[width=.49\textwidth]{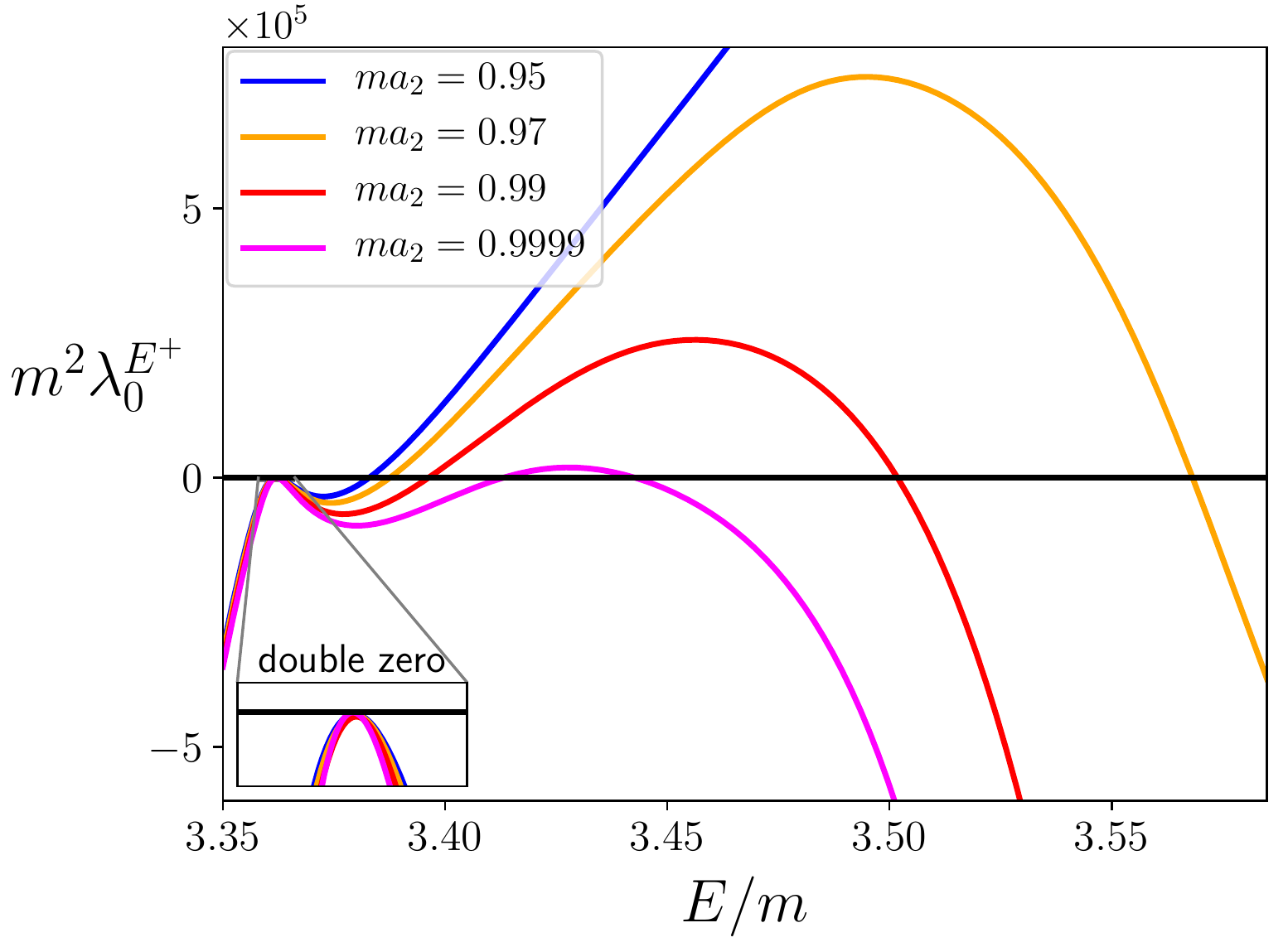}}
\caption{Smallest eigenvalue in magnitude of $F_3^{-1}$ in the $E^+$ irrep as a function of the energy for two different values of $mL$ . The parameters are $ma_0 = -0.1$ and $r_0 = P_0 = \Kdf=0$. Physical and unphysical solutions as well as a double pole at the free energy 
(to be discussed in Sec.~\ref{sec:free}) are indicated.   
\label{fig:unphys} }
\end{figure}

We first investigate whether unphysical solutions can be induced by adding $d$-wave
interactions alone, with $\Kdf=0$. We do not find such solutions
for large negative values of $m a_2$---the results obtained in Sec.~\ref{sec:effecta2}
all correspond to zero crossings in the correct direction.
However, as $m a_2$ approaches unity (which, as we saw in Sec.~\ref{sec:implementation},
is the upper bound allowed for the formalism), we do find examples of unphysical
solutions. Since we have seen in Secs.~\ref{sec:effecta2} and \ref{sec:pipipi} that
the impact of $d$-wave interactions is greater for irreps other than $A_1^+$, we focus
on the $E^+$ irrep, and work in the vicinity of the energy 
of the first noninteracting excited state, $E_1^{\text{free}}$.
In Fig.~\ref{fig:unphys}, we plot the smallest eigenvalue in magnitude of $F_3^{-1}+\Kdf=F_3^{-1}$ 
in the $E^+$ irrep 
as a function of energy, for two different values of $mL$ and a range of positive values of
$m a_2$ approaching unity. The only other nonvanishing scattering parameter is $ma_0=-0.1$.
Consider first the left panel, with $mL=8.1$. When $a_2=0$, there is a solution at 
$E\approx E_1^{\rm free}=3.53 m$, as shown by the lowest level in Fig.~\ref{fig:specta2B}.
As $a_2$ is increased, the energy shifts upwards, as expected since positive $a_2$ 
corresponds to a repulsive interaction.
When $m a_2=0.9$, the level is at $E_1 \approx 3.6 m$, and moves
to yet higher energies as $m a_2$ increases.
These solutions are physical, as shown in the bottom-left inset.
For $m a_2=0.9$ and $0.91$, however, 
there is also a single unphysical solution near $E=3.85m$,
which displays the additional unphysical behavior of having
a decreasing energy with increasingly repulsive $a_2$. Furthermore,
for $m a_2=0.92$, there is a triplet of solutions---two unphysical and one physical.
Since they are clearly related, we consider all three to be unphysical.
For even larger $m a_2$, there are no solutions in the energy range shown.

The right panel, Fig.~\ref{fig:unphysB}, displays a similar pattern, with an additional twist.
Here $mL=10$, so that $E_1^{\rm free}=3.36m$.
The energy of the physical solution lies above this, and increases with increasing $m a_2$.
There is also an unphysical solution at higher energy, whose energy decreases
with increasing $m a_2$.
The new feature is the presence of a double zero at $E_1^{\rm free}$. As discussed above,
this is manifestly unphysical since it leads to a double pole in $C_L(E)$. It is also unexpected,
as its energy lies at that of noninteracting particles. We discuss such solutions in detail
in the following section.

\begin{figure}[tb!]
\centering 
\includegraphics[width=.49\textwidth]{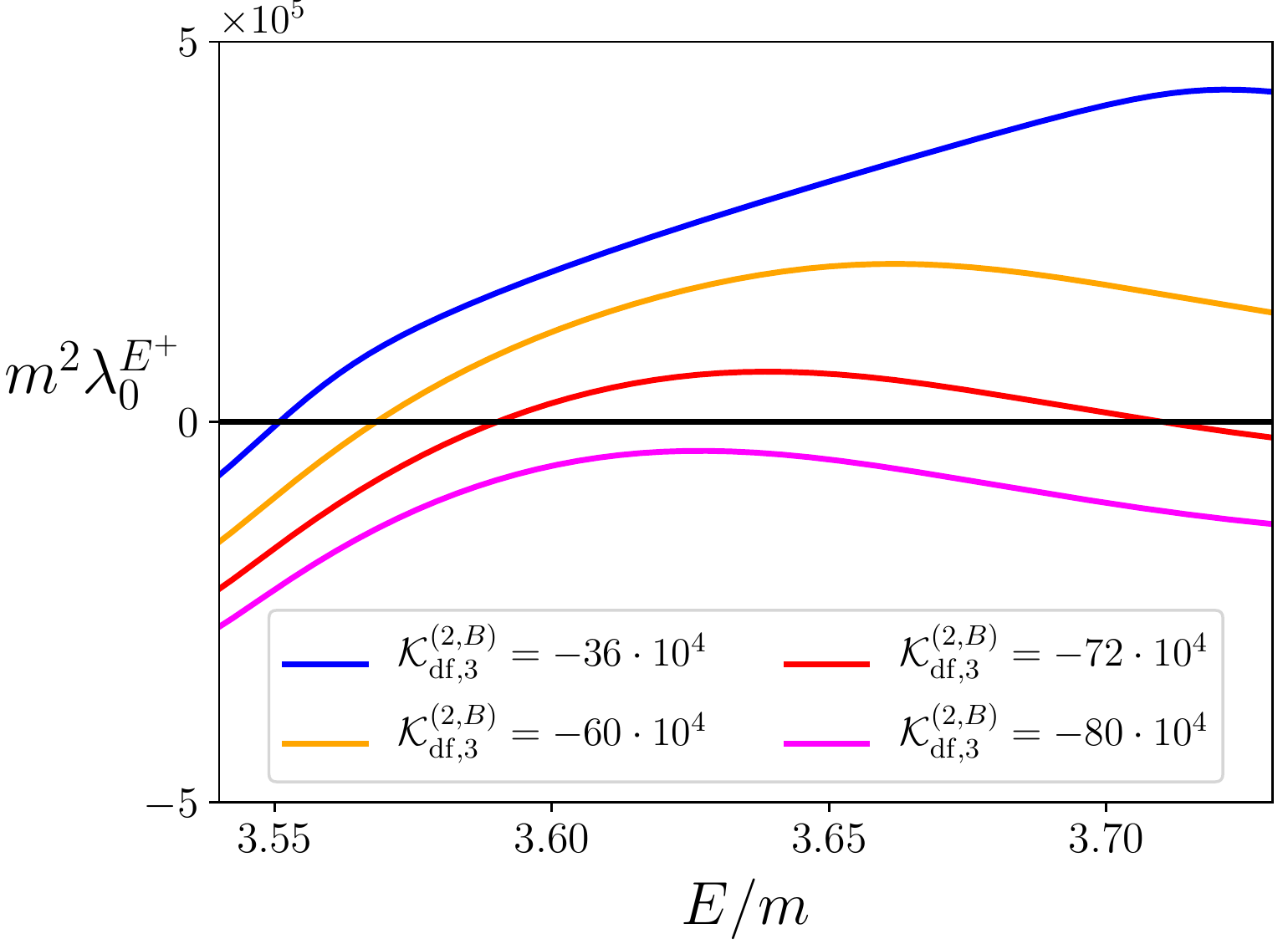}
\caption{\label{fig:unphysK3B} Eigenvalue of $F_3^{-1} + \Kdf$ with smallest magnitude
in the $E^+$ irrep as a function of the energy. The parameters are $mL=8.1$, 
$m a_0=m a_2 =0.1$, $r_0=P_0=0$, and $\Kdf=0$ for all terms except $\KB$.
}
\end{figure}

Another example of unphysical solutions in shown in Fig.~\ref{fig:unphysK3B},
this time induced by a large, negative value of $\KB$. 
Recall that, out of the parameters in $\Kdf$, the $E^+$ irrep is only sensitive to $\KB$.
Again, there are physical solutions that have the expected behavior of increasing energy
with increasingly negative $\KB$ (which corresponds to a repulsive interaction),
but there are also unphysical solutions at higher energy with opposite dependence on $\KB$.
Eventually, for large enough $|\KB|$ both solutions disappear.

We do not yet understand the source of these unphysical solutions,
i.e. which of the three possible sources mentioned at the end of the previous section are
most important. This is a topic for future study.
Our attitude is that, if a physical solution is well separated from an unphysical one, and
its behavior as interactions are made more attractive or repulsive is reasonable,
then we accept the physical solution and reject the unphysical one.
The examples we have shown occur when the interactions are strong and repulsive,
in which limit the two solutions come close together, and at some point become unreliable.
For attractive interactions, the two solutions are far apart, often with the unphysical
one lying outside the range in which the quantization condition is valid.
In this regime, which includes that
discussed in Sec.~\ref{sec:effecta2}, we trust the physical solutions.

We conclude by stressing that, in the case of three pions in QCD, 
the interactions are relatively weak, and we do not expect unphysical solutions to
be relevant.

\subsubsection{Solutions at free particle energies}
\label{sec:free}

This section concerns ``free solutions'': solutions to the quantization condition that,
even in the presence of interactions, lie at one of the energies given in Eq.~(\ref{eq:Efree}).
We expect that, in general, there will be no such solutions.
Exceptions can occur only if the symmetry of the finite-volume 
three-particle state is such that the chosen interactions do not couple to it. An example
in the two-particle sector is that, if $\vec P=0$, a finite-volume state 
lying in the $E^+$ irrep would not be shifted from its noninteracting value
if only $s$- and $p$-wave interactions were included, since the lowest wave contributing
to $E^+$ has $\ell=2$.
One question we address here is where such examples occur in the
three-particle sector.

\begin{figure}[tb!]
\subfigure[ \label{fig:free1a} $A_1^+$ irrep, $s$ wave]{\includegraphics[width=.49\textwidth]{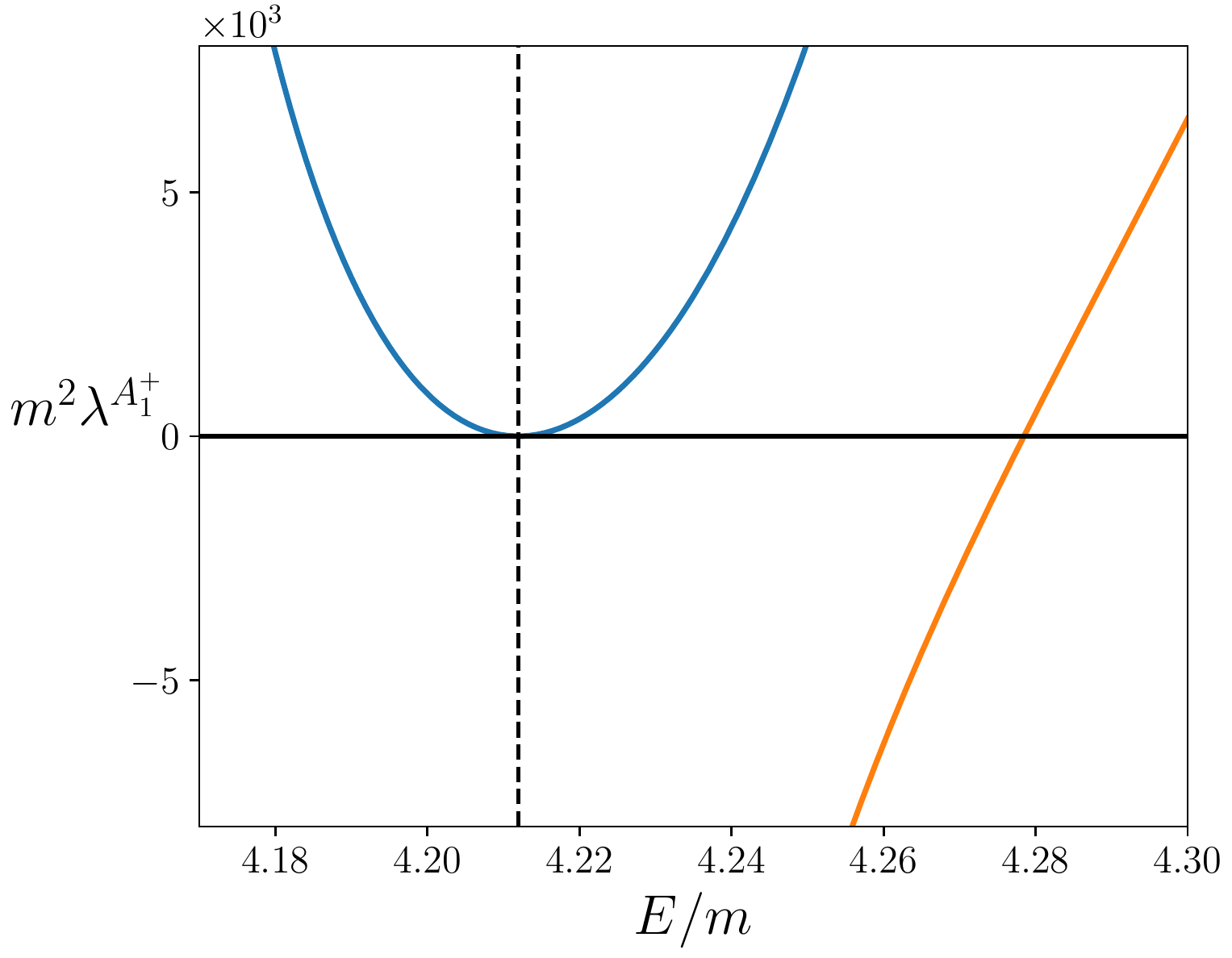}}
\hfill
\subfigure[ \label{fig:free1b} $E^+$ irrep, $s$ wave]{\includegraphics[width=.49\textwidth]{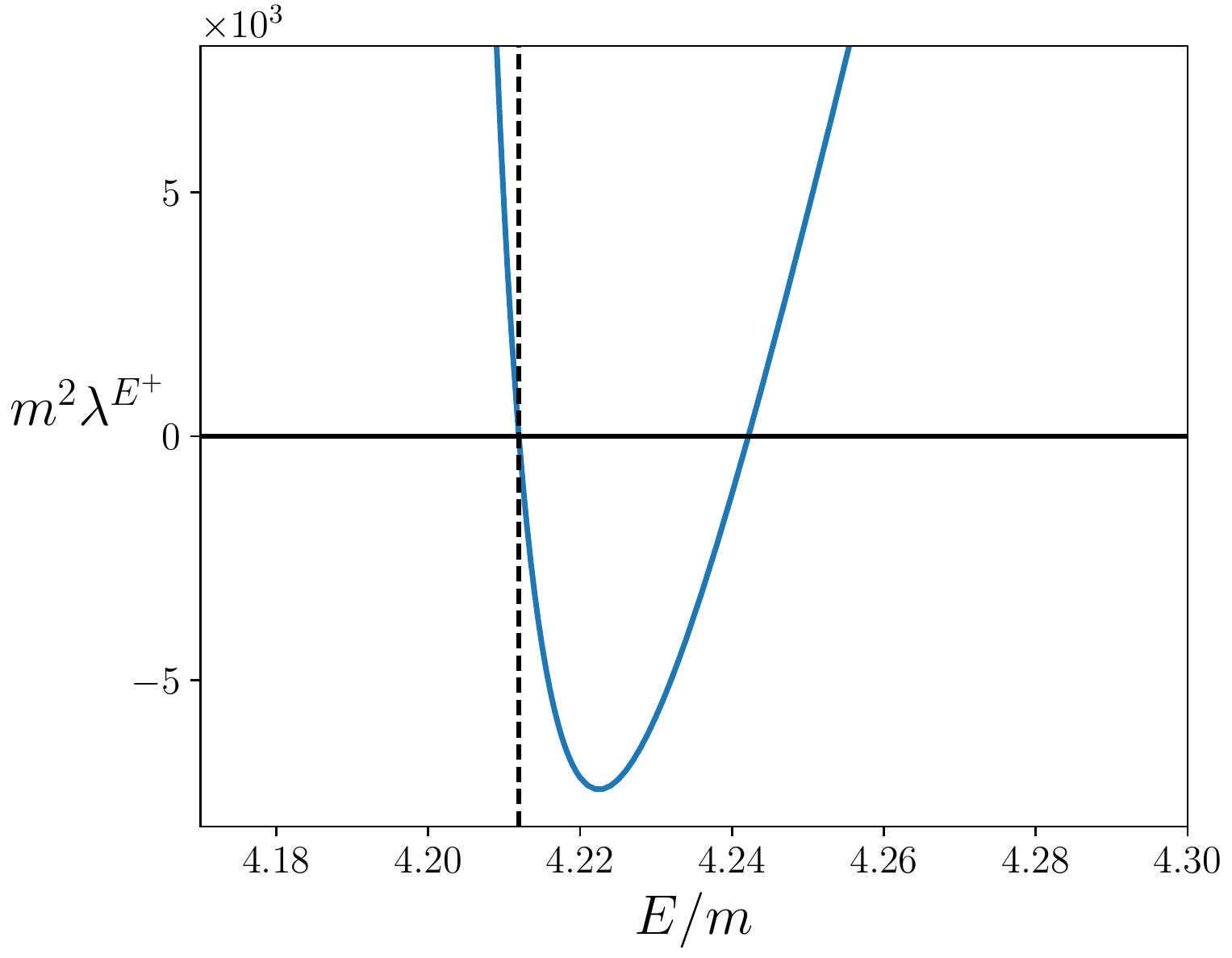}}
\hfill
\\
\hfill
\subfigure[ \label{fig:free1c} $E^+$ irrep, $s$ and $d$ waves]{\includegraphics[width=.49\textwidth]{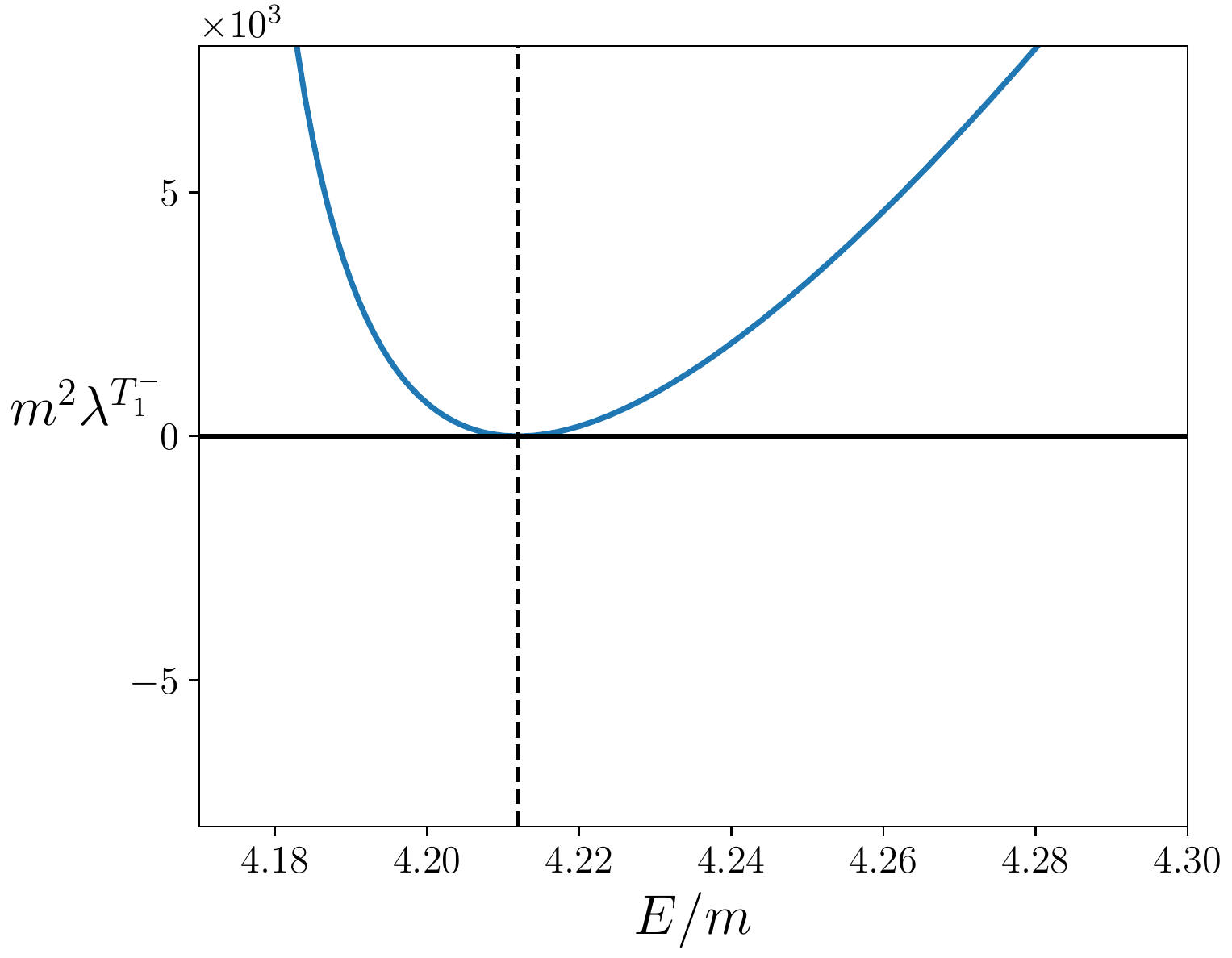}}
\hfill
\subfigure[ \label{fig:free1d} $T_1^-$ irrep, $s$ and $d$ waves]{\includegraphics[width=.49\textwidth]{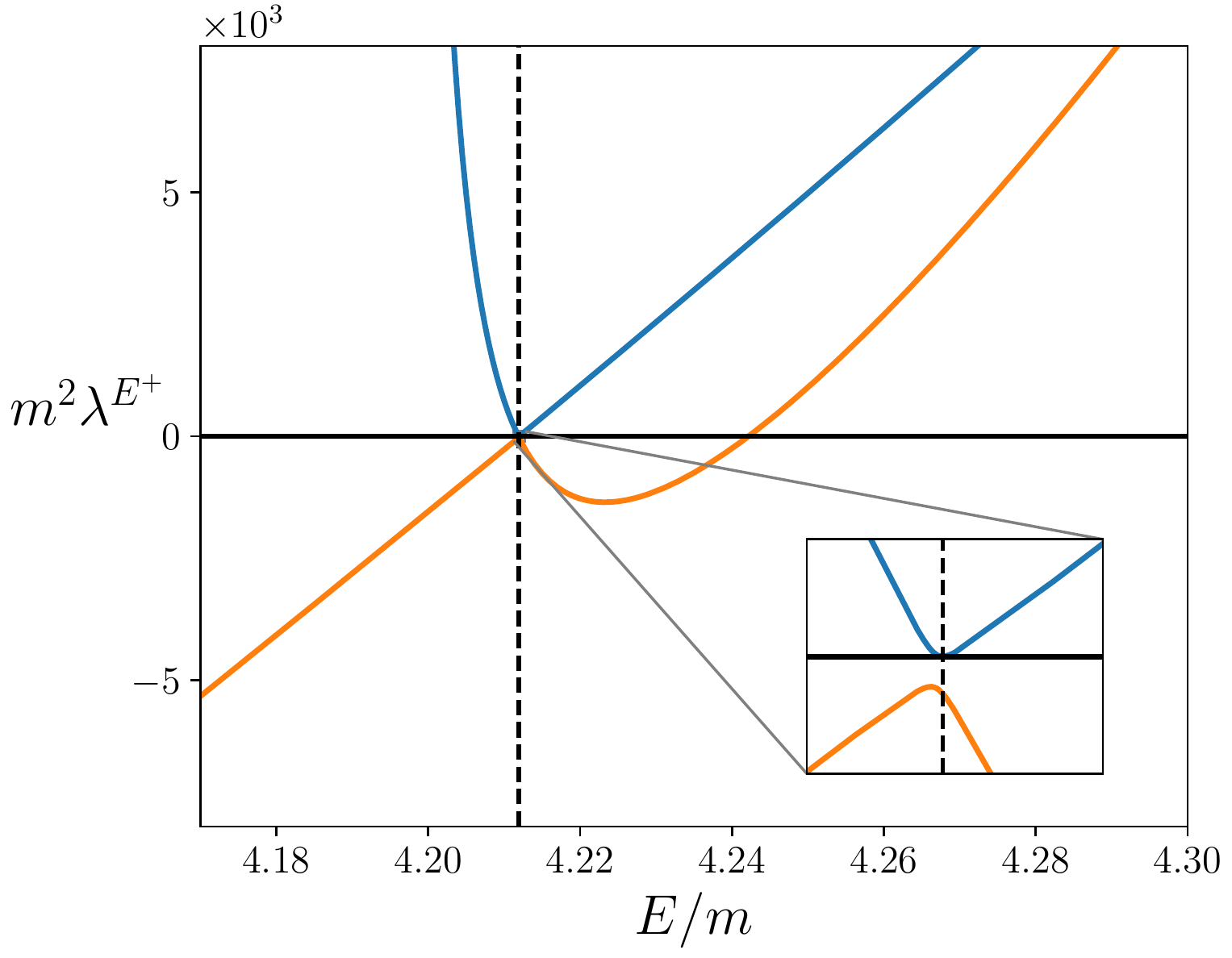}}
\hfill
\caption{Examples of solutions to the quantization condition for $\Kdf=0$ occurring
at the free energy $E_1^\fr$ (shown in all plots as the vertical dashed line).
Plots show eigenvalues of $F_3^{-1}$ as a function of $E/m$, 
with  $ma_0=0.1$, $r_0=P_0=0$ and $mL=5$. Solutions to the quantization
occur when an eigenvalue crosses zero.
(a) $A_1^+$ irrep with only $\ell=0$ channels;
(b) $E^+$ irrep, with only $\ell=0$ channels;
(c) $T_1^-$ irrep, with both $\ell=0$ and $2$, and $ma_2=0.1$;
(d) $E^+$ irrep, with both $\ell=0$ and $2$, and $ma_2=0.1$.
For the $E^+$/$T_1^-$ irreps, 
all eigenvalues are doubly/triply degenerate.
In (d), both apparent crossings are in fact avoided, as illustrated by the inset. }
\label{fig:free1}
\end{figure}

We were prompted to study this issue by finding examples of free solutions
in our numerical study.
One example has already been seen above, in Fig.~\ref{fig:unphysB},
and further examples are shown in Fig.~\ref{fig:free1}.
The first two plots show solutions with only $s$-wave channels included.
In Fig.~\ref{fig:free1a}, which shows results for the $A_1^+$ irrep, 
we see a double zero at the first excited free energy, $E_1^\fr$, as well as a solution
shifted to slightly higher energies. 
The latter is expected, 
since the repulsive interactions should raise the energy of the free state.
In the $E^+$ irrep, by contrast, there is a single zero at $E_1^\fr$, 
with the unphysical sign for the residue, as well as an interacting solution at higher energy.
The other two plots show examples of free zeros when $s$- and $d$-wave channels
are included. Both the  $T_1^-$ irrep, shown in Fig.~\ref{fig:free1c}, and
the $E^+$ irrep, shown in Fig.~\ref{fig:free1d}, have a double-zero at $E_1^\fr$.

We find similar results for higher excited free energy levels,
in which case they appear in an increasing number of irreps.
We list these irreps for the first two excited free energies in Table~\ref{tab:free}.
There are, however, no free solutions for the lowest free energy $E_0^\fr=3m$.\footnote{%
Strictly speaking, this is only true when one uses the improved form of the
quantization condition given in Eq.~(\ref{eq:QC3Q}), and described in Appendix~\ref{app:defns},
which removes spurious solutions to Eq.~(\ref{eq:QC3}).}

\begin{table}[t!]
\begin{center}
\begin{tabular}{ c c c c  c }
Level & $\ell$ & Irreps with zeros & Zeros removed by
\\ \hline
$E_1^\fr$ & 0 & $A_1^+$; $T_1^-$; $E^+$(1)& ($\KA$ or $\KB$); $\KB$; $\KB$
\\
$E_1^\fr$ & 0 \& 2 & $A_1^+$; $T_1^-$; $E^+$ & $\geq$ quartic for each
\\
$E_2^\fr$ & 0 & $A_1^+$; $T_1^-$; $T_2^-$ &($\KA$ or $\KB$); $\KB$; ($\Kdf^{(3,B)}$ or $\Kdf^{(3,E)}$)
\\
$E_2^\fr$ & 0 \& 2 & $A_1^+$; $E^+$; $T_2^+$; $T_1^-$; $T_2^-$ & $\geq$ quartic for each
\\
\hline
\end{tabular}
\end{center}
\caption{Irreps in which free zeros appear for the first two excited levels when $\Kdf=0$.
The ``(1)'' in the first row denotes that the $E_1^\fr$, $\ell=0$ free zeros in the $E^+$ irrep are single roots with unphysical residue; all other free zeros in the table are (unphysical) double roots.
Also noted are the lowest-order terms in the threshold expansion of $\Kdf$ that remove the free zeros.
The notation ``$\geq$ quartic'' indicates that a term of at least quartic order is needed.
Note that cubic-order terms are needed to remove the $E_2^\fr$, $\ell=0$ free zeros in the $T_2^-$ irrep, as neither of the quadratic terms $\KA$ and $\KB$ has nonzero eigenvalues in this irrep.
%
}
\label{tab:free}
\end{table}

In all the examples we have found, the free solutions are also unphysical---they
are either double zeros or single zeros with the wrong residue. We do not know if
this is a general result. 
Also, although the examples shown above are for $\Kdf=0$, free solutions also occur
when some components of $\Kdf$ are turned on.
Indeed, one of the questions we address in the following is which components of $\Kdf$
are required to either remove the free solutions or move them away from $E_n^\fr$.
Our first task, however, is to understand in more detail when and why free solutions occur.
All such solutions originate from the fact that $\wt F$ and $\wt G$ have
single poles at all the free energies.
These can lead to poles in $F_3$ and thus zeros in $F_3^{-1}$.
We analyze in detail only the lowest two free energies, i.e. those with level number $n=0$
and $1$ in the notation of Table~\ref{tab:freeirreps},
and then draw some general conclusions.

For $E \approx E_0^\fr=3m$,
the only elements of $\wt F$ and $\wt G$ that have poles at
$E_0^\fr$ have vanishing spectator momenta and $\ell=0$,\footnote{%
Pole contributions with $\ell=2$ and/or $\ell'=2$ vanish because, at the pole, 
$\vec a^*=\vec a\,'^*=0$.}
specifically
\begin{equation}
\wt F_{000;000} \sim \frac{1}{2}\wt G_{000;000} \sim p_0 \equiv \frac1{16 m^3 L^3 (E-3m)}
\,.
\end{equation}
Here we are using the symbol $\sim$ to indicate ``up to nonpole parts''. 
All other elements of these matrices, and of $\K_2$, either vanish or are of $\cO(1)$.
From Table~\ref{tab:dI} it now follows that poles in $\wt F$ and $\wt G$
only appear in the $A_1^+$ irrep, and the issue is whether these lead to
a pole in $F_3$.

To address this we consider the simplest case in which the volume is chosen such that only
the lowest two momentum shells are active, which is the case for  $mL\approx 5$. 
From Table~\ref{tab:dI} we then see that in the $A_1^+$ irrep
 the matrices are three dimensional,  with indices 
\begin{equation}
\left([\textrm{shell 1}, \ell=0], \
[\textrm{shell 2}, \ell=0],\
[\textrm{shell 2}, \ell=2]\right)\,.
\label{eq:3dim}
\end{equation}
We will use a $1+2$ block notation for the matrices, since this conveys all the necessary
information.
Close to $E_0^\fr$ the matrices have the form\footnote{%
There are also potential poles in the $\ell=2$ components arising from the
vanishing of $q_{2,k}^*$ and $q_{2,p}^*$ in $\wt G$ and $\wt F$,
Eqs.~(\ref{eq:defG}) and (\ref{eq:Ft0}). However,
as discussed at the end of Appendix~\ref{app:defns},
the quantization condition can be formulated such that these
purely kinematical poles are canceled, and it is legitimate to ignore them.}
\begin{equation}
\wt F = \begin{pmatrix} p_0 + \cO(1)  & 0 \\ 0 & \cO(1) \end{pmatrix}\,,
\ \ 
\wt G = \begin{pmatrix} 2p_0 + \cO(1) & \cO(1) \\ \cO(1)& \cO(1) \end{pmatrix}\,,
\end{equation}
where $\cO(1)$ elements are constrained only by the fact that $\wt F$ and $\wt G$
are symmetric. $\K_2$ is a diagonal matrix with $\cO(1)$ elements.
From this it follows that 
\begin{equation}
H = \wt F+\wt G + (2\omega \K_2)^{-1}
=\begin{pmatrix} 3p_0 + \cO(1) & \cO(1) \\ \cO(1) & \cO(1) \end{pmatrix}
\ \ \Rightarrow\ \
H^{-1} = \begin{pmatrix} \frac1{3p_0} + \cO(1/p_0^2) & \cO(1/p_0) \\ \cO(1/p_0) & \cO(1)
\end{pmatrix}
\end{equation}
and thus in turn that
\begin{equation}
\wt F H^{-1} \wt F= \begin{pmatrix} p_0/3 + \cO(1) & \cO(1) 
\\ \cO(1)& \cO(1) \end{pmatrix}
\ \ \Rightarrow \ \
F_3 = {\cal O}(1)\,.
\end{equation}
We thus find that free poles at $E_0^\fr$ cancel in $F_3$.
This argument generalizes to any number of active shells, since there are
no additional poles, and the only change is that the second block in the above analysis
is enlarged. 
The result agrees with  our numerical finding that there are no free poles at $E_0^\fr$.

Next we consider poles at the second free energy, $ E_1^\fr$. For $mL\approx 4-6$
there are
then three active shells, so the matrices to consider become larger, e.g. six-dimensional
in the $A_1^+$ irrep, and the analysis correspondingly more complicated.
We work out the case of the $A_1^+$ irrep in Appendix~\ref{app:free1},
both with $\ell=0$ channels only and with $\ell=0$ and $2$ channels included.
In both cases we find that $F_3^{-1}$ has a double zero at $E=E_1^\fr$.
This lies in a one-dimensional subspace of the full matrix space, and
what differs between the two cases is this subspace.
For $\ell=0$ only, the matrix indices are
\begin{equation}
\left([\textrm{shell 1}, \ell=0], \
[\textrm{shell 2}, \ell=0],\
[\textrm{shell 3}, \ell=0], \dots \right)\,.
\label{eq:index2}
\end{equation}
with the dimension depending on the choice of $L$.
The double zero of $F_3^{-1}$ lies, in this case, in the space spanned by
\begin{equation}
\langle x'_1|= \sqrt{\tfrac17}\left( \sqrt6, -1, 0,\dots \right)\,.
\label{eq:span2}
\end{equation}
For $\ell=0$ and $2$, the matrix indices are
\begin{equation}
\left([\textrm{shell 1}, \ell=0], \
[\textrm{shell 2}, \ell=0],\
[\textrm{shell 2}, \ell=2], 
[\textrm{shell 3}, \ell=0],\
\dots \right)\,,
\label{eq:index3}
\end{equation}
and the space of the double zero of $F_3^{-1}$ is spanned by
\begin{equation}
\langle x_1| = \sqrt{\tfrac1{12}} 
\left(\sqrt6, -1, -\sqrt5, 0, \dots \right)\,.
\label{eq:span3}
\end{equation}
The factors in Eqs.~(\ref{eq:span2}) and (\ref{eq:span3}) result from the
form of the spherical harmonics and the size of the first two shells.
They are thus kinematical.

These analytic results confirm what we find numerically.
For example, the double zero  at $E_1^\fr$ shown in Fig.~\ref{fig:free1a}
exactly matches that expected from the analysis of Appendix~\ref{app:free1},
and we have checked numerically that it lies in the predicted subspace.

We now discuss how the single zeros at free energies arise.
There is a particularly simple case in which we can easily understand
these analytically: the $E^+$ irrep when we keep only $s$-wave channels
{\em and} 
choose $mL$ such that only the first two shells are active.
We must also choose $mL$ such that $E_1^\fr < 5 m$
(so that the formalism applies);
one example is $mL=3.8$, for which $E_1^\fr = 4.86m$.
In fact, as shown in Table~\ref{tab:dI},
the first shell has no $E^+$ component for $\ell=0$, so this simple case
actually involves only the second shell, for which the $E^+$ irrep appears once.
Although the $E^+$ irrep is two-dimensional, within this space all matrices are 
proportional to the identity.
Thus the matrices are effectively one-dimensional.

The second shell consists of six elements, which we label by the direction of
the spectator momentum $\vec k$ in the following order
\begin{equation}
\vec k \in o_{001}= (2\pi/L) \{-\hat z,-\hat y, -\hat x, \hat x, \hat y, \hat z\}\,.
\end{equation}
In this basis, the $E^+$ eigenvectors can be chosen as
\begin{equation}
\tfrac12 (1,0,-1,-1,0,1) \ \ {\rm and}\ \ \sqrt{\tfrac1{12}} (-1, 2, -1,-1,2,-1)\,.
\end{equation}
It is then simple to calculate the pole terms to be
\begin{equation}
\wt F = {\mathbf 1} \left[p_1 +\cO(1) \right]  \ \ {\rm and}\ \
\wt G = {\mathbf 1} \left[p_1 +\cO(1) \right]  \,,
\end{equation}
where
\begin{equation}
p_1 \equiv \frac1{8 m \omega_1^2 L^3 (E-E_1^\fr)}\,.
\end{equation}
It immediately follows that
\begin{equation}
F_3 = 
\frac1{L^3} \left[ \frac{\wt F}3 - \wt F H^{-1} \wt F\right]
=
-\frac{p_1}{6L^3} {\mathbf 1}\left[ 1 + \cO(1/p_1)\right]\,.
\end{equation}
Thus $F_3$ indeed has a single pole at $E=E_1^\fr$, and $F_3^{-1}$ a single 
(doubly degenerate) zero.
Increasing $L$ so that there are more active shells does not change the pole structure
or the presence of the single zero.
We also see that the zero in $F_3^{-1}$ has a negative coefficient, implying that
it decreases through zero, consistent with the behavior seen in Fig.~\ref{fig:free1b}.

Thus we have understood in a few simple cases why the free zeros 
listed in Table~\ref{tab:free} appear.
It is interesting to contrast this to the results of Ref.~\cite{\BHSnum},
where the quantization condition was studied numerically in the
isotropic approximation. In that work no free zeros in $F_3^{-1}$ were found.
At first this may seem puzzling, because the isotropic approximation is a subset
of our analysis when we restrict to $\ell=0$ channels. 
The resolution is that the additional isotropic projection that is used is
orthogonal to the subspace in which the zeros live.
This is demonstrated in Appendix~\ref{app:iso},
along with a derivation of the precise relation between the isotropic approximation
and the analysis carried out here.

\bigskip
The final stage of our analysis is to study whether the inclusion of components of
 $\Kdf$ removes the free zeros. Here by ``remove'' we mean that there is no longer
 a solution to the quantization condition at a free energy. This can be accomplished
 either by removing the solution altogether (which is possible for a double zero,
 which only touches the axis) or by moving it away from the free energy
 (the likely solution for a single zero).
  We expect that if $\Kdf$ were not truncated
 then there would be no free zeros, since there would be some overlap between
 the state and the three-particle interaction. This is indeed consistent with what we find.
What turns out to be surprising, however, is which components of $\Kdf$ 
that are needed to remove the free zeros.
 
We first consider the $\ell=0$, $A_1^+$ case. To remove the double zero,
it must be that the projection of $\Kdf$ into the space of zeros
is nonvanishing:
\begin{equation}
[\Kdf(E_1^\fr)] |x'_1\rangle \ne 0\,,
\label{eq:nozeros}
\end{equation}
where $|x'_1\rangle$ is defined in Eq.~(\ref{eq:span2}).
Here the square brackets indicate the matrix that results when $\Kdf$ is decomposed
into the $k\ell m$ basis and projected into an irrep.
Note that this equation need only hold for $E=E_1^\fr$, i.e. at the energy of the
free zero.

The isotropic parts of $\Kdf$, Eq.~(\ref{eq:Kisoterm}), do not solve the problem.
These terms have the matrix form
\begin{equation}
[\K^\iso] \propto |1_K\rangle \langle 1_K|\,,
\end{equation}
where
\begin{equation}
\langle 1_K| = \left(1, \sqrt{6}, \sqrt{12}, \dots \right)
\,.
\label{eq:1K}
\end{equation}
Since this vector is orthogonal to $|x'_1\rangle$, it follows that, for all energies,
\begin{equation}
[\K^\iso] |x'_1\rangle = 0\,,
\end{equation}
so that Eq.~(\ref{eq:nozeros}) is not satisfied.
The form of $|1_K\rangle$ follows from the fact that $\K^\iso$ is independent
of the spectator momentum, so that the $A_1^+$ projection simply gives
factors of the square root of the multiplicity of the shells.
We thus expect that the inclusion of {\em any} dependence on the spectator momentum
will lead to a $[\Kdf]$ satisfying Eq.~(\ref{eq:nozeros}).
This is what we find in practice with both of the quadratic terms, 
i.e. those with coefficients $\KA$ and $\KB$ 
[see Eqs.~(\ref{eq:K3Aterm}) and (\ref{eq:K3Bterm})].

This result is an example of a general pattern: the part of $\Kdf$ that ``removes'' the
free zeros comes from terms that involve higher values of $\ell$ than those being
included in $F_3^{-1}$. Here, we need quadratic terms, which have both
$\ell=0$ and $2$ components,
in order to remove the free zeros from the $\ell=0$ part of $F_3^{-1}$.
To be clear, the $\ell=2$ components of the quadratic terms play no role;
it is simply that by going to higher order one obtains a more complicated
form of the $\ell=0$ parts, and this is sufficient to remove the unwanted free zeros.
Further examples of this are shown in the last column of Table~\ref{tab:free}, 
where we list, for all irreps that enter in a given free momentum shell, the terms
in $\Kdf$ that remove the free zero.

The second example we consider is the combined $\ell=0$ and $2$
part of $F_3^{-1}$ in the $A_1^+$ irrep. In this case, we need
\begin{equation}
[\Kdf(E_1^\fr)] |x_1\rangle \ne 0
\label{eq:nozeros2}
\end{equation}
[with $|x_1\rangle$ given in Eq.~(\ref{eq:span3})]
in order to remove the free zeros.
We find numerically that this equation is not satisfied by any of the quadratic
or cubic terms contributing to $\Kdf$, but that quartic terms do satisfy it.\footnote{%
In this case it is crucial to set the energy to $E_1^\fr$; for other energies 
Eq.~(\ref{eq:nozeros2}) is satisfied.}
This exemplifies the general pattern discussed above: quadratic and cubic
terms contain only $\ell=0$ and $2$, while quartic terms include also $\ell=4$ parts.
We were initially surprised by this result, because $\Kdf$ is an infinite-volume quantity,
while $|x_1\rangle$ arises from finite-volume considerations.
However, we show analytically in Appendix~\ref{app:needquartic}
that orthogonality follows solely from
the rotation invariance and particle-interchange symmetry of $\Kdf$, together with the
fact that quadratic and cubic terms  contain only $\ell=0$ and $2$ parts. 
Thus it is an example of the phenomenon described at the beginning of this section,
in which symmetries make the finite-volume state transparent to certain
interactions. It is also clear from the arguments in Appendix~\ref{app:needquartic}
that  all that is required for Eq.~(\ref{eq:nozeros2}) to be satisfied
is to use contributions to $\Kdf$ that involve $\ell\ge 4$, i.e. terms of quartic or higher order
in the threshold expansion.

Finally, we consider the case of the single zero in the $E^+$ irrep for $\ell=0$ channels only,
shown in Fig~\ref{fig:free1b}.
Here we aim to shift the zero away from the free energy.
This is accomplished by including a contribution from $\Kdf$ that lives in the
$E^+$ irrep. As noted in the final paragraph of Sec.~\ref{sec:implementation}, 
the lowest-order 
term in the threshold expansion for which this is the case is the $\KB$ term.
Thus, once again, we have to use a term in $\Kdf$ that contains higher values of
$\ell$ (here $\ell=2$) than are included in $F_3$.

\begin{figure}[tb!]
\centering
\hfill
\subfigure[ \label{fig:freeremovea} $E^+$ irrep, $s$ wave]{\includegraphics[width=.49\textwidth]{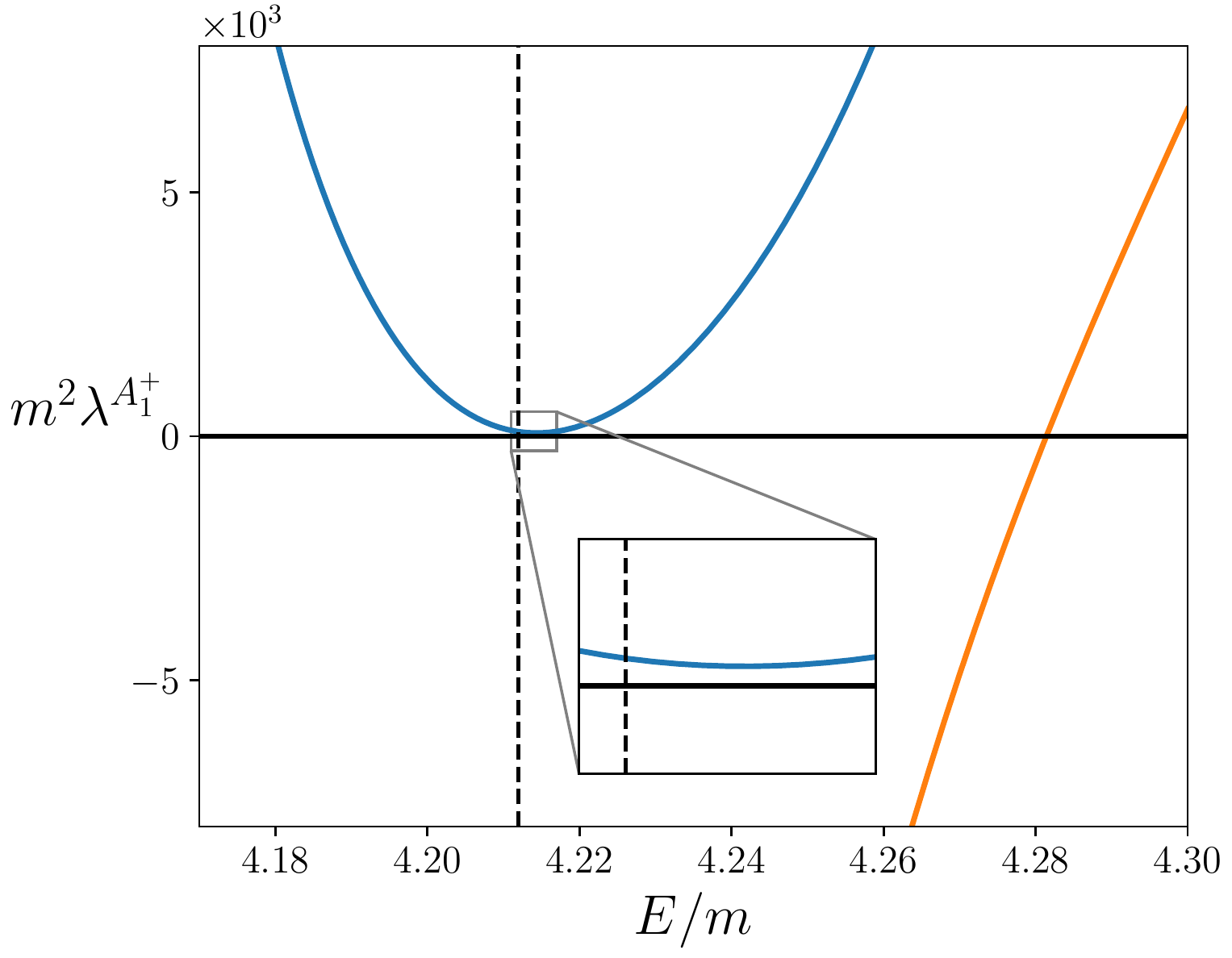}}
\hfill
\subfigure[ \label{fig:freeremoveb} $T_1^-$ irrep, $s$ wave]{\includegraphics[width=.49\textwidth]{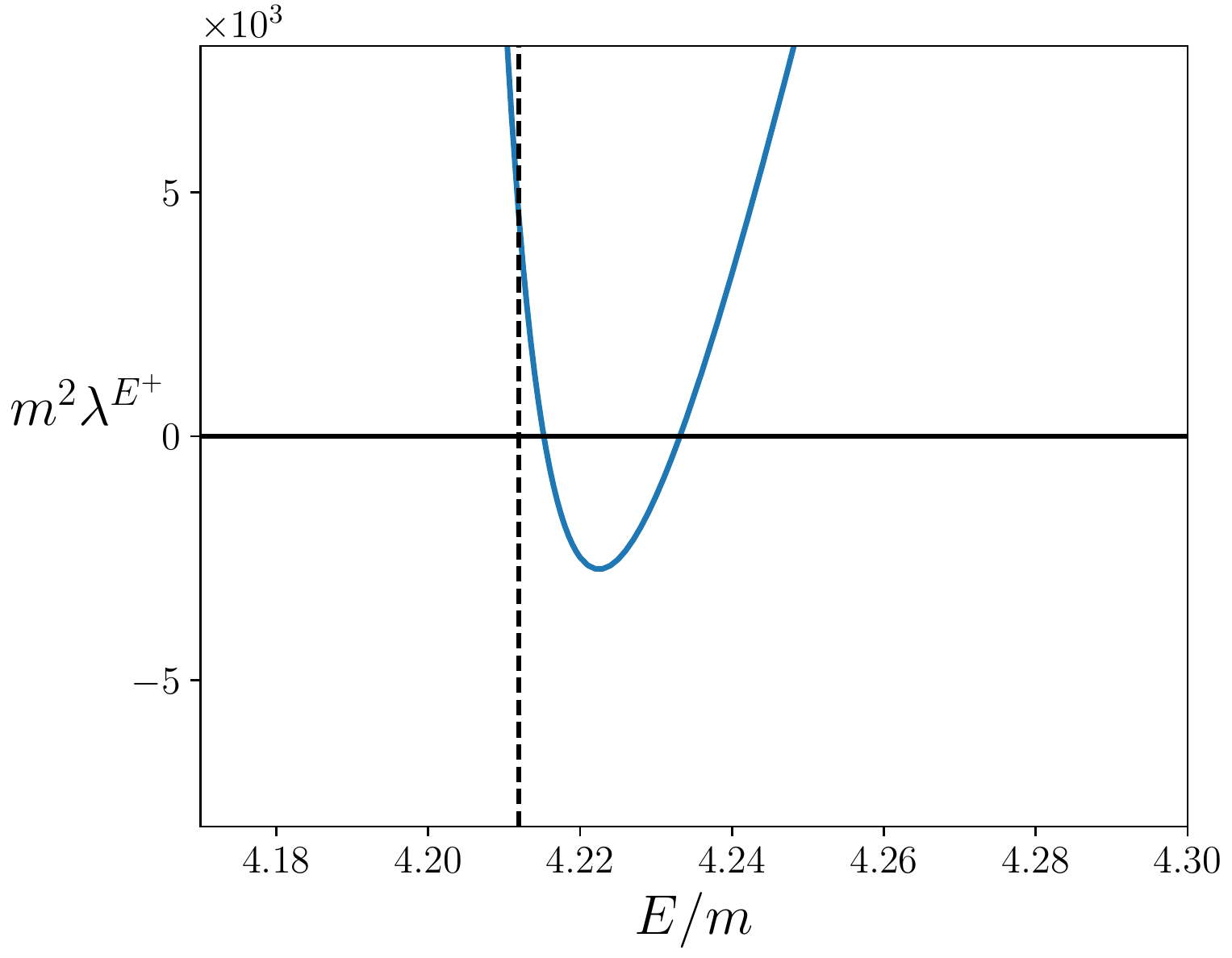}}
\hfill
\caption{
Effect of turning on $\Kdf$ on the free solutions shown in Fig.~\ref{fig:free1a}
and ~\ref{fig:free1b}, with all other parameters unchanged.
Eigenvalues are now those of $F_3^{-1}+\Kdf$.
(a) $A_1^+$ irrep with $\KA=8000$; 
(b) $E^+$ irrep with $\KB=8000$.}
\label{fig:freeremove}
\end{figure}

These theoretical arguments are supported by our numerical results.
We show two examples in Fig.~\ref{fig:freeremove}.
These correspond to the two cases shown in Figs.~\ref{fig:free1a} and \ref{fig:free1b},
except that we have turned on $\KA$ and $\KB$, respectively.
We expect the double-zero in the former case ($A_1^+$ irrep)
to removed by the addition of any quadratic term in $\Kdf$, 
and the figure shows that $\KA$ does the job. 
In Fig.~\ref{fig:freeremoveb}, corresponding to the $E^+$ irrep, we need to
use the $\KB$ term, since $\KA$ does not contain an $E^+$ component.
Since this is a single zero, it is not removed, but is rather shifted to a non-free energy. 
Note, however, that it remains unphysical because it decreases through zero.
In fact, for higher values of $\KB$, the zeros coalesce and then disappear.

\bigskip
We close this section with two general comments on the nature of the resolution
that we have presented to the problem of unwanted free solutions.
The first concerns the result that we need higher-order terms in the threshold
expansion of $\Kdf$ in order to remove the free zeros of a given order
in $F_3^{-1}$. On its face, this invalidates the threshold expansion, for we are
evaluating distinct terms in the quantization condition at different orders.
We do not think this is the case, however,
because we know that, above threshold,
{\em all terms} in the expansion of $\Kdf$ are present at some level,
and it only takes an infinitesimal value for the coefficient of 
the requisite higher-order term to remove the unwanted solution.
Thus we conclude that we can proceed, in practice, 
by truncating the expansion of all quantities at the same order in the threshold expansion, 
and simply ignore the free solutions.

The second comment concerns the fact that our resolution fails if the coefficient of
the required parts of $\Kdf$ vanish.
In fact, this would require the simultaneous 
vanishing of an infinite number of terms in the threshold expansion, 
since higher-order terms in the correct irrep can remove the free solutions.
Thus it would require an enormous fine-tuning, which seems highly implausible,
especially because there is no enhancement of the symmetry of $\Kdf$ at the
tuned point.

\section{Conclusions}

\label{sec:conc}

The work presented in this paper is the first step towards the systematic inclusion of higher partial waves in the three-particle quantization condition. We have used the generic relativistic field theory (RFT) approach, formulated so that the three-particle
scattering quantity, $\Kdf$, is Lorentz invariant.
This invariance proves very important in simplifying the threshold expansion of $\Kdf$.
Indeed, we find that, at quadratic order and for identical particles,
 only five parameters control the contribution from the three-particle sector, 
 of which only two describe dependence on angular degrees of freedom.  
This provides a simple starting point for studying the impact of $\Kdf$.
Working at quadratic order implies keeping both $s$- and $d$-wave two-particle channels (dimers).
We have numerically implemented the quantization condition at this order,
and obtained several new results that we now highlight.

The first of these is to determine the projection onto irreps of the cubic group including
higher partial waves. This has previously been done only for the case of $s$-wave 
dimers~\cite{Doring:2018xxx}. The generalization is nontrivial, 
since both the spectator momentum
and the parameters of the dimer transform. While we have worked this out explicitly only for 
coupled $s$- and $d$-wave dimers, the formalism holds for dimers with any angular momentum.

Second, we have understood how the two-particle scattering amplitudes in higher partial
waves enter in the $1/L$ expansion of the energy of the three-particle ground state.
We find that all even partial waves enter at $\cO(1/L^6)$, and have calculated analytically
the dependence on the $d$-wave amplitude in the weak-coupling limit and for $\Kdf=0$.
Although this contribution itself is likely too small to be seen in present simulations of
three-particle systems, we have used it as a nontrivial check of our implementation.

Third, we have shown that $d$-wave interactions, if they are moderately strong,
can have a sizable effect on the finite-volume three-particle spectrum.
For example, we have presented evidence for a generalized Efimov-like 
three-particle bound state
induced by a strongly attractive $d$-wave two-particle interaction.

Fourth, we have shown how the five parameters describing $\Kdf$ lead to distinguishable
effects on the spectrum of the $3\pi^+$ system, suggesting that they can be separately
determined in a dedicated lattice study. Indeed, this is the system within QCD
to which our truncated formalism is most applicable. 

Finally, we have characterized solutions to the quantization condition
that are unphysical. These presumably arise because of the truncation to a small number
of partial waves, and the fact that we have dropped terms that are exponentially
suppressed in $mL$. One class of solutions generally 
appears when either the two- or the three-particle interactions are strong and repulsive.
Our approach is to use parameters such that there are no 
unphysical solutions near to the physical solutions of interest.
The second class of solutions are those that occur at the energies of three noninteracting
particles. We have presented numerical evidence and analytical
 arguments that these are removed
if sufficiently high-order terms in $\Kdf$ are included.
We expect that other approaches to the three-particle quantization condition will face
similar issues, for which our observations may be relevant.

There remain many directions for future study. In order to make our implementation more useful,
it is important to generalize it to moving frames. The underlying formalism of 
Ref.~\cite{\HSQCa} applies in all finite-volume frames, but the projectors onto irreps
will need to be generalized to account for the reduced symmetry.
Another important generalization is to include subchannel resonances, i.e., dynamical
poles in $\K_2$. For this one must implement the formalism of 
Ref.~\cite{Briceno:2018aml}, and go beyond the threshold expansion.
Finally, we recall that $\Kdf$ is an intermediate quantity, related to the physical
three-particle scattering amplitude, $\cM_3$, by integral equations. Since it is only by
looking for complex poles in $\cM_3$ that one can study three-particle resonances,
it is crucial to develop methods to solve the necessary integral equations.

To conclude, we would like to restate that, as it is a relativistic approach, 
our implementation can simultaneously be useful to both the lattice QCD community 
and the field of cold atom physics.

\acknowledgments
We thank Ra\'ul Brice\~no, Hans-Werner Hammer, Max Hansen and Akaki Rusetsky
for discussions. 
The work of TDB and SRS was supported in part by the United States Department of Energy grant No. DE-SC0011637. FRL acknowledges the support of the European Project InvisiblesPlus H2020-MSCA-RISE-2015 and the MINECO projects FPA2017-85985-P and SEV-2014-0398. The work of FRL has also received funding from the European Union Horizon 2020 research and innovation program under the Marie Sklodowska-Curie grant agreement No. 713673 and "La Caixa" Foundation (ID 100010434).
The work of FRL and SRS was supported in part by the Munich Institute for Astro- and Particle Physics (MIAPP) of the DFG cluster of excellence ``Origin and Structure of the Universe''.
We are very grateful to MIAPP for its hospitality and stimulating environment.

\appendix
\section{Definitions}
\label{app:defns}
Here we collect definitions of quantities appearing in $F_3$, Eq.~\eqref{eq:F3},
that are not given in the main text.

We begin with the cutoff function:
\begin{align}
    H(\vec{k}) &= J(z)\,, \qquad
    z = \frac{E_{2,k}^{*2} - (1+\alpha_H)m^2}{(3-\alpha_H)m^2}\,, 
        \label{eq:H_k}
    \\
    J(z) &=
    \begin{cases}
        0, & z\leq 0 \\
        \exp{\left( -\frac{1}{z}\exp{\left[ -\frac{1}{1-z} \right]} \right)}, & 0<z<1 \\
        1, & 1\leq z
    \end{cases}
\end{align}
with $\alpha_H\in [-1,3)$ a constant. We choose $\alpha_H=-1$,
corresponding to the highest cutoff, in all our numerical investigations.

For $\wt G$ we use the relativistic form suggested in Ref.~\cite{\BHSQC},
\begin{equation}
    \widetilde{G}_{p\ell'm';k\ell m} \equiv \frac{1}{L^3} \frac{1}{2\omega_p} \frac{H(\vec{p}) H(\vec{k})}{b^2-m^2}
    \frac{4\pi \mc{Y}_{\ell'm'}(\vec{k}^*) \mc{Y}_{\ell m}(\vec{p}^{\,*}) }{q_{2,p}^{*\ell'}\,q_{2,k}^{*\ell} } \frac{1}{2\omega_k}\,, 
    \label{eq:defG}
\end{equation}
where $b= P-p - k$ is the momentum of the exchanged particle,
$\vec p^{\,*}$ is the result of boosting $p$ to the CM frame of the dimer for
which $k$ is the spectator momentum, and {\em vice versa}.
Explicitly, we have
\begin{equation}
    \vec{p}^{\,*} = (\gamma_k-1)(\hat{p}\cdot\hat{k})\hat{k} + \omega_p\gamma_k\beta_k\hat{k} + \vec{p}, \quad
    \beta_k = \frac{|\vec{k}\;|}{E-\omega_k}, \quad \gamma_k = (1-\beta_k^2)^{-1/2}
\,,
\end{equation}
with $\vec k^*$ given by $\vec p \leftrightarrow \vec k$.
Finally, $\mathcal Y_{\ell m}(\vec k)$ are harmonic polynomials,
\begin{equation}
    \mathcal{Y}_{\ell m}(\vec{k}) \equiv k^\ell Y_{\ell m}(\hat{k})\,,
\end{equation}
where $Y_{\ell m}$ are the \textit{real} spherical harmonics.
The elements of $\wt G$ are clearly straightforward to evaluate numerically.

For completeness, we quote the real $d$-wave harmonic polynomials
\begin{equation}
\begin{split}
    \sqrt{4\pi}\mathcal{Y}_{2-2}(\vec{k}) &= \sqrt{15}k_1k_2\,, \\
    \sqrt{4\pi}\mathcal{Y}_{2-1}(\vec{k}) &= \sqrt{15}k_2 k_3\,, \\
    \sqrt{4\pi}\mathcal{Y}_{20}(\vec{k}) &= \sqrt{5}(2k_3^2-k_1^2-k_2^2)/2\,, \\
    \sqrt{4\pi}\mathcal{Y}_{21}(\vec{k}) &= \sqrt{15}k_1 k_3\,, \\
    \sqrt{4\pi}\mathcal{Y}_{22}(\vec{k}) &= \sqrt{15}(k_1^2-k_2^2)/2\,.
\end{split}
\label{eq:Y22}
\end{equation}
The associated Wigner D-matrices are
\begin{align}
	\mc{D}_{\ell'm',\ell m}(R) &= \int d\Omega_{\hat{r}} Y_{\ell'm'}(R\hat{r}) Y_{\ell m}(\hat{r})
	= \delta_{\ell\ell'} \mc{D}^{(\ell)}_{m'm}(R)\,, 
	\label{eq:Wigner_D_1}
\end{align}
where $R$ is a rotation matrix. They are orthogonal matrices, and 
implement rotations of the spherical harmonics:
\begin{align}
	Y_{\ell m}(R\hat{r}) &= \sum_{m'=-\ell}^\ell \mc{D}^{(\ell)}_{mm'}(R) Y_{\ell m'}(\hat{r})\,.
 \label{eq:Wigner_D_2}
\end{align}

Finally, $\wt F(\vec k)$ is a sum-integral difference that 
is proportional to the zeta functions that appear in the two-particle
quantization condition~\cite{Luscher:1986n2,Luscher:1991n1}.
It requires ultraviolet (UV) regularization, and can be written in various forms that are
equivalent up to exponentially-suppressed corrections.
The form that follows from that presented in Ref.~\cite{\HSQCa} is
\begin{equation}
\wt F(\vec k)_{\ell' m';\ell m} =
\left[\frac1{L^3} \sum_{\vec a} - \text{PV}\int \frac{d^3a}{(2\pi)^3} \right]
\frac1{(q_{2,k}^{*})^{\ell'+\ell}}
\frac{H(\vec a) H(\vec b) 4\pi \cY_{\ell' m'}(\vec a^{\,*}) \cY_{\ell m}(\vec a^{\,*})}
{16 \omega_k \omega_a \omega _b(E-\omega_k-\omega_a -\omega_b)}
\,,
\label{eq:Ft0}
\end{equation}
where $b  = P-k -a$ here, and $\vec a^*$ is the result of boosting $a$ to the dimer
rest frame, with $k$ the spectator. Here the UV regularization is provided by the product
of $H$ functions, and the integral over the pole is defined by the principle value prescription
(leading to a real result).
Instead, we use a different form that is simpler to evaluate numerically.
Following the steps similar to those used in Ref.~\cite{Kim:2005gf}, 
we change variables and introduce a new regularization, 
finding that, up to exponentially-suppressed corrections, $\wt F$ can be rewritten as
\begin{align}
    \widetilde{F}(\vec{k})_{\ell'm';\ell m} &= 
    \frac{1}{32\pi^2 L \omega_k(E\!-\!\omega_k)} 
     \left[ \sum_{\vec{n}_a} \!- {\rm PV}\!\int d^3 {n}_a \right]
    \frac{e^{\alpha(x^2-r^2)}}{x^2-r^2} 
    \frac{4\pi \mathcal{Y}_{\ell'm'}(\vec{r}) \mathcal{Y}_{\ell m}(\vec{r})}{x^{\ell'+\ell}}\,,
    \label{eq:Ft1}
\end{align}
where $\vec a= \vec n_a (2\pi/L)$, $x=q_{2,k}^* L/(2\pi)$, and
\begin{align}
    \vec{r}(\vec{n}_k,\vec{n}_a) &= \vec{n}_a + \vec{n}_k \left[
    \frac{\vec{n}_a\cdot\vec{n}_k}{n_k^2}
    \left(\frac{1}{\gamma_k}-1\right) + \frac{1}{2\gamma_k}
    \right]\,,
    \label{eq:rdef}
\end{align}
with $\vec k= \vec n_k (2\pi/L)$.
The UV regularization is now provided by the exponential in the integrand,
and is parametrized by $\alpha > 0$. What is shown in Ref.~\cite{Kim:2005gf} is
that the $\alpha$ dependence is exponentially suppressed in $L$, and that, in practice,
one should choose a value that is small enough that the dependence on $\alpha$ 
lies below the accuracy required. We find that $\alpha \approx 0.5$ is usually small enough.

An important technical point is that, as seen from Eq.~(\ref{eq:Ft}), 
in the full matrix form $\wt F_{p\ell' m';k\ell m}$, 
$\wt F(\vec k)$ is always multiplied by $H(\vec k)$, 
from which it follows that $\gamma_k$ is always finite and real 
whenever $\wt F_{p\ell' m';k\ell m}$ is nonvanishing.

We close this appendix by commenting on the factors of $q^*$
(which we use generically for $q_{2,k}^*$ or $q_{2,p}^*$)
in the denominators of $\wt G$ and $\wt F$. These lead to poles 
for particular kinematic configurations,  which in turn can lead to
solutions to the quantization condition.
These solutions appear to be similar to free solutions discussed in 
Sec.~\ref{sec:free}, but are in fact spurious.
To understand this we need an argument given in Appendix A of Ref.~\cite{\HSQCa},
which shows that the factors of $q^*$ in the denominators
are always canceled by corresponding
factors in the numerators of $\K_2$, $\Kdf$, and the end cap factors
$A^\dagger$ and $A$ in the finite-volume correlation function $C_L(E)$
[see Eq.~(\ref{eq:CLEa})]. The presence of the necessary factors of
$q^*$ in $\K_2$ can be seen from Eq.~(\ref{eq:K2dwave}),
while those in $\Kdf$ arise from the quadratic dependence on $\vec a^{\,*}$ and $\vec a\,\!'^*$
described in Sec.~\ref{sec:proj}.
Indeed, one can derive a version of the quantization condition
in which all such factors are absent. To do so, we define the matrix
\begin{equation}
Q_{p\ell' m';k\ell m} = \delta_{pk}\delta_{\ell' \ell}\delta_{m'm} q_{2,k}^{*\ell}\,.
\end{equation}
Then, from the arguments of Ref.~\cite{\HSQCa} we know that we can write
the end caps as $A=Q \wt A$ and $A^\dagger= \wt A^\dagger Q$, with
$\wt A$ and $\wt A^\dagger$ nonsingular. Thus an alternative, improved form of the
quantization condition is
\begin{equation}
\det[(QF_3Q)^{-1}+ Q^{-1}\Kdf Q^{-1}] = 0\,.
\label{eq:QC3Q}
\end{equation}
Now we observe that, by simple algebraic manipulations, we
can rewrite this form of the quantization condition in terms of
$Q \wt F Q$, $Q \wt G Q$, $Q^{-1} \K_2 Q^{-1}$ and $Q^{-1} \Kdf Q^{-1}$,
in all of which the factors of $q^*$ cancel.
Since the difference between the two quantization conditions is a factor of $\det(Q^2)$,
it follows that the solutions to the new form, Eq.~(\ref{eq:QC3Q}),
are the same as those to Eq.~(\ref{eq:QC3}),
except that spurious solutions to the latter, arising from the factors of $q^*$, are removed.
In conclusion, we can
use the original form of the quantization condition,
Eq.~(\ref{eq:QC3}), as long as we ignore the spurious solutions.

\section{Numerical evaluation of $\widetilde{F}$}
\label{app:Ft}

In this appendix we describe some technical details 
concerning the evaluation of $\wt F(\vec k)$.

\subsection{Evaluating the integrals} \label{app:integrals}
An advantage of the form Eq.~(\ref{eq:Ft1}) is that the integrals can be
evaluated analytically. Dropping overall factors, the integral that is needed is
\begin{equation}
I^F_{\ell' m';\ell m} =
 {\rm PV}\!\int d^3 {n}_a 
    \frac{e^{\alpha(x^2-r^2)}}{x^2-r^2} 
    4\pi \mathcal{Y}_{\ell'm'}(\vec{r}) \mathcal{Y}_{\ell m}(\vec{r})\,.
\label{eq:IFt}
\end{equation}
Changing variables to $\vec r$, we find
\begin{align}
I^F_{\ell' m';\ell m} &=	
\gamma\text{PV}\int d^3r \frac{e^{\alpha(x^2-r^2)}}{x^2-r^2} 
4\pi\mc{Y}_{\ell'm'}(\vec{r})\mc{Y}_{\ell m}(\vec{r})
= \delta_{\ell'\ell}\delta_{m'm} I_\ell^F \,,
\\
I_\ell^F &= 4\pi\gamma\text{PV}\int r^2 dr \frac{e^{\alpha(x^2-r^2)}}{x^2-r^2} r^{2\ell}\,.
\end{align}
The remaining integral can be evaluated analytically for all $\ell$.
The explicit result for $\ell=0$ was worked out in Ref.~\cite{\BHSnum}, and
we have extended this to the $\ell=2$ case. For convenience, we quote both results
\begin{align}
	I_0^F &=
	4\pi\gamma \left[ -\sqrt{\frac{\pi}{\alpha}}\frac{1}{2}e^{\alpha x^2} + \frac{\pi x}{2}\text{Erfi}\left(\sqrt{\alpha x^2}\right)  \right]
	\\
	I_2^F &=
	4\pi\gamma \left[ -\sqrt{\frac{\pi}{\alpha^5}} \frac{3+2\alpha x^2 + 4\alpha^2x^4}{8}e^{\alpha x^2} + \frac{\pi x^5}{2}\text{Erfi}\left(\sqrt{\alpha x^2}\right)  \right]\,.
\end{align}

\subsection{Cutting off the sum} \label{app:sum_cutoff}
The sum in Eq.~(\ref{eq:Ft1}) is convergent, but in practice we
must introduce a cutoff in order to evaluate it numerically.
We use a spherical cutoff, $|\vec n_a|<n_{\rm max}$, and in this section explain how
we choose $n_{\rm max}$. 

The basic idea is to split the sum $S$ as
\begin{align}
	S= S_< + S_>\,,
\end{align}
where $S_<$ is the contribution from below the cutoff, and $S_>$ the
remainder. Assuming that the pole in the summand lies well below the cutoff,
then $S_>$ can be well-approximated by a remainder integral, $R_>$. 
We evaluate this integral,
and then choose $n_{\rm max}$ such that $R_>$ lies below our desired accuracy.
The resulting $n_{\rm max}$ depends on $E$, $L$ and the orbit of $\vec k$.

Dropping overall factors, and changing the overall sign,
the sum of interest from Eq.~(\ref{eq:Ft1}) is
\begin{align}
	S = 
H(\vec k) \sum_{\vec{n}_a} 
\frac{e^{\alpha(x^2-r^2)}}{r^2-x^2} r^{\ell'+\ell} 4\pi Y_{\ell'm'}(\hat{r}) Y_{\ell m}(\hat{r})
\,.
\label{eq:sumdef}
\end{align}
Here we have included the cutoff function $H(\vec{k})$ that enters
in the expression for $\wt F_{p\ell' m';k\ell m}$, Eq.~(\ref{eq:Ft}).
Although this is an overall factor, it will play an important role in 
the determination of $n_{\rm max}$.

The integral $R_>$ that results when replacing the sum over $\vec n_a$ with an integral 
is more easily evaluated by changing variables to $\vec r$.
The relation between $\vec n_a$ and $\vec r$, Eq.~(\ref{eq:rdef}), can be rewritten as
\begin{align}
	\gamma_k r_{\parallel} = n_{a,\parallel} - \frac{n_k}{2}\,, \qquad r_\perp = n_{a,\perp}\,,
\end{align}
with $\parallel$ and $\perp$ defined relative to $\vec{k}$.
The cutoff is chosen such that $n_{\rm max} \gg n_k$, implying that
the $n_k/2$ term in the expression for $r_\parallel$ is subleading.
Dropping this term, we find that a spherical cutoff on $\vec n_a$ corresponds to
an ellipsoidal cutoff on $\vec r$. 
This makes the integral difficult to evaluate, so we replace this 
with a spherical cutoff, $|\vec r|< \Lambda$, choosing $\Lambda = n_{\rm max}/\gamma_k$.
We call the resulting integral $R_\Lambda$.
The resulting spherical region is a superset of the original ellipsiodal region, so that
we overestimate the remainder, $R_\Lambda > R_>$, since the integrand is positive.

To evaluate $R_\Lambda$ we make two further approximations. 
First, we drop the $x^2$ term
in the denominator, which is subleading since $r^2 \gg x^2$ within the region of integration.
Second, we make the replacement $4\pi Y_{\ell'm'}(\hat{r}) Y_{\ell m}(\hat{r})\to 1$,
which leads to an overestimate of the integral. Then we find
\begin{align}
	R_\Lambda &\approx \overline R_\lambda \equiv
	\gamma_k H(\vec{k}) 4\pi \int_\Lambda^\infty dr ~ e^{\alpha(x^2-r^2)} r^{\ell'+\ell}
	\\
	&= \begin{cases}
		\gamma_k H(\vec{k}) e^{\alpha x^2} 2\pi \sqrt{\frac{\pi}{\alpha}} \text{Erfc}\left[\sqrt{\alpha}\Lambda\right], & \ell'=\ell=0
		\\
		\gamma_k H(\vec{k}) e^{\alpha x^2} \frac{\pi}{\alpha}  \lbrace 2\Lambda e^{-\alpha\Lambda^2} + \sqrt{\frac{\pi}{\alpha}} \text{Erfc}\left[\sqrt{\alpha}\Lambda\right] \rbrace, & \ell'+\ell=2 
		\\
		\gamma_k H(\vec{k}) e^{\alpha x^2} \frac{\pi}{\alpha^2} \lbrace (3\Lambda+2\alpha\Lambda^3) e^{-\alpha\Lambda^2} + \frac{3}{2}\sqrt{\frac{\pi}{\alpha}} \text{Erfc}\left[\sqrt{\alpha}\Lambda\right] \rbrace, & \ell'+\ell=4.
	\end{cases}
\end{align}
The overall factor of $\gamma_k$ is the Jacobian from changing the integration variable 
from $\vec{n}_a$ to $\vec{r}$.
We choose the $\Lambda$ by specifying a tolerance $\epsilon$ (we use $\epsilon=10^{-9}$) and numerically solving $\overline{R}_\Lambda = \epsilon$.\footnote{%
In practice we use the $\ell'=\ell=0$ result for $\overline{R}_\Lambda$ in all cases, 
which is a further approximation, but one that we find makes a small numerical impact.}
Given $\Lambda$, we then obtain the cutoff for the sum using 
$n_{\rm max}=\gamma_k\Lambda$.

We can now explain why we include the factor of $H(\vec k)$ in $S$.
As $|\vec k|$ approaches the value where $H(\vec k)$ vanishes, $\gamma_k$ diverges.
This leads to an increase in $n_{\rm max}$, 
both from the factor of $\gamma_k$ in $\overline{R}_\Lambda$,
and because $n_{\rm max}/\Lambda =\gamma_k$.
However, this increase is more than compensated by the very rapid drop in $H(\vec k)$ near
the end point, so that $n_{\rm max}$ is always finite.

\subsection{Using cubic symmetries}
\label{eq:propFt}

Symmetries can be exploited to optimize the computation of $\widetilde{F}$.
It follows from Eq.~(\ref{eq:Ft}) that
$\widetilde{F}(R\vec{k})$  can be obtained from $\widetilde{F}(\vec{k})$ 
via an orthogonal transformation for any cubic-group transformation $R\in O_h$,
\begin{align}
    \widetilde{F}(R\vec{k}) = \mc{D}(R) \widetilde{F}(\vec{k}) \mc{D}(R)^T\,. 
    \label{eq:Ftransf}
\end{align}
Here $\mc{D}(R)$ is the Wigner D-matrix defined in Eq.~(\ref{eq:Wigner_D_1}).
Thus once one has computed $\widetilde{F}(\vec{k})$ for some finite-volume momentum 
$\vec k$, one can use Eq.~(\ref{eq:Ftransf}) to
obtain $\widetilde{F}(\vec{k}')$ for all $\vec{k}'$ in the same momentum shell.
Furthermore, for each initial $\widetilde{F}(\vec{k})$ that one computes directly, any symmetries
of $\vec{k}$ can be used to simplify the construction of $\widetilde{F}(\vec{k})$.
In particular, if $R$ is in the little group of $\vec{k}$ (so that $R\vec{k}=\vec{k}$), 
then Eq.~\eqref{eq:Ftransf} says that 
$\widetilde{F}(\vec{k})$ is invariant under the transformation.
This often leads to linear relationships between several matrix elements
 $\widetilde{F}_{\ell'm',\ell m}(\vec{k})$, 
 in which case one need only compute the linearly-independent elements 
 in order to construct the full matrix.

\section{Further details of the projection onto cubic group irreps}
\label{app:projections}

We collect here some results that we have found useful in the computation of the
projection matrices and the determination of their properties.

\subsection{Computing $P_I$ efficiently} \label{app:P^I_comp}

The projector $P_I$ is defined in Eq.~(\ref{eq:P_I}). As explained in the main
text, it is block diagonal in momentum shells and in angular momentum,
with blocks $P_{I,o(\ell)}$.
Here we explain how to simplify the computation of $P_{I,o(\ell)}$
by reducing the sum in Eq.~(\ref{eq:P_I}), which
runs over all 48 elements of $O_h$, to a sum over the elements of the little
group of an element of the shell under consideration.

Let $\vec k'$ and $\vec k''$ be two elements of the orbit. Then, from Eqs.~(\ref{eq:P_I})
and (\ref{eq:URT}), we have 
\begin{align}
    \left[P_{I,o(\ell)}\right]_{k''k'} 
    &= \frac{d_I}{[O_h]} \sum_{R\in O_h} \chi_I(R) \delta_{k'_R k''} \mathcal{D}^{(\ell)}(R)\,,
    \label{eq:PIoell}
\end{align}
where $\delta_{k'_R k''}$ is unity if $R \vec k'= \vec k''$ and zero otherwise.
Thus the sum is restricted to those elements of $O_h$ that rotate $\vec k'$ into $\vec k''$. 
A convenient representation of these elements makes use of an (arbitrarily chosen)
canonical element of the orbit, denoted $\vec k$. Let $R_{L_k}$ be an element of the little
group $L_k$ of $\vec k$. Then all the elements of $O_h$ that rotate $\vec k'$ to $\vec k''$
can be written as $R_{k''k}R_{L_k}R_{kk'}$, 
where $R_{kk'}$ is any choice of transformation from $\vec{k}'$ to $\vec{k}$, 
and $R_{k''k}$ is any choice of transformation from $\vec{k}$ to $\vec{k}''$.
Thus the number of elements contributing to the sum in Eq.~(\ref{eq:PIoell}) is $[L_k]$,
the dimension of $L_k$. This allows us to rewrite the projector as
\begin{align}
    \left[P_{I,o(\ell)}\right]_{k''k'} 
    &= \frac{d_I}{[O_h]} 
    \sum_{R\in L_k} \chi_I(R_{k''k}RR_{kk'}) \mathcal{D}^{(\ell)}(R_{k''k}RR_{kk'})
    \\
    &= \frac{d_I}{N_o} 
    \mathcal{D}^{(\ell)}(R_{k''k}) \left[ \frac{1}{[L_k]} \sum_{R\in L_k} \chi_I(R_{k''k}RR_{kk'}) 
    \mathcal{D}^{(\ell)}(R) \right] \mathcal{D}^{(\ell)}(R_{kk'})\,, 
    \label{eq:P_general}
\end{align}
where $N_o=[O_h]/[L_k]$ is the number of elements in the orbit.

Once we have constructed the block projectors, we combine them into $P_I$
using Eq.~(\ref{eq:Pblocks}).
In practice, we want to reduce our original matrices ($M=\wt F$ etc.)
down to the part that lives in the projected subspace, which has dimension $d(P_I)$.
To do so, we evaluate the eigenvalues and eigenvectors of $P_I$.
Since $P_I$ is a projector, its eigenvalues $\lambda_i$ are either zero or unity.
We keep only the eigenvectors with unit eigenvalues,
for these span the projection subspace.
We orthonormalize the eigenvectors, and label them $\{\vec{v}_i\}_{i=1}^{d(P_I)}$.
The reduced matrix is then given by
\begin{equation}
M^{\rm red}_{ij} = \vec{v}_i^{\Tr} \cdot M \cdot \vec{v}_j\qquad (i,j\in 1-d(P_I))\,.
\label{eq:Mred}
\end{equation}

\subsection{Dimensions of irrep projection subspaces}
 \label{app:subspace}

As explained in the main text, in order to determine the number of eigenvalues of $M$ that
fall into a given irrep we need to compute the dimensions of the sub-block projectors,
\begin{equation}
d(P_{I,o(\ell)}) = \text{Tr }P_{I,o(\ell)}\,.
\label{eq:dimsubblock}
\end{equation}
Using the result for the projector, Eq.~(\ref{eq:P_general}), we find
\begin{align}
    d(P_{I,o(\ell)}) &= \sum_{\vec{k}'\in o} \text{Tr}\left[P_{I,o(\ell)}\right]_{k'k'}\,,
    \label{eq:dimsubblocka}
    \\
    &= \frac{d_I}{[L_k]}\sum_{R\in L_k} \chi_I(R) ~\text{Tr }\mathcal{D}^{(\ell)}(R)\,,
    \label{eq:dimsubblockb}
\end{align}
where the trace is only over the angular momentum indices, $m$, and
to obtain the second line we have used the cyclicity of the trace,
the fact that $R_{k'k}=R_{kk'}^{-1}$,
and the standard group-theoretic result $\chi(R' R R'^{-1})=\chi(R)$.
The resulting dimensions are collected in Table~\ref{tab:dI}.

\section{$a_2$ dependence of $\cM_{3,\thr}$}
\label{app:thra2}

In Sec.~\ref{sec:threshold}, we show that to determine
the $a_2$ dependence of the three-particle threshold energy,
we need to calculate the corresponding dependence of $\cM_{3,\thr}$.
The calculation is described in this appendix.

We begin by recalling from Ref.~\cite{\HSTH} that $\cM_{3,\thr}$
is defined by doing the minimal subtractions
necessary to have a finite quantity at threshold,
\begin{align}
\begin{split}
&\mathcal{M}_{3,\text{thr}} = \lim_{\delta \to 0} \Big[ \mathcal{M}_3(0, \left. {\hat{a}' }\right.^* ; 0, \hat a^*) \\ &- I_{0,\delta}(0, \left. {\hat{a}' }\right.^* ; 0, \hat a^*) - \int_\delta \frac{d^3k_1}{(2\pi)^3} \Xi_1(\vec k_1) - \int_\delta \frac{d^3k_1}{(2\pi)^3} \frac{d^3k_2}{(2\pi)^3} \Xi_2(\vec k_1, \vec k_2) \Big]\,. 
\label{eq:M3th}
\end{split}
\end{align}
Here $\cM_3$ is the three-particle scattering amplitude, expressed in terms of the
same variables used for $\Kdf$ in Eq.~(\ref{eq:Kdfnewvariables}).
The infrared (IR) divergence of $\cM_3$ at threshold is regularized using the $\delta$-scheme of
Ref.~\cite{\HSTH}, and three subtractions are needed in order to obtain a finite result.
The explicit expressions for $I_0$, $\Xi_1$ and $\Xi_2$ are given in Sec.~D of 
Ref.~\cite{\HSTH}, but will not be needed. All we need to know here
is that the subtractions depend on $a_0$, but not on $a_2$. Thus 
dependence on $a_2$ can only enter through $\cM_3$ itself.

To determine this dependence it is useful to recall the definition of the divergence-free 
scattering amplitude from Ref. \cite{Hansen:2015zga}, 
\begin{equation}
\mathcal{M}_{\text{df},3} (\vec p, \left. {\hat{a}' }\right.^* ; \vec k, \hat a^*)  
= \mathcal{M}_3(\vec p, \left. {\hat{a}' }\right.^* ; \vec k, \hat a^*)  
- \mathcal{D}(\vec p, \left. {\hat{a}' }\right.^* ; \vec k, \hat a^*) \,. 
\label{eq:Mdf}
\end{equation}
Here $\cD$ is a quantity that depends only on the two-particle scattering
amplitude $\cM_2$, whose expression will be given below.
It is chosen so as to subtract IR divergences from $\cM_3$ not only at threshold,
but also above. 
Reordering Eq.~(\ref{eq:Mdf}) as $\cM_3 = \cM_{\df,3}+\cD$, we note that, in general,
both contributions to $\cM_3$ depend on $a_2$. However, we also know from
Ref.~\cite{\HSQCb} that $\cM_{\df,3}$ vanishes when $\Kdf=0$. So, in this limit,
which is the case we consider numerically, $\cM_3=\cD$. This allows us to
calculate the $a_2$ dependence of $\cM_3$. 
We know that this dependence is finite at threshold because no $a_2$-dependent
subtraction was needed in Eq.~(\ref{eq:M3th}).

Before calculating the $a_2$ dependence of $\cM_3$, it is instructive to
relate the two subtracted versions of $\cM_3$,
\begin{equation}
\mathcal{M}_{3,\text{thr}} = 
\mathcal{M}_{\text{df},3} (0, \left. {\hat{a}' }\right.^* ; 0, \hat a^*) \Big|_{E=3m} 
+ \text{IR finite terms}\,.
\label{eq:IRterms}
\end{equation}
Since, as already noted, $\cM_{\df,3}$ vanishes when $\Kdf=0$, 
we see that it is the IR finite terms
that must contain the contribution to $\cM_{3,\thr}$ from higher partial waves.


What we have learned so far is that, for $\Kdf=0$,
the $a_2$ dependence of $\cM_{3,\thr}$ is given by that of $\cD$ evaluated at threshold.
Here we are interested in determining the leading dependence, which, as discussed
in the main text, is proportional to $a_2^5$.
This is given by
\begin{equation}
\cM_{3,\thr} \supset {a_2^5} \frac{d \cD_\thr}{d (a_2^5)}\Bigg|_{a_2=0}\,.
\label{eq:deltaMthr}
\end{equation}
Here $\cD_\thr$ is $\cD(\vec p,\hat a'^*;\vec k,\hat a^*)$
evaluated at $E=3m$ and $\vec p=\vec k=0$, so that there is
no dependence on $\hat a^*$ and $\hat a'^*$.
In fact, $\cD$ itself diverges in this limit, but the derivative in Eq.~(\ref{eq:deltaMthr}) does not.

To proceed, we need the explicit expression for $\cD$, given in Ref.~\cite{\HSQCb}.
It is obtained by symmetrizing over initial and final momenta 
the quantity $\cD^{(u,u)}$, which is given by 
\begin{equation}
\mathcal{D}^{(u,u)}(\vec p, \vec k) =  
-\mathcal{M}_2(\vec{p}) G^\infty(\vec{p},\vec k) \mathcal{M}_2( \vec p)  
+ \int_s \frac{1}{2\omega_s} \mathcal{M}_2(\vec{p})
G^\infty(\vec{p},\vec s)  \mathcal{M}_2( \vec s) G^\infty(\vec s,\vec k) \cM_2(\vec k)+ \dots\,.
 \label{eq:Duudef}
\end{equation}
Here $\int_s\equiv\int d^3s/(2\pi)^3$,
and the $\hat a^*$ and $\hat a'^*$ dependence has been
decomposed into partial waves, so that all quantities are implicitly matrices
in angular momentum space.
The spectator-momentum dependence is, however, kept explicit.
$\cM_2(\vec p)$ is the two-particle scattering amplitude for the dimer
when the spectator-momentum is $\vec p$. 
As for $\cK_2$ [see Eq.~(\ref{eq:K2mat})], it is diagonal in angular momentum 
\begin{equation}
\cM_2(\vec p)_{\ell' m';\ell m} = \delta_{\ell' \ell} \delta_{m' m} \cM_2^{(\ell)}\,.
\label{eq:M2decomp}
\end{equation}
It contains all (even) partial waves, including, in particular, the $d$-wave amplitude.
Finally, $G^\infty$ is given by
\begin{equation}
G^\infty_{\ell' m';\ell m}(\vec p, \vec k) \equiv
\frac{H(\vec{p}) H(\vec{k})}{b^2-m^2}
    \frac{4 \pi \mc{Y}_{\ell'm'}(\vec{k}^*) 
    \mc{Y}_{\ell m}(\vec{p}^{\,*}) }{q_{2,p}^{*\ell'}\,q_{2,k}^{*\ell} } 
    \,,
\label{eq:Ginf}
\end{equation}
where the kinematic quantities are the same as those appearing in Eq.~(\ref{eq:defG}).
Equation~(\ref{eq:Ginf}) is the relativistically-invariant version of the definition
given in Eq.~(81) of Ref.~\cite{\HSQCb}.

At threshold, only the $s$-wave part of $\cD^{(u,u)}$ is nonzero,
and symmetrization simply leads to an overall factor of $9$:
\begin{equation}
\frac{d \cD_\thr}{d (a_2^5)}\Bigg|_{a_2=0} 
= 9 \frac{d \cD^{(u,u)}_{00;00}(\vec 0, \vec 0)}{d (a_2^5)}\Bigg|_{a_2=0} \,.
\label{eq:DtoDuu}
\end{equation}
Looking at Eq.~(\ref{eq:Duudef}), we see that the $s$-wave projection implies that
the factors of $\cM_2$ on both ends are pure $s$-wave, so the first appearance of
$d$-wave scattering occurs in the second term. This gives the leading $a_2$-dependent
part of $\cD^{(u,u)}$:
\begin{equation}
\mathcal{D}^{(u,u)}_{00;00}(\vec 0,\vec 0) \supset  
I_d = \int_s \sum_{m=-2}^2 \frac1{2 \omega_s} \mathcal{M}_2^{(0)}(\vec 0) 
G_{00;2m}^\infty(\vec 0,\vec s\,)  \mathcal{M}_2^{(2)}(\vec s\,) 
 G_{2m;00}^\infty(\vec s,\vec 0) \mathcal{M}_2^{(0)}(\vec 0) \,.
 \label{eq:terma2}
\end{equation}
At leading order in perturbation theory in $a_0$ and $a_2$, $\cM_2^{(\ell)}=\K_2^{(\ell)}$,
with $\K_2^{(\ell)}$ given by the leading terms in Eqs.~(\ref{eq:K2swave}) and
(\ref{eq:K2dwave}). Inserting these results, we find that $I_d$ is IR and UV convergent,
so we do not need to actually take the derivative in Eq.~(\ref{eq:DtoDuu}).
By numerical evaluation we find 
\begin{equation}
\cM_{3,\thr} \supset 9 I_d = - \frac{14109.6}{m^2} \times (m a_0)^2 (m a_2)^5 
[1 + \cO(a_0) + \cO(a_2^5) ]\,.
\label{eq:M3thrres}
\end{equation}
This gives the leading term in the result (\ref{eq:M3thra2}) quoted in the main text.
The corrections in (\ref{eq:M3thrres}) 
arise from the subleading terms in the expressions for $\K_2^{(\ell)}$. 

We close with two further observations. First, a similar calculation with $\cM_2^{(2)}$
in $I_d$ replaced by any (even) higher-order amplitude leads to a nonzero contribution to 
$\cM_{3,\thr}$. Thus all higher partial waves contribute to $\Delta E_3$ at
$\cO(L^{-6})$. Second, higher-order terms in $\cD^{(u,u)}$ will also contribute to
$\cM_{3,\thr}$, although suppressed by powers of $a_\ell$. For example, the
first term not shown in Eq.~(\ref{eq:Duudef}), which has four factors of $\cM_2$,
leads to contributions to $\Delta E_3$ proportional to $a_0^3 a_2^5/L^6$
and $a_0^2 a_2^{10}/L^6$. These are of the same order 
as the corrections in Eq.~(\ref{eq:M3thrres}).

\section{Free solutions at the first excited energy}
\label{app:free1}
In this appendix we analyze free solutions to the quantization
condition in the $A_1^+$ irrep at the energy of the first excited 
noninteracting state, $E_1^\fr=m+2\omega_1$ (with $\omega_1=\sqrt{m^2 + k_L^2}$
and $k_L=2 \pi /L$).
Our aim is to understand when $F_3^{-1}$ has zeros at this energy, and to
determine their properties.
We work with box lengths $4 \lesssim mL \lesssim 6$
such that there are three active shells, although the final
result generalizes straightforwardly to any number of shells.

\subsection{$A_1^+$ irrep with $s$ and $d$ waves}

We first consider the case in which both $\ell=0$ and $\ell=2$ channels are included.
The matrices that enter into the quantization condition
are then six dimensional: the first three indices as in Eq.~(\ref{eq:3dim}),
and the remaining three from the third shell (one with $\ell=0$, and two with $\ell=2$;
see Table~\ref{tab:dI}). The free poles enter only in the first two shells, and are
proportional to 
\begin{equation}
p= \frac3{8 L^3m \omega_1^2 \left(E-E_1^\fr\right)}
\label{eq:p1}
\,.
\end{equation}
It will be useful to introduce the  vectors
\begin{equation}
\langle v_1 | = (1,0,0,0,0,0)\,,\quad
\langle v_2 | = \left(0, \sqrt{\frac16}, \sqrt{\frac56}, 0, 0, 0\right)\,,
\end{equation}
in terms of which the pole parts are given by 
[using Eqs.~(\ref{eq:Ft0}) and (\ref{eq:defG})]
\begin{align}
\wt F &= p \left(|v_1\rangle \langle v_1| + 2 | v_2\rangle \langle v_2 |\right)  + \cO(1) \,, 
\label{eq:app:F}
\\
\wt G &= 2p \left(|v_1\rangle \langle v_2| + 
 | v_2\rangle \langle v_1 | +  | v_2\rangle \langle v_2 |\right)  + \cO(1) \,.
 \label{eq:app:G}
\end{align}
As in the example discussed in Sec.~\ref{sec:free}, all we need
to know about the $\cO(1)$ contributions are that they are real and symmetric.
The relative factor of $\sqrt5$ between the two terms in $\langle v_2 |$ arises
from $Y_{20}(\hat z) = \sqrt5  Y_{00}(\hat z)$.
Combining the results for $\wt F$ and $\wt G$ we find
\begin{equation}
H = 5 p |w_1\rangle \langle w_1 |  + \cO(1)\,,\quad
|w_1\rangle = \sqrt{\tfrac15} \left( |v_1\rangle + 2 | v_2\rangle\right)\,.
\end{equation}
Thus, while the pole parts of $\wt F$ and $\wt G$ are both of rank 2, that of
$H$ is of rank 1, due to a partial cancelation.

In the following, we determine the pole structure of $F_3$, aiming to find a basis
in which this structure is simple.
We begin by changing to a more convenient basis, namely $|w_1\rangle$ combined with
\begin{equation}
|w_2\rangle = \sqrt{\tfrac15}\left(2 |v_1\rangle - | v_2\rangle\right)\,,
\end{equation}
and any choice of four other vectors filling out the orthonormal set.
We use a $1+1+4$ block notation, in which
\begin{equation}
H = p\begin{pmatrix} 5 & 0 & 0\\ 0 & 0 & 0\\ 0 & 0 & 0\end{pmatrix} 
+ \cO(1)
\ \ {\rm and}\ \ 
\wt F = p\begin{pmatrix} 
9/5 & -2/5 & 0\\
-2/5& 6/5 & 0 \\
 0 & 0 & 0 
\end{pmatrix} + \cO(1)\,.
\end{equation}
The inverse of $H$ has the form
\begin{equation}
H^{-1} =
\begin{pmatrix}
1/(5p) + \cO(1/p^2) & {\alpha_{12}}/{p} + \cO(1/p^2)  
  &{\vec \alpha_{13}}/{p} + \cO(1/p^2) \\
{\alpha_{12}}/{p}+\cO(1/p^2) & \alpha_{22} + \beta_{22}/p+\cO(1/p^2) & 
\vec \alpha_{23} + \cO(1/p) \\
{\vec \alpha_{13}^{\rm Tr}}/{p}+ \cO(1/p^2) &
\vec \alpha_{23}^{\rm Tr} + \cO(1/p) &
\stackrel{\leftrightarrow}\alpha_{33} + \cO(1/p)
\end{pmatrix}\,,
\end{equation}
where the quantities $\alpha_{12}$, $\alpha_{22}$, $\beta_{22}$ etc. are given in terms of
the $\cO(1)$ parts of $H$ in a way that is not pertinent. At this stage we can see that
$\wt F H^{-1} \wt F$ will contain a double pole proportional to $\alpha_{22}$ that will
have the form of an outer product, as well as a complicated single-pole term.
Performing the algebra we find
\begin{equation}
L^3 F_3 = p^2 \frac{4 \alpha_{22}}{25}
\begin{pmatrix}
-1 & 3 & 0\\
3 & -9 & 0 \\
0 & 0 & 0 
\end{pmatrix}
+
p 
\begin{pmatrix} a & b & -\vec z \\
b & -9a-6b & 3 \vec z\\
- \vec z^{\rm Tr}& 3 \vec z^{\rm Tr} & 0 
\end{pmatrix}
+ \cO(1)\,,
\end{equation}
where $a$, $b$ and $\vec z$ are given in terms of the $\alpha_{ij}$ and $\beta_{22}$.

Thus we have learned that $F_3$ contains a free double pole that can be written
\begin{equation}
- p^2 \frac{8 \alpha_{22}}{5 L^3} |x_1\rangle \langle x_1|\,,\qquad
\langle x_1 | 
=\sqrt{\tfrac1{10}} \left(-1,3,0\right)
= \sqrt{\tfrac12} \left(\langle v_1| - \langle v_2|\right)\,.
\label{eq:x1}
\end{equation}
The form of $|x_1\rangle$ is determined entirely by the pole structure of
$\wt F$ and $H$, although the overall coefficient is determined by the $\cO(1)$ parts.
Qualitatively we can say that although $F$ contains two independent poles
in this irrep, the $H^{-1}$ factor cancels one of them, leading to a left-over double pole.

We conclude by discussing the impact of the single pole contribution to $F_3$.
First we note that the coefficient of $p$ can be written as
\begin{equation}
-(8a+6b) |x_1\rangle\langle x_1|
- \frac1{\sqrt{10} N_2}\left( |x_1\rangle\langle x_2| + |x_2\rangle\langle x_1|\right)
\end{equation}
where the new normalized basis vector is
\begin{equation}
\langle x_2 | = N_2 \left(9a+3b, 3a+b, -10 \vec z\right)\,.
\end{equation}
Thus in the basis consisting of $|x_1\rangle$, $|x_2\rangle$ and four other orthonormal
vectors, $F_3$ has the $1+1+4$ block form 
\begin{equation}
F_3 = \begin{pmatrix}
f p^2 + g p & h p & 0 \\
h p  & 0  & 0 \\
0  & 0 & 0
\end{pmatrix} 
+ \cO(1)\,,
\end{equation}
where $f$, $g$ and $h$ are known constants.
This matrix can be diagonalized using a final, fourth basis. All we need to know here is
that, close to the pole, when $|p|\gg 1$, the shift in the eigenvalues due to the off-diagonal
$h p$ term is $\pm (hp)^2/(fp^2+gp) \sim  \cO(1)$. Thus in the final basis we have
 \begin{equation}
 F_3 = {\rm diag} 
 \left[f p^2 + g p + \cO(1) , \cO(1) , \cO(1) , \cO(1) , \cO(1) , \cO(1) \right]\,,
\end{equation}
and thus
 \begin{equation}
 F_3^{-1} = {\rm diag} 
 \left[ 1/(f p^2) + \cO(1/p^3) , \cO(1) , \cO(1) , \cO(1) , \cO(1) , \cO(1) \right]\,.
\end{equation}
Note that the size of the change to this final basis is proportional to $1/p$,
and thus vanishes at the zero of $F_3^{-1}$, so that the double zero lies
in the subspace spanned by $|x_1\rangle$.

In summary, we find that the single pole in $F_3$ is hidden beneath the double pole, 
such that in the inverse there is simply a double zero.
As $L$ is increased, there are more active shells, but the only change to the result
of this section 
is that the number of vanishing components of  $|x_1\rangle$ increases
[see Eq.~(\ref{eq:x1})]. The nonvanishing components are unchanged.

\subsection{$A_1^+$ irrep with only $s$ waves}

We have repeated the previous analysis for the case of only $\ell=0$ contributions
and three active shells.\footnote{%
This builds upon, and corrects, the analysis given in Appendix~C of Ref.~\cite{\HSQCa}.}
The matrices are now three dimensional, with one entry
per shell. We do not present the details, except to note that we follow the same steps as
in the previous subsection,
and find very similar conclusions aside from some changes in factors.
In particular, $F_3^{-1}$ still has a double zero, but this now lives
 in the space spanned by the vector
\begin{equation}
\langle x'_1| = \left(\sqrt{\tfrac67}, -\sqrt{\tfrac17}, 0\right) \,,
\end{equation}
where entries are ordered  as in Eq.~(\ref{eq:index2}).

\section{Properties of the isotropic approximation}
\label{app:iso}

This appendix recalls the definition of the isotropic approximation, 
describes its relation to the work of this paper,
and explains why the free solutions discussed in Sec.~\ref{sec:free}
are absent in this approximation.

The isotropic approximation was introduced in Ref.~\cite{\HSQCa} and
used in the numerical investigation of Ref.~\cite{\BHSnum}. 
It involves three components: 
(1) Only $\ell=0$ dimer channels are included in $\wt F$, $\wt G$, $\K_2$ and $\Kdf$;
(2) The resulting $\Kdf$ is taken to be independent of the spectator momentum, 
although dependence on the total energy $E$ is allowed;
(3) $F_3$ is projected onto the isotropic vector $| 1_K\rangle$,
which has a unit entry for every available choice of spectator momentum.
Note that the third step automatically picks out solutions in the $A_1^+$ irrep.

The isotropic approximation is thus a subset of an approach we
use several times in this paper,
namely restricting dimers to $\ell=0$, keeping only the isotropic part of $\Kdf$
in the expansion about threshold,
and projecting onto the $A_1^+$ irrep. 
We  refer to this as the ``low-energy $A_1^+$ approximation''.
The major difference is the absence of the third step---we do not project onto $|1_K\rangle$.
A minor difference is that, for $\Kdf$ to be purely isotropic,
 we must work only at linear order in the threshold expansion. 
Thus we can have at most a linear dependence of $\Kdf$ on $E^2$, 
as opposed to the arbitrary dependence allowed in the isotropic approximation. 

To explain the relationship between the two approximations,
we begin in the low-energy $A_1^+$ approximation.
All matrices, including $F_3$, are 
labeled by an index denoting the shell of the spectator momentum, 
as shown in Eq.~(\ref{eq:index2}).
All matrices have the same finite dimension given by the number of shells lying below our cutoff.
Since $\Kdf$ is isotropic, the quantization condition is\footnote{%
Note that $[F_3]^{-1} = [F_3^{-1}] $ because of the cubic symmetry of the components of 
$F_3$.}
\begin{equation}
\det\left( [F_3]^{-1} + |1_K\rangle \K^\iso \langle 1_K | \right)=0\,,
\label{eq:QClow}
\end{equation}
where the square braces indicate the $A_1^+$, $\ell=0$ matrix, and
\begin{equation}
\langle 1_K| = \left(1, \sqrt{6}, \sqrt{12}, \dots \right)
\end{equation}
in this basis. 
The entries here are the square roots of the sizes of the shells.
We can rewrite the determinant in the quantization condition as
\begin{equation}
\det\left( [F_3]^{-1} \right) \det\left(1 + [F_3]  |1_K\rangle \K^\iso \langle 1_K | \right)
= 
\frac{1 + \langle 1_K| [F_3]  |1_K\rangle \K^\iso}
{\det [F_3] }
\,,
\label{eq:QClow2}
\end{equation}
where we have used $\det(1+M) = \exp \tr \ln (1+M)$,
expanded in $M$, used the cyclicity of the trace, and resummed.
The isotropic approximation consists of keeping only the solutions arising from
the numerator on the right-hand side of Eq.~(\ref{eq:QClow2}), 
i.e. those satisfying
\begin{equation}
F_3^\iso \equiv \langle 1_K| [F_3]  |1_K\rangle  = - 1/\K^\iso \,.
\label{eq:QCiso}
\end{equation}
It follows from Eq.~(\ref{eq:QClow2})
 that any solution in the isotropic approximation is also a solution in
the low-energy $A_1^+$ approximation, barring an accidental, and unexpected,
juxtaposition with a zero of $\det([F_3])$.\footnote{%
 This holds also when $\K^\iso\to 0$, for then a solution to 
Eq.~(\ref{eq:QCiso}) implies that $[F_3]$ has a diverging eigenvalue, and thus that
$\det([F_3^{-1}]) \to 0$. }
Thus, aside from this caveat, which appears to be irrelevant in practice, 
all solutions to the low-energy $A_1^+$ approximation that require a nonzero $\K^\iso$
are also obtained in the isotropic approximation.

What are lost in the isotropic approximation are solutions to the quantization condition
(\ref{eq:QClow}) that arise when an eigenvector of $F_3$ diverges
(so that $\det([F_3])\to \infty$) while $F_3^\iso$ remains finite.
This requires that the corresponding eigenvector of $F_3$ is orthogonal to $|1_K\rangle$.
In our experience, this only happens for solutions that occur at free energies
(which, we recall, means one of the energies of three
noninteracting particles in the given volume),
although we do not know of a fundamental reason why this should be so.
Furthermore, it was found numerically in Ref.~\cite{\BHSnum}
that there are no free solutions in the isotropic approximation.
Taken together, these observations suggest
 that the isotropic approximation picks out all the non-free
solutions to the quantization condition 
obtained in the low-energy $A_1^+$ approximation. 
%
%

In the remainder of this appendix we explain analytically 
the result found numerically in Ref.~\cite{\BHSnum}, namely
that there are no free solutions in the  isotropic approximation.
 As discussed in Sec.~\ref{sec:free}, such solutions
occur first at $E=E_1^\fr$, and there yield a double pole in $\det(F_3)$ lying in the space
spanned by $| x'_1\rangle$, Eq.~(\ref{eq:span2}).
This pole is, however,  absent in the isotropic approximation because 
$\langle 1_K | x'_1\rangle = 0$, so the pole is removed from $F_3^\iso$.

Our aim is to generalize this argument to any excited free energy.
We will do so for $\vec P=0$, and for an excited state in which the three momenta,
labeled $\vec k$, $\vec p$ and $\vec b= -\vec k -\vec a$, 
lie in different shells, e.g. $\vec k= k_L (0,0,1)$, $\vec p= k_L (1,1,0)$
and $\vec b=k_L (-1,-1,-1)$, with $k_L=2\pi/L$. 
We denote the degeneracies of these shells by $N_1$, $N_2$, and $N_3$, respectively
($6$, $12$ and $8$ in our example).
For each choice of $\vec k$ from shell 1, we define $N_{12}$ 
as the number of choices of $\vec p$ from shell 2 
that can lead to a free solution, and define $N_{13}$ analogously.
By cubic symmetry $N_{12}$ and $N_{13}$ do not depend on the choice of $\vec k$
from shell 1. Clearly we have $N_{12}=N_{13}$, since each solution contains both
a $\vec p$ and $\vec b$. We define $N_{23}=N_{21}$ and $N_{31}=N_{32}$ 
analogously. The total degeneracy of free-particle solutions is then
\begin{equation}
N_{\rm sol} = N_1 N_{12} = N_2 N_{23} = N_3 N_{31}\,.
\end{equation}

As above, we denote the $\ell=0$, $A_1^+$ parts of $\wt F$ and $\wt G$ by
$[\wt F]$ and $[\wt G]$, which are indexed by the shell number.
The poles in these matrices occur only when both indices lie in one of the
three shells discussed above, and thus we can focus on this three-dimensional
subspace. 
The matrices in this subspace have the form
\begin{equation}
[\wt F] = p \begin{pmatrix} N_{12} & 0 & 0\\
0 & N_{23} & 0 \\ 0 & 0 & N_{31} \end{pmatrix} + \cO(1)\ \ {\rm and}\ \
[\wt G] = p \begin{pmatrix} 0 & \sqrt{N_{12} N_{23}} & \sqrt{N_{12} N_{31}}\\
\sqrt{N_{23} N_{12}}& 0 & \sqrt{N_{23}N_{31}} \\ 
 \sqrt{N_{31}N_{12}} & \sqrt{N_{31} N_{23}}& 0 \end{pmatrix} + \cO(1)\,,
\end{equation}
where
\begin{equation}
p = \frac1{8 L^3 \omega_k\omega_p\omega_b(E-\omega_k -\omega_p -\omega_b)}
\,.
\end{equation}
The coefficients in $[\wt F]$ count the number of choices of $\vec a$ in
Eq.~(\ref{eq:Ft0}) that lead to the pole.
For example, for the $(1,1)$ element, there are
$N_{12}+ N_{13}=2 N_{12}$ choices, which combines with the overall factor of $1/2$ in 
$\wt F$ to give the quoted result $N_{12}$.
To understand the form of $[\wt G]$ consider the $(1,2)$ element of the pole part. 
This arises from each of the $N_{\rm sol}$ solutions, multiplied by the
normalization factors for the $A_1^+$ projections, $1/\sqrt{N_1 N_2}$.
Then we use
\begin{equation}
\frac{N_{\rm sol}}{\sqrt{N_1 N_2}}
=
\sqrt{\frac{N_1 N_{12} N_2 N_{23}}{N_1 N_2}}
=
\sqrt{N_{12} N_{23}}
\end{equation}
to obtain the quoted result.

Combining, we find that the pole part of $H$ lives in a one-dimensional subspace,
\begin{align}
[H] &=[\wt F] + [\wt G] + [1/(2\omega \K_2)]
=  | W_1\rangle \lambda p \langle W_1 | + \cO(1)\,, 
\label{eq:Hresiso}
\\
\langle W_1 | &= \left(\sqrt{\frac{N_{12}}{\lambda}}, 
\sqrt{\frac{N_{23}}{\lambda}}, \sqrt{\frac{N_{31}}{\lambda}}\right)\,,
\quad
\lambda= N_{12}\!+\! N_{23} \!+\! N_{31}\,.
\end{align}
Here we are assuming that $\K_2$ does not have a zero at $E=E_1^\fr$.
It follows from Eq.~(\ref{eq:Hresiso})  that $[H]^{-1}$ has the form
(see, e.g., Eq.~(C14) of Ref.~\cite{\HSQCa}):
\begin{align}
[H]^{-1} &= | W_1\rangle \frac{1+\cO(1/p)}{\lambda p} \langle W_1 |
+ \cO(1/p) \sum_{i\ne 1} \left( | W_1\rangle \langle W_i |+  | W_i\rangle \langle W_1 |\right)
+ \sum_{i,j\ne 1}  | W_i\rangle \cO(1) \langle W_j |\,.
\end{align}
Here $|W_2\rangle$ and $|W_3\rangle$ are any choice for the
 other two members of an orthonormal basis of
which $|W_1\rangle$ is a member. Note that only the coefficient of the first term is known;
for all other terms only the power of $p$ is known.

We can now calculate the pole part of $F_3^\iso$, which requires projection
with $\langle 1_K |$. Within our subspace
\begin{equation}
\langle 1_K | \longrightarrow \left(\sqrt{N_1},\sqrt{N_2},\sqrt{N_3}\right)\,,
\end{equation}
from which it follows that
\begin{align}
\langle 1_K | [\wt F] &= p \sqrt{\lambda N_{\rm sol}} \langle W_1 | + \cO(1)\,, 
\\
\langle 1_K | [\wt F] |1_K\rangle
&= 3p N_{sol} + \cO(1) \,,
\\
\langle 1_K | [\wt F] [H]^{-1} [\wt F] |1_K\rangle
&= p N_{\rm sol} + \cO(1)\,,
\end{align}
and thus that
\begin{equation}
F_3^\iso = \frac1{L^3} \langle 1_K |\left( \frac{[\wt F]}3 -[\wt F] [H]^{-1} [\wt F] 
\right) |1_K\rangle
= \cO(1)\,.
\end{equation}
As claimed, all poles have canceled from $F_3^\iso$.

It is straightforward to generalize
this result to the case that two or more shells are the same,
and also to moving frames, i.e. $\vec P\ne 0$,
although we do not present the details here.

\section{Failure of Eq.~(\ref{eq:nozeros2}) for quadratic and cubic terms in the threshold expansion}
\label{app:needquartic}

As noted in the main text,
we find numerically that the following results hold,
\begin{equation}
[\Kdf^{(2)} (E_1^\fr)] |x_1\rangle = [\Kdf^{(3)} (E_1^\fr)] |x_1\rangle = 0 \,,
\label{eq:toshow}
\end{equation}
where  the superscript on $\Kdf$ indicates the order in the threshold expansion of $\Kdf$.
The vector $|x_1\rangle$ is given in Eq.~(\ref{eq:span3}), and the square brackets
indicate the $A_1^+$ projection of $\Kdf$ expressed in the $k\ell m$ basis. 
Our aim here is to give an analytic explanation for these results.

We can rewrite Eq.~(\ref{eq:toshow}), using the symmetry of $\Kdf$ 
and the form of $|x_1\rangle$, as
\begin{equation}
[\Kdf^{(2,3)}]_{1i} = \sqrt{\tfrac16}[\Kdf^{(2,3)}]_{2i} + \sqrt{\tfrac56} [\Kdf^{(2,3)}]_{3i} 
\ \ {\rm at} \ \ E=E_1^\fr\,, \ \ \forall i\,.
\label{eq:master0}
\end{equation}
The ordering of the indices is given in  Eq.~(\ref{eq:index3}).
We recall that the $\sqrt 6$ here arises because the first shell has $6$ elements,
while the $\sqrt 5$ arises because $Y_{20}(\hat z)=\sqrt5 Y_{00}$.
The superscript on $\Kdf$ indicates that the equation should hold for both the
quadratic and cubic terms in the threshold expansion.

We wish to demonstrate Eq.~(\ref{eq:master0}) for any choice of $i$.
To do so we first change notation, recalling from Sec.~\ref{subsec:K3_decomp}
that the $\vec k, \ell, m$ indices can be replaced by dependence on  $\vec k, \hat a^*$.
Here we are also replacing the spectator-momentum index
$k$ with $\vec k$, both in order to be more explicit, and because $\Kdf$ is an infinite-volume
quantity that is defined for all $\vec k$.
At first, we make this change only for the initial-state indices, leading to the
hybrid notation $\Kdf(E;\vec p, \ell',m'; \vec k ,\hat a^*)$.\footnote{%
We are abusing notation by using the same name, $\Kdf$, for the function expressed
in terms of different variables, but the number of indices uniquely determines which
choice of basis we are using.}
In terms of this new quantity, we claim that Eq.~(\ref{eq:master0}) holds for any choice of
the index $i$ if 
\begin{multline}
\Kdf^{(2,3)}(E_1^\fr; \vec 0, 0,0;  \vec k ,\hat a^*) 
+ c\, \Kdf^{(2,3)}(E_1^\fr; \vec 0, 2,0;  \vec k ,\hat a^*) =
\\
\Kdf^{(2,3)}(E_1^\fr; k_L \hat z, 0,0;\vec k ,\hat a^*) 
+ \sqrt5 \Kdf^{(2,3)}(E_1^\fr; k_L \hat z, 2,0; \vec k ,\hat a^*) \,,
\label{eq:masterall}
\end{multline}
is valid for all $\vec k$ and $\hat a^*$, and for one choice of $c$. 
To understand this, first note that (\ref{eq:masterall}) applies for an
arbitrary initial state, and this subsumes all possible values of
the finite-volume index $i$.
As for the final state, to obtain Eq.~(\ref{eq:master0}) we need to project onto the $A_1^+$
irrep. Doing so,  the second term on the left-hand side of Eq.~(\ref{eq:masterall}) 
vanishes, as can be
seen from the absence of an $\ell=2$ entry in the $A_1^+$ row of the $(000)$
shell column in Table~\ref{tab:dI}.
This is why it is sufficient if Eq.~(\ref{eq:masterall}) holds for one value of $c$.
The $A_1^+$ projections of the remaining three terms in Eq.~(\ref{eq:masterall})
leads to the three terms in Eq.~(\ref{eq:master0}).
The averaging over the first shell leads to the factors of $\sqrt 6$ in the latter result.
Note that to perform this averaging one must also use the rotation invariance of
$\Kdf$.
It is also important that $m'=0$ in the last term in Eq.~(\ref{eq:masterall}), 
since this is the component that lives in the $A_1^+$ irrep when the
spectator momentum lies in the $\hat z$ direction.

In the following, we demonstrate that Eq.~(\ref{eq:masterall}) holds if $c=\sqrt 5$.
There are three inputs needed for this demonstration.
The first is the observation that the same configuration of final-state particles
can contribute to both sides of Eq.~(\ref{eq:masterall}).
To explain this we need to write both initial and final states in the form
used prior to their decomposition into harmonics, so that we have
$ \Kdf(E;\vec p,\hat a'^*; \vec k,\hat a^*)$.
Then one can show, using permutation symmetry alone, that
\begin{equation}
\Kdf(E_1^\fr;\vec 0, \hat z; \vec k, \hat a^*)
=
\Kdf(E_1^\fr;k_L\hat z, \hat z; \vec k, \hat a^*)
\,.
\label{eq:equivalence}
\end{equation}
This result holds for any term in the threshold expansion of $\Kdf$
(or, indeed, for the entire quantity), 
and thus we do not include a superscript.
To understand Eq.~(\ref{eq:equivalence}),
note that the three particles in the final state 
have momenta $\vec 0$, $k_L\hat z$ and $-k_L \hat z$. 
Calling $\vec 0$ the spectator momentum yields the left-hand side of Eq.~(\ref{eq:equivalence}),
while calling $k_L \hat z$ the spectator yields the right-hand side.
Since both choices describe the same momentum configuration, they must be
equivalent due to the permutation symmetry of $\Kdf$.

The second input is that $\Kdf^{(2,3)}$ is either independent of, or quadratic in,
$\hat a'^*$.
This is explained in Sec.~\ref{subsec:K3_decomp},
and is in one-to-one correspondence with the fact that only $s$- and $d$-waves
contribute.

The final key input concerns angular averaging of a quadratic form:
\begin{equation}
\left(\hat n_i \hat n_j V_{ij}\right)\big|_{\ell=0} + 
\sqrt5\left(\hat n_i \hat n_j V_{ij}\right)\big|_{\ell=2,m=0}
=
\tfrac13 V_{ii} + \tfrac13 \left(2 V_{33}- V_{11}-V_{22}\right)
\\
=
V_{33}\,,
\label{eq:key2}
\end{equation}
where $V_{ij}$ is an arbitrary tensor.
In other words, the combination appearing on the left-hand side can be evaluated
by setting $\hat n=\hat z$. 
The same is trivially true for a quantity that is independent of $\hat n$.

Combining the second and third key inputs, we deduce that 
\begin{equation}
\Kdf^{(2,3)}(E; \vec p, 0,0;\vec k ,\hat a^*) 
+ \sqrt5 \Kdf^{(2,3)}(E; \vec p, 2,0; \vec k ,\hat a^*) 
=
\Kdf^{(2,3)}(E; \vec p, \hat a'^*\!=\!\hat z;\vec k ,\hat a^*) 
\label{eq:key3}
\end{equation}
holds for any choice of $E$ and $\vec p$. Applying this to both sides of Eq.~(\ref{eq:masterall}),
with $E=E_1^\fr$, and $\vec p=\vec 0$ for the left-hand side and $\vec p=k_1\hat z$
for the right-hand side, we find that Eq.~(\ref{eq:masterall}) with $c=\sqrt5$
is equivalent to the first key identity Eq.~(\ref{eq:equivalence}).
This establishes the desired result.

This derivation will fail for terms of quartic and higher order in $\Kdf$, since the
combination of $\ell'=0$ and $2$ parts that appears in Eq.~(\ref{eq:key3}) 
will no longer allow the replacement of $\hat a'^*$ with $\hat z$, implying that
Eq.~(\ref{eq:equivalence}) cannot be used.
For example, considering one of the terms that arises in quartic terms, we find
\begin{equation}
(\hat a'^* \cdot \hat n)^4\big|_{\ell'=0} + \sqrt5 (\hat a'^* \cdot \hat n)^4 \big|_{\ell'=2,m'=0}
\ne \hat n_z^4
\,.
\end{equation}
We have checked this numerically by decomposing the simplest of the quartic terms
and finding that Eq.~(\ref{eq:toshow}) does not hold.



\bibliographystyle{JHEP}      
\bibliography{bibtexref.bib}

\providecommand{\href}[2]{#2}\begingroup\raggedright\begin{thebibliography}{10}

\bibitem{Hansen:2014eka}
M.~T. Hansen and S.~R. Sharpe, \emph{{Relativistic, model-independent,
  three-particle quantization condition}},
  \href{https://doi.org/10.1103/PhysRevD.90.116003}{\emph{Phys. Rev.}
  {\bfseries D90} (2014) 116003}
  [\href{https://arxiv.org/abs/1408.5933}{{\ttfamily 1408.5933}}].

\bibitem{Hansen:2015zga}
M.~T. Hansen and S.~R. Sharpe, \emph{{Expressing the three-particle
  finite-volume spectrum in terms of the three-to-three scattering amplitude}},
  \href{https://doi.org/10.1103/PhysRevD.92.114509}{\emph{Phys. Rev.}
  {\bfseries D92} (2015) 114509}
  [\href{https://arxiv.org/abs/1504.04248}{{\ttfamily 1504.04248}}].

\bibitem{Hammer:2017uqm}
H.-W. Hammer, J.-Y. Pang and A.~Rusetsky, \emph{{Three-particle quantization
  condition in a finite volume: 1. The role of the three-particle force}},
  \href{https://doi.org/10.1007/JHEP09(2017)109}{\emph{JHEP} {\bfseries 09}
  (2017) 109} [\href{https://arxiv.org/abs/1706.07700}{{\ttfamily
  1706.07700}}].

\bibitem{Hammer:2017kms}
H.~W. Hammer, J.~Y. Pang and A.~Rusetsky, \emph{{Three particle quantization
  condition in a finite volume: 2. general formalism and the analysis of
  data}}, \href{https://doi.org/10.1007/JHEP10(2017)115}{\emph{JHEP} {\bfseries
  10} (2017) 115} [\href{https://arxiv.org/abs/1707.02176}{{\ttfamily
  1707.02176}}].

\bibitem{Briceno:2017tce}
R.~A. Brice\~no, M.~T. Hansen and S.~R. Sharpe, \emph{{Relating the
  finite-volume spectrum and the two-and-three-particle $S$ matrix for
  relativistic systems of identical scalar particles}},
  \href{https://doi.org/10.1103/PhysRevD.95.074510}{\emph{Phys. Rev.}
  {\bfseries D95} (2017) 074510}
  [\href{https://arxiv.org/abs/1701.07465}{{\ttfamily 1701.07465}}].

\bibitem{Mai:2017bge}
M.~Mai and M.~D{\"o}ring, \emph{{Three-body Unitarity in the Finite Volume}},
  \href{https://doi.org/10.1140/epja/i2017-12440-1}{\emph{Eur. Phys. J.}
  {\bfseries A53} (2017) 240}
  [\href{https://arxiv.org/abs/1709.08222}{{\ttfamily 1709.08222}}].

\bibitem{Briceno:2018aml}
R.~A. Brice\~no, M.~T. Hansen and S.~R. Sharpe, \emph{{Three-particle systems
  with resonant subprocesses in a finite volume}},
  \href{https://arxiv.org/abs/1810.01429}{{\ttfamily 1810.01429}}.

\bibitem{HSreview}
M.~T. Hansen and S.~R. Sharpe, \emph{{Lattice QCD and Three-particle Decays of
  Resonances}},  \href{https://arxiv.org/abs/1901.00483}{{\ttfamily
  1901.00483}}.

\bibitem{Dudek:2013yja}
{\scshape Hadron Spectrum} collaboration, J.~J. Dudek, R.~G. Edwards, P.~Guo
  and C.~E. Thomas, \emph{{Toward the excited isoscalar meson spectrum from
  lattice QCD}}, \href{https://doi.org/10.1103/PhysRevD.88.094505}{\emph{Phys.
  Rev.} {\bfseries D88} (2013) 094505}
  [\href{https://arxiv.org/abs/1309.2608}{{\ttfamily 1309.2608}}].

\bibitem{Bulava:2016mks}
J.~Bulava, B.~Fahy, B.~H{\"o}rz, K.~J. Juge, C.~Morningstar and C.~H. Wong,
  \emph{{$I=1$ and $I=2$ $\pi-\pi$ scattering phase shifts from $N_{\mathrm{f}}
  = 2+1$ lattice QCD}},
  \href{https://doi.org/10.1016/j.nuclphysb.2016.07.024}{\emph{Nucl. Phys.}
  {\bfseries B910} (2016) 842}
  [\href{https://arxiv.org/abs/1604.05593}{{\ttfamily 1604.05593}}].

\bibitem{Romero-Lopez:2018rcb}
F.~Romero-L\'opez, A.~Rusetsky and C.~Urbach, \emph{{Two- and three-body
  interactions in $\varphi ^4$ theory from lattice simulations}},
  \href{https://doi.org/10.1140/epjc/s10052-018-6325-8}{\emph{Eur. Phys. J.}
  {\bfseries C78} (2018) 846}
  [\href{https://arxiv.org/abs/1806.02367}{{\ttfamily 1806.02367}}].

\bibitem{Briceno:2017max}
R.~A. Briceno, J.~J. Dudek and R.~D. Young, \emph{{Scattering processes and
  resonances from lattice QCD}},
  \href{https://doi.org/10.1103/RevModPhys.90.025001}{\emph{Rev. Mod. Phys.}
  {\bfseries 90} (2018) 025001}
  [\href{https://arxiv.org/abs/1706.06223}{{\ttfamily 1706.06223}}].

\bibitem{Briceno:2018mlh}
R.~A. Brice\~no, M.~T. Hansen and S.~R. Sharpe, \emph{{Numerical study of the
  relativistic three-body quantization condition in the isotropic
  approximation}},
  \href{https://doi.org/10.1103/PhysRevD.98.014506}{\emph{Phys. Rev.}
  {\bfseries D98} (2018) 014506}
  [\href{https://arxiv.org/abs/1803.04169}{{\ttfamily 1803.04169}}].

\bibitem{Doring:2018xxx}
M.~D{\"o}ring, H.~W. Hammer, M.~Mai, J.~Y. Pang, A.~Rusetsky and J.~Wu,
  \emph{{Three-body spectrum in a finite volume: the role of cubic symmetry}},
  \href{https://doi.org/10.1103/PhysRevD.97.114508}{\emph{Phys. Rev.}
  {\bfseries D97} (2018) 114508}
  [\href{https://arxiv.org/abs/1802.03362}{{\ttfamily 1802.03362}}].

\bibitem{Mai:2018djl}
M.~Mai and M.~D{\"o}ring, \emph{{Finite-volume spectrum of $\pi^+\pi^+$ and
  $\pi^+\pi^+\pi^+$ systems}},
  \href{https://arxiv.org/abs/1807.04746}{{\ttfamily 1807.04746}}.

\bibitem{Luscher:1986n2}
M.~Luscher, \emph{{Volume Dependence of the Energy Spectrum in Massive Quantum
  Field Theories. 2. Scattering States}},
  \href{https://doi.org/10.1007/BF01211097}{\emph{Commun.Math.Phys.} {\bfseries
  105} (1986) 153}.

\bibitem{Luscher:1991n1}
M.~Luscher, \emph{{Two particle states on a torus and their relation to the
  scattering matrix}},
  \href{https://doi.org/10.1016/0550-3213(91)90366-6}{\emph{Nucl.Phys.}
  {\bfseries B354} (1991) 531}.

\bibitem{Andersen:2017una}
C.~W. Andersen, J.~Bulava, B.~Horz and C.~Morningstar, \emph{{The elastic
  $I=3/2$ $p$-wave nucleon-pion scattering amplitude and the $\Delta(1232)$
  resonance from $N_{\mathrm{f}}=2+1$ lattice QCD}},
  \href{https://arxiv.org/abs/1710.01557}{{\ttfamily 1710.01557}}.

\bibitem{Woss:2018irj}
A.~J. Woss, C.~E. Thomas, J.~J. Dudek, R.~G. Edwards and D.~J. Wilson,
  \emph{{Dynamically-coupled partial-waves in $\rho\pi$ isospin-2 scattering
  from lattice QCD}},  \href{https://arxiv.org/abs/1802.05580}{{\ttfamily
  1802.05580}}.

\bibitem{Blanton:2018guq}
T.~D. Blanton, R.~A. Brice\~{n}o, M.~T. Hansen, F.~Romero-L\'opez and S.~R.
  Sharpe, \emph{{Progress report on the relativistic three-particle
  quantization condition}},  in \emph{{36th International Symposium on Lattice
  Field Theory (Lattice 2018) East Lansing, MI, United States, July 22-28,
  2018}}, 2018, \href{https://arxiv.org/abs/1810.06634}{{\ttfamily
  1810.06634}}.

\bibitem{Hansen:2015zta}
M.~T. Hansen and S.~R. Sharpe, \emph{{Perturbative results for two and three
  particle threshold energies in finite volume}},
  \href{https://doi.org/10.1103/PhysRevD.93.014506}{\emph{Phys. Rev.}
  {\bfseries D93} (2016) 014506}
  [\href{https://arxiv.org/abs/1509.07929}{{\ttfamily 1509.07929}}].

\bibitem{Sharpe:2017jej}
S.~R. Sharpe, \emph{{Testing the threshold expansion for three-particle
  energies at fourth order in $\phi^4$ theory}},
  \href{https://doi.org/10.1103/PhysRevD.96.054515,
  10.1103/PhysRevD.98.099901}{\emph{Phys. Rev.} {\bfseries D96} (2017) 054515}
  [\href{https://arxiv.org/abs/1707.04279}{{\ttfamily 1707.04279}}].

\bibitem{Hansen:2016fzj}
M.~T. Hansen and S.~R. Sharpe, \emph{{Threshold expansion of the three-particle
  quantization condition}}, \href{https://doi.org/10.1103/PhysRevD.96.039901,
  10.1103/PhysRevD.93.096006}{\emph{Phys. Rev.} {\bfseries D93} (2016) 096006}
  [\href{https://arxiv.org/abs/1602.00324}{{\ttfamily 1602.00324}}].

\bibitem{atkins1970tables}
P.~W. Atkins, M.~S. Child and C.~S.~G. Phillips, \emph{Tables for group
  theory}, vol.~6. Oxford University Press Oxford, 1970.

\bibitem{Georgi:1982jb}
H.~Georgi, \emph{{Lie Algebras In Particle Physics. From Isospin To Unified
  Theories}}, {\emph{Front. Phys.} {\bfseries 54} (1982) 1}.

\bibitem{Beane:2007qr}
S.~R. Beane, W.~Detmold and M.~J. Savage, \emph{{n-Boson Energies at Finite
  Volume and Three-Boson Interactions}},
  \href{https://doi.org/10.1103/PhysRevD.76.074507}{\emph{Phys. Rev.}
  {\bfseries D76} (2007) 074507}
  [\href{https://arxiv.org/abs/0707.1670}{{\ttfamily 0707.1670}}].

\bibitem{Tan:2007bg}
S.~Tan, \emph{{Three-boson problem at low energy and implications for dilute
  Bose-Einstein condensates}},
  \href{https://doi.org/10.1103/PhysRevA.78.013636}{\emph{Phys. Rev.}
  {\bfseries A78} (2008) 013636}
  [\href{https://arxiv.org/abs/0709.2530}{{\ttfamily 0709.2530}}].

\bibitem{Luu:2011ep}
T.~Luu and M.~J. Savage, \emph{{Extracting Scattering Phase-Shifts in Higher
  Partial-Waves from Lattice QCD Calculations}},
  \href{https://doi.org/10.1103/PhysRevD.83.114508}{\emph{Phys. Rev.}
  {\bfseries D83} (2011) 114508}
  [\href{https://arxiv.org/abs/1101.3347}{{\ttfamily 1101.3347}}].

\bibitem{PhysRevA.95.032707}
P.~M.~A. Mestrom, J.~Wang, C.~H. Greene and J.~P. D'Incao, \emph{Efimov--van
  der waals universality for ultracold atoms with positive scattering lengths},
  \href{https://doi.org/10.1103/PhysRevA.95.032707}{\emph{Phys. Rev. A}
  {\bfseries 95} (2017) 032707}.

\bibitem{PhysRevA.86.062511}
J.~Wang, J.~P. D'Incao, Y.~Wang and C.~H. Greene, \emph{Universal three-body
  recombination via resonant $d$-wave interactions},
  \href{https://doi.org/10.1103/PhysRevA.86.062511}{\emph{Phys. Rev. A}
  {\bfseries 86} (2012) 062511}.

\bibitem{EFIMOV1970563}
V.~Efimov, \emph{Energy levels arising from resonant two-body forces in a
  three-body system},
  \href{https://doi.org/https://doi.org/10.1016/0370-2693(70)90349-7}{\emph{Physics
  Letters B} {\bfseries 33} (1970) 563 }.

\bibitem{Meissner:2014dea}
U.-G. Mei{\ss}ner, G.~R\'ios and A.~Rusetsky, \emph{{Spectrum of three-body
  bound states in a finite volume}},
  \href{https://doi.org/10.1103/PhysRevLett.117.069902,
  10.1103/PhysRevLett.114.091602}{\emph{Phys. Rev. Lett.} {\bfseries 114}
  (2015) 091602} [\href{https://arxiv.org/abs/1412.4969}{{\ttfamily
  1412.4969}}].

\bibitem{Bedaque:1998kg}
P.~F. Bedaque, H.~W. Hammer and U.~van Kolck, \emph{{Renormalization of the
  three-body system with short range interactions}},
  \href{https://doi.org/10.1103/PhysRevLett.82.463}{\emph{Phys. Rev. Lett.}
  {\bfseries 82} (1999) 463}
  [\href{https://arxiv.org/abs/nucl-th/9809025}{{\ttfamily nucl-th/9809025}}].

\bibitem{Yndurain:2002ud}
F.~J. Yndurain, \emph{{Low-energy pion physics}},
  \href{https://arxiv.org/abs/hep-ph/0212282}{{\ttfamily hep-ph/0212282}}.

\bibitem{Beane:2011sc}
{\scshape NPLQCD} collaboration, S.~R. Beane, E.~Chang, W.~Detmold, H.~W. Lin,
  T.~C. Luu, K.~Orginos et~al., \emph{{The I=2 pipi S-wave Scattering Phase
  Shift from Lattice QCD}},
  \href{https://doi.org/10.1103/PhysRevD.85.034505}{\emph{Phys. Rev.}
  {\bfseries D85} (2012) 034505}
  [\href{https://arxiv.org/abs/1107.5023}{{\ttfamily 1107.5023}}].

\bibitem{Dudek:2012gj}
J.~J. Dudek, R.~G. Edwards and C.~E. Thomas, \emph{{S and D-wave phase shifts
  in isospin-2 pi pi scattering from lattice QCD}},
  \href{https://doi.org/10.1103/PhysRevD.86.034031}{\emph{Phys. Rev.}
  {\bfseries D86} (2012) 034031}
  [\href{https://arxiv.org/abs/1203.6041}{{\ttfamily 1203.6041}}].

\bibitem{Fu:2013ffa}
Z.~Fu, \emph{{Lattice QCD study of the s-wave $\pi\pi $ scattering lengths in
  the I=0 and 2 channels}},
  \href{https://doi.org/10.1103/PhysRevD.87.074501}{\emph{Phys. Rev.}
  {\bfseries D87} (2013) 074501}
  [\href{https://arxiv.org/abs/1303.0517}{{\ttfamily 1303.0517}}].

\bibitem{Kurth:2013tua}
T.~Kurth, N.~Ishii, T.~Doi, S.~Aoki and T.~Hatsuda, \emph{{Phase shifts in $I=2
  \ {\pi}{\pi}$-scattering from two lattice approaches}},
  \href{https://doi.org/10.1007/JHEP12(2013)015}{\emph{JHEP} {\bfseries 12}
  (2013) 015} [\href{https://arxiv.org/abs/1305.4462}{{\ttfamily 1305.4462}}].

\bibitem{Helmes:2015gla}
{\scshape ETM} collaboration, C.~Helmes, C.~Jost, B.~Knippschild, C.~Liu,
  J.~Liu, L.~Liu et~al., \emph{{Hadron-hadron interactions from N$_{f}$ = 2 + 1
  + 1 lattice QCD: isospin-2 $\pi \pi$ scattering length}},
  \href{https://doi.org/10.1007/JHEP09(2015)109}{\emph{JHEP} {\bfseries 09}
  (2015) 109} [\href{https://arxiv.org/abs/1506.00408}{{\ttfamily
  1506.00408}}].

\bibitem{Akakiprivate}
J.-Y. Pang, J.-J. Wu, H.-W. Hammer, U.-G. Mei{\ss}ner and A.~Rusetsky,
  \emph{{Energy shift of the three-particle system in a finite volume}},
  \href{https://arxiv.org/abs/1902.01111}{{\ttfamily 1902.01111}}.

\bibitem{Detmold:2008fn}
W.~Detmold, M.~J. Savage, A.~Torok, S.~R. Beane, T.~C. Luu, K.~Orginos et~al.,
  \emph{{Multi-Pion States in Lattice QCD and the Charged-Pion Condensate}},
  \href{https://doi.org/10.1103/PhysRevD.78.014507}{\emph{Phys. Rev.}
  {\bfseries D78} (2008) 014507}
  [\href{https://arxiv.org/abs/0803.2728}{{\ttfamily 0803.2728}}].

\bibitem{Kim:2005gf}
C.~h. Kim, C.~T. Sachrajda and S.~R. Sharpe, \emph{{Finite-volume effects for
  two-hadron states in moving frames}},
  \href{https://doi.org/10.1016/j.nuclphysb.2005.08.029}{\emph{Nucl. Phys.}
  {\bfseries B727} (2005) 218}
  [\href{https://arxiv.org/abs/hep-lat/0507006}{{\ttfamily hep-lat/0507006}}].

\end{thebibliography}\endgroup



\end{document}